\documentclass[12pt,a4paper]{article}

\usepackage{algorithm}
\usepackage{tabularx} 
\usepackage{algpseudocode}
\usepackage{geometry}
\usepackage{bm}        
\usepackage{amssymb}   
\usepackage{array}
\usepackage{caption}
\RequirePackage{amsmath,amssymb}
\RequirePackage{mathrsfs}
\RequirePackage{amsfonts,bm,bbm}
\usepackage[unicode]{hyperref}
\hypersetup{hidelinks}
\RequirePackage{dsfont}
\RequirePackage{braket}
\RequirePackage{tabularx}

\RequirePackage{nicefrac}
\RequirePackage{graphicx}

\usepackage{booktabs}
\usepackage{longtable}
\usepackage{multirow}

\RequirePackage{colortbl}
\RequirePackage{xcolor}
\RequirePackage[symbol]{footmisc}
\usepackage{mathtools}
\usepackage{comment}
\usepackage{blkarray}
\usepackage{stackengine}
\usepackage{float}
\usepackage{rotating}
\usepackage{hhline}
\usepackage{tikz}
\usepackage{pdflscape}
\usepackage{tabularx}
\usepackage{makecell}

\usepackage[section]{placeins}

\usepackage[natbib=true,maxbibnames=99,giveninits=true,sorting=none,style=numeric]{biblatex}
\def\beqn{\begin{eqnarray}}
\def\eeqn{\end{eqnarray}}
\newcommand{\beas}{\begin{align}\begin{split}}
\newcommand{\eeas}{\end{split}\end{align}}

\usepackage{empheq}
\usepackage[most]{tcolorbox}
\usepackage{blkarray}
\newtcbox{\mymath}[1][]{%
    nobeforeafter, math upper, tcbox raise base,
    enhanced, colframe=blue!30!black,
    colback=blue!30, boxrule=1pt,
    #1}

\newcommand{\CC}[2]{C{#1\atopwithdelims[]#2}}
\newcommand{\Z}[2]{Z{#1\atopwithdelims[]#2}}

\newcommand{\ba}{\begin{eqnarray}}
\newcommand{\ea}{\end{eqnarray}}
\DeclareRobustCommand{\sqbinom}{\genfrac[]{0pt}{}}

\newcommand{\vth}{\vartheta} 
\newcommand{\vthb}{\bar{\vartheta}} 
\newcommand{\smb}[2]{\hspace{-0.075cm}\left[\begin{smallmatrix}#1\\#2\end{smallmatrix} \right]}

\DeclareMathSymbol{\mg}{\mathrel}{symbols}{"1D}

\newcommand{\one}{\mathds{1}}
\newcommand{\Sv}{\bm{S}}
\newcommand{\Bx}{\bm{x}}
\newcommand{\Bv}{\bm{b}}

\newcommand{\Bz}{\bm{z}}

\newcommand{\cN}{{\cal N}}

\newcommand{\beq}{\begin{equation}}
\newcommand{\eeq}{\end{equation}}
\newcommand{\barr}{\begin{array}}
\newcommand{\earr}{\end{array}}
\newcommand{\equ}[1]{\begin{gather} #1 \end{gather}}

\newcommand{\sfrac}[2]{\mbox{$\frac{#1}{#2}$}}
\newcounter{oldcounter}

\newcommand{\bder}{\bar\partial}
\newcommand{\bff}{{\bar{f}}}

\newcommand{\bw}{{\bar w}}

\newcommand{\byy}{{\bar y}}

\newcommand{\bgb}{{\bar\beta}}

\newcommand{\bgf}{{\bar\phi}}

\newcommand{\bgl}{{\bar\lambda}}

\newcommand{\bgps}{{\bar\psi}}
\newcommand{\bget}{{\bar\eta}}

\newcommand{\Bga}{{\boldsymbol \alpha}}
\newcommand{\Bgb}{{\boldsymbol \beta}}

\newcommand{\Bgx}{{\boldsymbol \xi}}

\newcommand{\Intr}{\mathbbm{Z}}

\newcommand{\ztwo}{\Intr_2 \! \times \! \Intr_2 }

\numberwithin{equation}{section}

\begin{document}
\begin{titlepage}
\setcounter{page}{1}

\rightline{LTH-1414}
\rightline{April 2026}

\vspace*{\fill}

\begin{center}
{\Large\bfseries
Classification of Pati--Salam\\[0.5em]
Asymmetric $\mathbb{Z}_2 \times \mathbb{Z}_2$ Heterotic String Orbifolds
\par}

\vspace{1.2cm}

{\large
Luke A. Detraux\textsuperscript{1}\footnote{E-mail address: ldetraux@liverpool.ac.uk},
Alon E. Faraggi\textsuperscript{1}\footnote{E-mail address: alon.faraggi@liverpool.ac.uk}
and
Benjamin Percival\textsuperscript{2}\footnote{E-mail address: b.percival@mmu.ac.uk}
\par}

\vspace{0.9cm}

{\itshape
\begin{tabular}{@{}c@{}}
\textsuperscript{1} Dept.\ of Mathematical Sciences, University of Liverpool, Liverpool L69 7ZL, UK\\[0.35em]
\textsuperscript{2} Dept.\ of Natural Sciences, Manchester Metropolitan University, M1 5GD, UK
\end{tabular}
\par}
\end{center}

\vspace*{\fill}

\begin{abstract}
\noindent
We develop a systematic classification of asymmetric $\Intr_2$ orbifold actions in Pati--Salam heterotic string vacua constructed in the free fermionic formulation. Starting from symmetric $\ztwo$ orbifold vacua with an $SO(10)$ GUT, we allow the Pati--Salam breaking vector to act asymmetrically on the internal degrees of freedom. 
The 
asymmetric orbifold action freezes 
geometrical moduli whilst inducing doublet--triplet splitting in the untwisted sector. 
Notably, this doublet--triplet splitting operates for any
asymmetric action, including pure asymmetric shifts that preserve
all geometric moduli, and is therefore independent of moduli
stabilisation. 
Classifying the breaking vector according to its twist action, we find six inequivalent classes of geometric moduli spaces characterised by 12, 8, 4 or 0 real untwisted moduli. Through combining these asymmetric twists with all compatible asymmetric shifts, 24 inequivalent cases are identified and characterised by their residual moduli content and internal Narain lattice. For each case we construct representative basis sets admitting three chiral generations, 
providing the starting point for further classification within each class.
We perform explicit GGSO (generalized Gliozzi–Scherk–Olive) phase enumerations in representative model classes with 12, 8, 4 and 0 moduli, classify the resulting $\cN=1$ and $\cN=0$ vacua according to phenomenological criteria and identify exophobic, phenomenologically viable models. 
We compute the partition function and corresponding one-loop vacuum energy at the free fermionic point in moduli space for each phenomenologically viable model across the four classes. As the number of geometrical moduli decreases, the number of distinct partition functions for these vacua collapses to a small number, reflecting a pronounced degeneracy under GGSO phase variations.
\end{abstract}

\vspace*{\fill}
\end{titlepage}

 \section{Introduction}\label{intro}

The Standard Model of particle physics is consistent with all the observational
subatomic data to date. 
Yet, the Standard Model 
is not easily reconciled with the gravitational data. 
The most glaring disagreement is with the
observed vacuum energy, which is predicted in the Standard Model
to be many orders of magnitude above observations. Similarly, 
the Standard Model does not provide sufficient matter to account for
the observational galactic and cosmological data. Furthermore, the Standard
Model is composed of a collection of ad hoc gauge and matter sectors
whose properties are parametrised in terms of a large number
of arbitrary parameters. The fundamental origin of these
parameters can only be sought by synthesising the
Standard Model with gravity. String Theory is a conceptually
mundane extension of the concept of a quantum point particle, 
with well defined physical attributes, that provides a
self--consistent framework for the synthesis of the
Standard Model with gravity. In that context, phenomenological
string models have been constructed since the mid--1980s
\cite{Candelas:1985en}.

Since the late eighties, the heterotic--string models in the
free fermionic formulation are among the most
studied string models
\cite{fsu5, fny, Antoniadis1990, Faraggi:1991jr, slm2, cfn}.
These free fermionic heterotic--string models correspond to
toroidal $\ztwo$ 
orbifolds \cite{z2z21, z2z22, z2z23, z2z24} 
and detailed dictionaries exist between the bosonic and fermionic
representations of these string vacua \cite{z2z25}.
The free fermionic $\ztwo$ 
orbifolds provide
benchmark models to explore the unification of
the Standard Model with gravity. While the early models
consisted of isolated and sporadic examples,
since the early 2000s, systematic computerised methods to classify
and analyse large classes of fermionic $\ztwo$ 
orbifold
models have been developed. 
These methods give rise to an abundance
of three-generation models with an underlying $SO(10)$ GUT group,
which is broken at the string scale to one of its subgroups
and contains the necessary Higgs representations
to reduce the gauge symmetry to that of the Standard Model.

The free fermionic classification methods are developed progressively.
The initial classification was performed with an unbroken
$SO(10)$ GUT group and solely with respect to the spinorial $\mathbf{16}$
and anti--spinorial $\overline{\mathbf{16}}$ representations of $SO(10)$
\cite{Faraggi:2004rq}.
It was subsequently extended to include vectorial $\mathbf{10}$ $SO(10)$
representations and led to the discovery of Spinor--Vector Duality
(SVD) \cite{fkr1, fkr2}
in the space of (2,0) heterotic--string compactifications.
Classifications of vacua in which SO(10) is broken to the Pati–Salam \cite{acfkr2}, Flipped SU(5) \cite{frs}, Standard-like \cite{slmclass} and left–right symmetric \cite{lrsclass} subgroups were subsequently developed.
The breaking of the $SO(10)$ GUT symmetry by discrete
Wilson 
lines in the string models results in the
appearance of exotic states in the spectrum
that carry fractional electric charge
\cite{exots1,exots2, exots3, alonexots, exots4, exots5},
which are severely constrained by observations
(see {\it e.g.} \cite{vhalyo}).
Exotic fractionally charged states therefore have to
be sufficiently
massive and/or sufficiently rare. 
While such states necessarily appear in the string spectrum \cite{exots3},
they may be confined to the massive spectrum. 
Such models were dubbed 
exophobic string
vacua in the string classifications. Exophobic string vacua
with three chiral generations appeared
in the Pati--Salam classification \cite{acfkr1} but not in any of the
other cases. In these cases, it is further required to impose that
the fractionally charged massless states arise
in vector--like representations and are not chiral, to ensure that they
can receive superheavy 
masses along supersymmetric 
flat directions.
The classification methodology is extended
here to analyse all the sectors that produce exotic states and
ensure that they are projected from the massless spectrum.

The free fermionic classification programme was extended to
non-supersymmetric heterotic string vacua
\cite{so10tclass, PStclass, type0,type0bar}.
This includes vacua with explicit supersymmetry (SUSY) breaking,
as well as vacua with spontaneously broken supersymmetry by
a stringy Scherk--Schwarz mechanism \cite{SS1,SS2,SS3,CDC2,CDC4}. 
The new element in these classifications is the proliferation of
sectors that can \textit{a priori} produce spacetime tachyons \cite{aafs}. 
The tachyonic states can be projected out by the GGSO (generalized Gliozzi–Scherk–Olive) projections;
and vacua that are tachyon free exist in ten dimensions
\cite{AGMV,Kawai:1986vd,DH}, as well as in four
dimensions \cite{ADM, aafs}. 
The task in the classification methodology is the efficient
enumeration of the tachyon producing sectors and the elimination
of tachyonic states from the physical spectrum.
Non-supersymmetric string constructions open numerous interesting questions. 
Their vacuum energy is generically non-vanishing and their perturbative stability must therefore be treated with caution. The vacuum energy can be analysed numerically as a function of the moduli \cite{ADM, fr1, fr2, fr3, AFMP1}, albeit in a restricted manner, revealing extrema with either negative or positive vacuum energy. In these analyses additional geometric moduli are typically fixed at the free fermionic point in moduli space and remain unstabilised; moving away from these points may reintroduce tachyonic instabilities. Vacua in which all geometric moduli are fixed do exist with positive vacuum energy, though typically of the order of the string scale \cite{BDDFP}. Furthermore, massless states may acquire one-loop corrections and become tachyonic.

Nevertheless, special constructions exhibit improved perturbative behaviour. A perturbatively stable vacuum at one loop was recently demonstrated in a type II model \cite{ADF}, albeit with negative vacuum energy. Earlier asymmetric type II orbifolds were shown to possess vanishing perturbative vacuum energy \cite{Kachru_1999,Harvey_1998,Shiu_1999}, with related heterotic dual descriptions highlighting the broader relevance of these constructions. More recently, toroidal asymmetric orbifolds with vanishing one-loop vacuum energy have been systematically constructed in \cite{larotonda2026asymmetricorbifoldsvanishingoneloop}. These constructions illustrate how asymmetric orbifold actions can significantly constrain, and in some cases cancel, perturbative contributions to the cosmological constant in non-supersymmetric string vacua.
In all these constructions, however, the dilaton vacuum expectation value typically remains unfixed and exhibits runaway behaviour, leaving open the question of the existence of a fully stable non-supersymmetric string vacuum.

Fixing the moduli in string vacua is a crucial issue in string phenomenology.
Moduli stabilisation can be done in the effective field theory limit
({\it e.g.} \cite{McAllister:2023vgy} and references therein). 
Alternatively, moduli can be fixed in string constructions
by using asymmetric boundary conditions with respect to the
internal worldsheet fermions \cite{moduli}. In a bosonic representation of the
internal degrees of freedom, such asymmetric assignments correspond
to asymmetric action on the internal coordinates. Such asymmetric
action is not possible in the effective field theory limits,
that treat the internal compactified space as a classical
geometry, and is intrinsically stringy, {\it i.e.} the internal geometry should be viewed as a quantum geometry. Another
view of the asymmetric action is in terms of $T$--folds that
mod out by duality symmetries of the internal
toroidal space \cite{tfolds, SGNPKV, twebs, Nibbelink_2021}.
Assignments
of asymmetric boundary conditions have additional phenomenological
consequences that we describe below.

The free fermionic classifications performed to date
have largely involved symmetric boundary conditions.
In such constructions there are twelve real untwisted moduli of the $\ztwo$ orbifold or, equivalently, the three complex structure and three K\"ahler moduli.
In the fermionic language these moduli correspond to quartic
Thirring interactions among the worldsheet fermions associated
with the six internal coordinates \cite{Bagger:1986cd, ChangKumar}. 
Each invariant Thirring interaction corresponds to a massless scalar that parametrises deformations away from the enhanced symmetry point \cite{LNY}.
The subset of Thirring terms that are invariant is reduced by asymmetric boundary conditions, which project out the corresponding moduli fields \cite{moduli}.
%
The asymmetric extension of the $\ztwo$ classification programme
was previously carried out for vacua with an unbroken
$SU(5)\times U(1)$ subgroup of $SO(10)$ \cite{asymmclass}.
In this case, a distinctive phenomenological feature of asymmetric
assignments is the cubic-level top--bottom Yukawa selection rule
\cite{Faraggi:1991be, yukawa},
which operates in flipped $SU(5)$ and Standard-like models.

In this paper we extend the classification methodology to Pati--Salam vacua with asymmetric boundary conditions. In these constructions, the asymmetric Pati--Salam breaking vector not only reduces the gauge symmetry but also induces a doublet--triplet splitting mechanism that operates in Pati--Salam and Standard--like models \cite{Faraggi:1994cv, dtsm}.
In symmetric models, all untwisted electroweak doublets are projected out while the corresponding colour triplets are retained. The Higgs doublets must therefore arise from twisted sectors, and the Yukawa couplings of the chiral generations originate from cubic interactions of the form $T_iT_jT_k$ with $i\ne j\ne k$. By contrast, asymmetric boundary conditions acting on one or more orbifold planes project the untwisted colour triplets while retaining the electroweak doublets. This gives rise to couplings of the form $T_iT_iU_i$ and suggests the possible existence of models in which vectorial twisted states are entirely projected out, in particular eliminating massless colour triplets from the spectrum.

Incorporating asymmetric boundary conditions introduces additional complications into the classification programme. In the symmetric case, the basis vectors can be fixed from the outset and distinct vacua are generated by varying the (generalised) GSO projection coefficients. In contrast, asymmetric constructions require a preliminary classification of the internal action of the Pati--Salam breaking vector itself. Different asymmetric twist assignments lead to inequivalent residual moduli spaces and determine which Narain--like symmetric shift vectors are compatible with modular invariance. The admissible basis set therefore depends on the specific asymmetric action.

Asymmetric models are thus organised into distinct subspaces characterised by their number of real geometric moduli (12, 8, 4, or 0) and by the manner in which asymmetric shifts modify their internal Narain lattice. The first task of the present analysis is to classify these inequivalent (modular invariant) asymmetric configurations. The second task is to develop a systematic methodology to determine a basis set and scan the GGSO phase space within each model class. As we show below, increasing asymmetry, in particular reducing the number of untwisted moduli, leads to a pronounced degeneracy of partition functions under GGSO phase choices. Highly asymmetric subspaces therefore exhibit a collapse in the number of phenomenologically distinct configurations.

Our paper begins in Section \ref{FFModelBuilding} with a discussion on how four dimensional heterotic string models are constructed in the free fermionic formalism. Building on this, Section \ref{sc:asymm_PS_models} introduces the general Pati--Salam 
breaking vector and the constraints it imposes. 
Section \ref{ModuliPS} discusses the orbifold action and moduli space of the asymmetric Pati--Salam vectors, which we use to classify each case in Section \ref{sc:Bga_class}. For each case we provide a representative Pati--Salam vector and an example basis set to build three-generation models in Section \ref{sc:3genBasis}. In Section \ref{Phenofeatures} we discuss how phenomenological models can be built from these vectors and their shared features. Sections \ref{ClassA}, \ref{ClassB}, \ref{ClassC} and \ref{ClassD} give explicit examples of constructions which retain 12, 8, 4 and 0 moduli, respectively. For each case we classify the resulting models via their phenomenological features and analyse the vacuum energy of the phenomenologically viable models. Finally we draw our conclusions in Section \ref{Conc}.





\section{Heterotic strings in the free fermionic formulation}\label{FFModelBuilding}
\subsection{The ten-dimensional free fermionic heterotic string}
The worldsheet conformal field theory of the ten dimensional heterotic superstring has a supersymmetric 
 side with the spacetime bosonic fields $X^\mu$ and their superpartners $\Psi^\mu$, with $\mu=1,...,8$ in the lightcone gauge. On the non-supersymmetric (right-moving) side of the theory, the spacetime bosonic fields $\bar{X}^\mu$ require an additional $c=16$ central charge contribution to cancel the conformal anomaly. This defines the rank 16 gauge sector, which can be realised by 32 chiral real fermions $\bff^a$, $a=1,...,32$, or 16 complex fermions $\bgl^A$, $A=1,...,16$. 
By convention, we denote with a bar the fields that we refer to as anti-holomorphic and right-moving. With an eye towards GUT model building we label these 16 complex gauge sector fermions as:
\beq 
 \bgl^A: \bgps^{1,...,5}, \ \bget^{1,2,3}, \ \bgf^{1,...,8},
\eeq 
and view the $\bgf^{1,...,8}$ as associated to the rank 8 hidden gauge group, the $\bgps^{1,...,5}$ generating an SO(10) GUT and the $\bget^{1,2,3}$ generating three U(1) currents. 

The free fermionic realisation requires, as essential input, the specification of boundary conditions for the fermionic fields as they propagate around non-contractible cycles of the genus-g worldsheet. To specify the boundary conditions, we construct a set of $n$ basis vectors, $\{\bm{v}_i\}^n_{i=1}= \mathscr{B}$, such that components $v(f)\in (-1,1]$ denote the boundary condition of each fermion, $f$. 

The other ingredient required to define a free fermionic model is a set of generalised GSO (GGSO) phases, $\CC{\bm{v}_i}{\bm{v}_j}$, which are the coefficients within the partition function when summing over the full spin structures defined on the worldsheet amplitude. The partition function of the free fermionic model is concisely expressed as
\begin{equation}
Z_f=\sum_{\bm{\alpha},\bm{\beta}\in \Xi} \CC{\bm{\alpha}}{\bm{\beta}} \Z{\bm{\alpha}}{\bm{\beta}},
\end{equation} 
where $\Xi$ is the (Abelian) additive group spanned by the basis vectors, $\bm{\alpha},\bm{\beta}=m_i \bm{v}_i$, and $m_i\in \Intr_2$ within the class of models investigated in this work. 
Furthermore, 
$Z  {\bm{\alpha} \atopwithdelims [] \bm{\beta}}$ are products of Jacobi theta functions, such that
\[
\Z{\bm{\alpha}}{\bm{\beta}}(\tau,\bar\tau)
=\prod_{f}
\left(
\frac{\vartheta\!\begin{bmatrix}
\alpha(f) \\[2pt] \beta(f)
\end{bmatrix}}{\eta(\tau)}
\right)^{\tfrac{1}{2}}
\;\;
\prod_{\bar f}
\left(
\frac{\overline{\vartheta}\begin{bmatrix}
\alpha(\bar f) \\[2pt] \beta(\bar f)
\end{bmatrix}}{\eta(\bar{\tau})}
\right)^{\tfrac{1}{2}} .
\]
Constraints from modular invariance on the basis vectors and GGSO phases, first derived in \cite{ABK1,ABK2,KLT}, are given in Appendix \ref{app:ABKrules}.

For all $\bm{\alpha}\in \Xi$,  there is a Hilbert space obtained by acting on the NS vacuum or the (doubly-degenerate) Ramond vacua. 
Once the basis vectors and GGSO phases are specified, the modular invariant Hilbert space $\mathcal{H}$ of states $|{S_{\bm{\alpha}}}>$ is found through implementing the one-loop GGSO projection according to:
\begin{equation}
\label{HilbertSpace}
    \mathcal{H}=\bigoplus_{\bm{\alpha}\in \Xi}\prod^{N}_{i=1}
    \left\{ e^{i\pi \bm{v_i}\cdot F_{\bm{\alpha}}}|S_{\bm{\alpha}}>=\delta_{\bm{\alpha}}
    \CC{\bm{\alpha}}{\bm{v_i}}^*
    |S_{\bm{\alpha}}>\right\}\mathcal{H}_{\bm{\alpha}},
\end{equation}
where $F_{\bm{\alpha}}$ is the fermion number operator and $\delta_{\bm{\alpha}}=\exp{(i\pi \alpha(\psi^\mu))}\in\{1,-1\}$ is the spin-statistics index.
The sector masses are given by
\begin{align}\label{massform}
\begin{split}
M_L^2&=-\frac{1}{2}+\frac{\bm{\alpha}_L \cdot\bm{\alpha}_L}{8}+N_L\\
M_R^2 &=-1+\frac{\bm{\alpha}_R \cdot \bm{\alpha}_R}{8}+N_R ,
\end{split}
\end{align}
where $N_L$ and $N_R$ are sums over left and right moving oscillator frequencies,
respectively
\begin{align}
    N_L&=\sum_{\lambda}\nu_\lambda+\sum_{\lambda^*}\nu_{\lambda^*}, \\
    N_R&=\sum_{\bar{\lambda}}\nu_{\bar{\lambda}}+\sum_{\bar{\lambda}^*}\nu_{\bar{\lambda}^* } .
\end{align}
Here $\lambda$ is a holomophic oscillator and $\bar{\lambda}$ is an antiholomorphic oscillator and the frequency is defined through the boundary condition in the sector $\bm{\alpha}$
\beq \label{freq}
\nu_\lambda = \frac{1+\alpha(\lambda)}{2}, \ \ \ \
\nu_{{\lambda^*}} = \frac{1-\alpha({\lambda^*})}{2}.
\eeq 
Physical states must satisfy the
Virasoro matching condition, $M_L^2=M_R^2$, such that massless states are those with $M_L^2=M_R^2=0$ and on-shell tachyons arise for sectors with $M_L^2=M_R^2<0$.

\subsection{Four dimensional heterotic model building}

In four dimensions, we write the spacetime fields as $X^\mu(z,\bar{z}),\psi^\mu(z)$, with $\mu=1,2$ in the lightcone gauge. An additional $c=9$ contribution is required on the holomorphic side, which can be generated by introducing 18 free real fermions 
\begin{equation}
\chi^{1,...,6},y^{1,...,6},w^{1,...,6},
\end{equation}
where we interpret the $\chi^{1,...,6}$ as the superpartners of the six compact bosonic coordinate fields $X^I$. Meanwhile, on the antiholomorphic side an additional $\bar{c}=6$ contribution can be generated by the real free fermions
\beq 
\byy^{1,...,6},\bw^{1,...,6}.
\eeq 
Model building then follows the same modular invariance rules as the 10D case, but applied to basis vectors which include boundary condition assignments for these additional free fermions. 

In order to maintain the $N=1$ SCFT algebra on the holomorphic side,
we ensure the preservation of the supercurrent 
\beq \label{eq:SCurrent_Bga}
T_F(z)=i\psi^\mu \partial X^\mu(z)+i\sum_{I=1}^6 \chi^Iy^Iw^I(z).
\eeq 

This four dimensional heterotic string can be viewed as the ten dimensional theory expanded around a different background given by a 4D spacetime and an internal 6D torus generated by the fermionic degrees of freedom $\{y^I,w^I \ | \ \bar{y}^I,\bar{w}^I\}$. 

\subsection{Bosonisation}\label{sc:bosonisation}

Through bosonisation of the fermionic fields, free fermionic superstrings can be translated to (Narain) orbifold compactifications. For the rank 16 gauge sector, the 16 complex fermions $\bgl^A$ can be bosonised into 16 chiral bosons $\bar{X}^A$ and we can write the equivalence between the 16 Cartan currents as: 
\beq  
\label{eq:AntiHolomorphicBosonisation}
\overline{J}^{A} =\ :\! (\bgl^{A})^* \bgl^{A} \!:\  \cong i\, \bder \bar{X}^{A}~.
\eeq
which generate the Kac–Moody algebra for $SO(32)$.

The internal real fermionic degrees of freedom, $\{y^I, w^I \ | \ \byy^I, \bw^I\}$, when bosonised, become compact chiral bosons describing the torus with radii fixed at the special self-dual (`free fermionic') point. Through bosonisation we have $i\partial X^I_L =:y^Iw^I: $, with this choice of fermion bilinear demanded by the preservation of the supercurrent. On the right-moving side 
the ``standard bosonisation'' is
\beq \label{eq:standard_bosonisation}
i\partial X_R^I \cong:\byy^I\bw^I:,
\eeq  
mirroring the holomorphic labelling.

\section{Pati--Salam models from asymmetric orbifolds} \label{sc:asymm_PS_models}

Asymmetric free fermionic models are appealing  for a number of reasons. 
First, they provide a partial solution to the issue of moduli stabilisation since some, or all, geometric moduli may be frozen at the free fermionic point.
Second, they give rise to striking phenomenological implications not present within symmetric models. 
A central example examined in this work is the Doublet-Triplet splitting mechanism \cite{dtsm,Faraggi:1994cv}, in which Higgs doublets can arise from the untwisted (NS) sector of the theory.
This is in contrast to symmetric models where only triplet states are obtained from the NS sector, which can lead to issues with proton decay \cite{Faraggi:1994cv}. 
A further example of phenomenological characteristic of asymmetric models is in generating Yukawa coupling selection, as discussed in refs. \cite{asymmclass,yukawa}.

\subsection{\texorpdfstring{$\mathbf{\Intr_2\times \Intr_2}$ NAHE models}{Z2 x Z2 NAHE models}}

The $\ztwo$ symmetric twist orbifold proves pivotal within realistic free fermionic model building. 
The `NAHE' basis set \cite{Ferrara:1987jr,Antoniadis:1987tv,NAHE2} is given by  
\begin{align}
\begin{split}\label{eq:NAHE}
    \one&=\{\text{all}\},\\
    \Sv&=\{\psi^\mu,\chi^{1,...,6}\}\\
	\Bv_1&= \{\psi^\mu, \chi^{12}, y^{3,4,5,6} \ | \ \byy^{3,4,5,6};\bgps^{1,...,5},\bget^1\}\\
	\Bv_2&= \{\psi^\mu,\chi^{34},y^{1,2},w^{5,6} \ | \ \byy^{1,2}, \bw^{5,6}; \bgps^{1,...,5},\bget^2\},\\
    \Bv_3&= \{\psi^\mu,\chi^{56},w^{1,2,3,4} \ | \ \bw^{1,2,3,4} ; \bgps^{1,...,5},\bget^3\}. 
\end{split}
\end{align}
With just the basis set $\{\one,\Sv\}$ we would have $\cN=4$ spacetime supersymmetry in four dimensions and an $SO(44)$ gauge group from the right-moving side.
It is useful to identify the NAHE basis vector combination
\beq\label{eq:bH} 
\bm{H}=\one+\Bv_1+\Bv_2+\Bv_3=\{\bgf^{1,...,8}\},
\eeq 
which is a $SO(16)$ hidden sector-generating element and could be used to replace $\Bv_3$ in the basis to help reinterpret the NAHE-derived models as bosonic orbifolds.
The $\{\Bv_1,\Bv_2\}$ vectors are then associated to symmetric $\ztwo$ twists reducing spacetime supersymmetry $\cN=4\rightarrow 1$, producing an $SO(10)$ gauge symmetry and three fixed orbifold planes that produce a geometric origin for obtaining three chiral generations \cite{NAHE2, z2z21}. At the level of the NAHE set, the sectors $\Bv_k$, $k=1,2,3$, each produce 16 copies of the spinorial $\mathbf{16}/\overline{\mathbf{16}}$ of the $SO(10)$ gauge factor.
The resultant NAHE--set NS--sector gauge bosons generate a $ SO(6)^3\times SO(10) \times SO(16)$ gauge group.

\subsection{\texorpdfstring{The Pati-Salam breaking vector $\Bga$}{The Pati-Salam breaking vector alpha}}

The novel part of our model building methodology is the addition of a Pati-Salam breaking vector, $\bm{\alpha}$, which allows for asymmetric internal fermions. Our aim is to construct a single vector of maximum utility that simultaneously:
breaks the $SO(10)$ gauge symmetry to the Pati-Salam subgroup; 
freezes geometric moduli and 
projects Higgs triplets.
This vector will take the following form:

\beq\label{eq:Bga_vec}
    \Bga = \bm{A}+\{ \bgps^{1,2,3},\bgf^{1,2}\} 
\eeq
where the vector $\bm{A}$ assigns the boundary conditions for the internal fermions
\begin{align}\label{eq:A_vec}
    \bm{A}= 
    \{A(y^{1,...,6},w^{1,...,6}) \ | \ A(\byy^{1,...,6},\bw^{1,...,6}); A(\bget^1),A(\bget^2),A(\bget^3)\}  , 
\end{align}
and the supercurrent constraint imposes $A(y^i)=A(w^i)$ for i=1,...,6.
In total we find the $\Bga$ boundary condition assignments are specified by $21$ free (binary) parameters. We note that we can also construct general $\bm{\alpha}$ vectors with four hidden fermions; however all orbifold actions in Table \ref{tb:Bga_class_B_2} are covered by the case with two hidden sector fermions. We give an example models with both two and four hidden fermions in Sections \ref{ClassA} - \ref{ClassD}.

\section{Moduli from $\ztwo$ Pati--Salam
  models}\label{ModuliPS}

\subsection{Symmetric $\ztwo$ orbifold moduli}

Restricting attention to the untwisted subspace given by
$\mathscr{B}=\{\one, \Sv\}$, the NS sector produces $6\times 22$
scalar fields (in addition to the dilaton):
\begin{align} \label{eq:h_moduli}
h_{ij}&=|\chi^i\rangle_L\otimes |\bar{y}^j \bar{w}^j\rangle_R,
  \quad i,j=1,\ldots,6,\\
h_{iA}&= |\chi^i\rangle_L\otimes |\bgl^A\bgl^{A*}\rangle_R,
  \quad A=6,\ldots,22,
\end{align}
which are the massless states in the Cartan subalgebras of
$\mathrm{SO}(12)$ and $\mathrm{SO}(32)$, respectively.  The $6\times
6$ scalar fields $h_{ij}$ are in one-to-one correspondence with the
metric $G_{ij}$ and anti-symmetric tensor $B_{ij}$ of a $T^6$
compactification, whilst the $6\times 16$ scalar fields $h_{iA}$
correspond to Wilson lines; together these moduli scalars parametrise
the coset space \cite{narain_86}
\beq
\frac{\mathrm{SO}(6,22)}{\mathrm{SO}(6)\times \mathrm{SO}(22)}.
\eeq
The two-dimensional worldsheet action for Abelian Thirring
interactions is written as \cite{
  Bagger:1986cd,ChangKumar, Chang:1988ci}
\begin{equation}
    S = \int d^2 z \; h_{ik}(X) \; J^i_L(z)\, J^k_R(\bar{z}),
\end{equation}
for $k=1,\ldots,22$.  We note that the moduli scalar fields $h_{ik}(X)$
are in one-to-one correspondence with the marginal operators
\beq\label{eq:Thirring_marginal_ops}
J^i_L J^j_R
  = {:}y^iw^i{:}\;{:}\byy^j \bw^j{:}
  \cong \partial X^i \,\bder \bar{X}^j
\eeq
for $j=1,\ldots,6$, and similarly for the Wilson lines generated from
the gauge-sector currents.


To understand the internal geometry, we consider the bosonisation discussed in section \ref{sc:bosonisation} and note that the internal space fields are bosonised as
\beq
\xi^i
  = \sqrt{\tfrac{1}{2}}(y^i+iw^i)
  = e^{iX_L^i},
\qquad
\bar{\xi}^i
  = \sqrt{\tfrac{1}{2}}(\byy^i+i\bw^i)
  = e^{iX_R^i},
\eeq
such that the internal bosonic coordinates are
$X(z,\bar{z})=X_L(z)+X_R(\bar{z})$.  These fields can then be complexified as
\beq\label{eq:Zpm_def}
Z^{\pm}_k = X^{2k-1} \pm iX^{2k},
\qquad
\psi^{\pm}_k = \chi^{2k-1} \pm i\chi^{2k},
\qquad (k = 1, 2, 3),
\eeq
where $Z^{\pm}_k$ are the complex coordinates of the six compactified
dimensions, now viewed as three complex planes, and $\psi^{\pm}_k$
are the corresponding superpartners\footnote{We note that an additional $1/\sqrt{2}$
normalisation factor in the definition of $Z^{\pm}_k$ may be added that would propagate into the normalisation of the
worldsheet moduli action.}.

From the bosonisation equations it can be observed how the boundary-condition assignments for the fermions determine the bosonic
coordinate transformations. Considering real boundary conditions
($\Intr_2$ orbifolds), the transformation rules are summarised in
Table~\ref{tb:Z2transfns} for symmetric actions.

\begin{table}[!ht]
\caption{\label{tb:Z2transfns}
  Summary of the relation between L--R symmetric internal real fermion
  boundary-condition assignments and symmetric bosonic $\Intr_2$
  coordinate transformations.  The period convention is fixed so that
  the lattice identification $X\sim X + 2\pi$ holds at the
  free-fermionic self-dual radius.}
\centering
\begin{tabular}{|c|c c|c|}
\hline
$(y^i,w^i)=(\byy^i,\bw^i)$
  & $X_L^i$ & $X_R^i$
  & $X^i = X^i_L + X^i_R$ \\
\hline
$(0,0)$ & $X_L + \pi$ & $X_R + \pi$ & $X + 2\pi$ \\
$(1,0)$ & $-X_L$       & $-X_R$       & $-X$ \\
$(0,1)$ & $-X_L + \pi$ & $-X_R + \pi$ & $-X + 2\pi$ \\
$(1,1)$ & $X_L$        & $X_R$        & $X$ \\
\hline
\end{tabular}
\end{table}

In terms of the complexified coordinates $Z^\pm_k$ of
eq.~(\ref{eq:Zpm_def}), the symmetric actions of
Table~\ref{tb:Z2transfns} translate into
\begin{equation}\label{eq:Zk_transform}
Z^\pm_k \;\to\; \pm\, Z^\pm_k + 2\pi \delta_1  + 2\pi i \delta_2 ,
\end{equation}
where $\delta_{1,2} = 0, 1$ signify possible shifts in the real and imaginary directions.

For the NAHE set (\ref{eq:NAHE}), we now consider the effect of the
basis vectors $\{\Bv_1,\Bv_2,\Bv_3\}$ on the internal geometry and moduli space.  Each acts as a twist in two complex planes whilst leaving one inert; for example $\Bv_1$ maps $Z_1\to Z_1$ and
$Z_{2,3}\to -Z_{2,3}$. The moduli space can be understood by investigating the action of these vectors on the Thirring marginal
operators (\ref{eq:Thirring_marginal_ops}) or, equivalently, on the
NS scalar fields of the type (\ref{eq:h_moduli}). We find that only the 12 Thirring terms
\beq
J^{i}_L\bar{J}^{j}_R,\; i,j\in\{1,2\};\qquad
J^{i}_L\bar{J}^{j}_R,\; i,j\in\{3,4\};\qquad
J^{i}_L\bar{J}^{j}_R,\; i,j\in\{5,6\},
\eeq
are invariant, which match up with the subset of the scalar fields
\beq
h_{ij}=|\chi^i\rangle_L\otimes |\bar{y}^j \bar{w}^j\rangle_R
\begin{cases}
  i,j=1,2 \\
  i,j=3,4 \\
  i,j=5,6
\end{cases}.
\eeq
The resultant geometric moduli space for the $\ztwo$ orbifold is then
\beq\label{eq:sym_modspace}
\left(\frac{\mathrm{SO}(2,2)}{\mathrm{SO}(2)\times \mathrm{SO}(2)}\right)^3.
\eeq

This 12-dimensional coset accounts only for the untwisted geometric
moduli within the three orbifold planes; the Thirring terms associated
to Wilson lines are GGSO projected out by the $\{\Bv_1,\Bv_2,\Bv_3\}$ basis vectors \cite{LNY}.

To translate these 12 real scalar fields $h_{ij}$ to the three K\"{a}hler and three complex-structure moduli of the $\ztwo$ orbifold, we complexify the fields in each plane. The first complex plane gives:
\begin{align}
    \begin{split}
        H_1^{(1)}&=\frac{1}{\sqrt{2}}(h_{11}+ih_{21})
          =\frac{1}{\sqrt{2}}|\chi^1+i\chi^2\rangle_L
           \otimes |\bar{y}^1 \bar{w}^1\rangle_R,\\
        H_2^{(1)}&=\frac{1}{\sqrt{2}}(h_{12}+ih_{22})
          =\frac{1}{\sqrt{2}}|\chi^1+i\chi^2\rangle_L
           \otimes |\bar{y}^2 \bar{w}^2\rangle_R.
    \end{split}
\end{align}
These are then used to construct the K\"{a}hler and complex-structure moduli
\begin{align}\label{eq:TU_states}
    \begin{split}
        T_1 &= \frac{1}{\sqrt{2}}\bigl(H_1^{(1)}-iH_2^{(1)}\bigr)
              = \frac{1}{\sqrt{2}}|\chi^1+i\chi^2\rangle_L
                \otimes |\bar{y}^1 \bar{w}^1-i\bar{y}^2 \bar{w}^2\rangle_R,\\
        U_1 &= \frac{1}{\sqrt{2}}\bigl(H_1^{(1)}+iH_2^{(1)}\bigr)
              = \frac{1}{\sqrt{2}}|\chi^1+i\chi^2\rangle_L
                \otimes |\bar{y}^1 \bar{w}^1+i\bar{y}^2 \bar{w}^2\rangle_R,
    \end{split}
\end{align}
and analogously for $T_{2,3}$ and $U_{2,3}$. We note that these complex moduli span the coset moduli space
\beq\label{eq:TU_modspace}
\left(\frac{\mathrm{SU}(1,1)}{U(1)}\otimes
      \frac{\mathrm{SU}(1,1)}{U(1)}\right)^3.
\eeq
This matches the standard worldsheet action for the untwisted moduli
of the $\ztwo$ orbifold, which reads
\begin{equation}\label{eq:moduli_action}
S' = \int d^2 z \sum_{i=1}^3
  \Bigl(
    T_{i}\,\partial Z_i^{+}\,\bder Z_i^{-}
  + U_{i}\,\partial Z_i^{+}\,\bder Z_i^{+}
  \Bigr) + \text{h.c.},
\end{equation}
in terms of the three $T$-type (K\"{a}hler) and three $U$-type
(complex-structure) moduli of the $\ztwo$ orbifold.  Note the \emph{distinct} right-moving chiralities: $T_i$ couples the
left-mover $\partial Z^+_i$ to the right-mover $\bder Z^-_i$, whereas $U_i$ couples $\partial Z^+_i$ to
$\bder Z^+_i$. This distinction follows directly
from the state construction in eq. (\ref{eq:TU_states}), as verified
by the bosonisation relation ${:}\byy^j\bw^j{:}=i\bder X^j_R$. 
We remark that the precise assignment of which chirality pairs with $T_i$ versus $U_i$ depends
on the sign convention adopted in the bosonisation; the convention here is consistent with the state definitions (\ref{eq:TU_states})
and matches ref. \cite{Halyo:1994kn}.

We note that the K\"{a}hler potential \cite{LNY} for $T_i$ and $U_i$ is given by
\begin{equation}\label{eq:Kahler_pot}
K(T_i,\bar{T}_i,U_i,\bar{U}_i)
= -\sum_{i=1}^{3} \log\!\left(1 - T_i \bar{T}_i\right)
  -\sum_{i=1}^{3} \log\!\left(1 - U_i \bar{U}_i\right).
\end{equation}

Here $T_i$ and $U_i$ are Poincar\'{e} \emph{disk} coordinates for
$\mathrm{SU}(1,1)/\mathrm{U}(1)$, valid for $|T_i|,|U_i|<1$, in
which the Thirring coupling constants serve as the moduli fields
\cite{Halyo:1994kn, LNY}.  The standard upper half-plane coordinates $t_i$ of the effective supergravity description are related to these
by the Cayley transform $t_i = i(1-T_i)/(1+T_i)$, under which the
K\"{a}hler potential takes the familiar form $K=-\log(2\,\mathrm{Im}\,t_i)$.

\subsection[\texorpdfstring{(Non-)Geometric picture and moduli space
  from an asymmetric Pati--Salam vector}{}]%
{\texorpdfstring{(Non-)Geometric picture and moduli space from an
  asymmetric Pati--Salam vector}{(Non-)Geometric picture and moduli
  space from an asymmetric Pati--Salam
  vector}}\label{sc:Non_Geom_Pic}

We now consider the impact of a Pati--Salam breaking asymmetric vector
(\ref{eq:Bga_vec}) on the geometry and resultant moduli space. Since
$\Bga$ does not act on the RNS fermions $\psi^{\mu=1,2},\chi^{1,\cdots,6}$,
the holomorphic supercurrent ensures that $J^i_L\to J^i_L$ for all $i=1,\ldots,6$. With this in mind, all possible $\Bga$ asymmetric
actions on the right-moving internal coordinates are summarised in Table \ref{tb:Asymm_Bga_actions}.

Before presenting that table, we fix terminology.  An
\emph{asymmetric twist} on direction $i$ means $X_R^i\to -X_R^i$
(right-movers reflected, left-movers inert); an \emph{asymmetric shift} means $X_R^i\to X_R^i+\pi$ (right-movers shifted by a half-period); an \emph{asymmetric roto-translation} combines both.
These generalise the symmetric actions of Table \ref{tb:Z2transfns}, where left and right transform identically.

\begin{table}[!ht]
\caption{\label{tb:Asymm_Bga_actions}
  Summary of the relation between L--R asymmetric internal real fermion
  boundary-condition assignments and asymmetric bosonic $\Intr_2$
  transformations.  Coordinates are normalised so that the compact
  period of $X^i$ is $2\pi$ at the free-fermionic self-dual radius;
  the $\pi$ shifts in $X_L^i$ and $X_R^i$ then correspond to
  half-period lattice shifts.  The $Z^\pm$ column shows the net shift
  of the \emph{combined} coordinate $Z^\pm_k = X^{2k-1}\pm iX^{2k}$
  when the same boundary-condition pair is applied to \emph{both}
  directions $2k{-}1$ and $2k$ of a complex plane.  For asymmetric actions (rows 1, 2, 4, 5), because $X_L^i$ and $X_R^i$ transform
  differently, the coordinate $X^i = X_L^i + X_R^i$ is not a
  geometric eigenstate: the action generically mixes $Z^+$ and $Z^-$
  (i.e.\ left- and right-moving modes), reflecting the non-geometric
  (T-fold) nature of the construction~\cite{tfolds,Nibbelink_2021}.
  The table entries show only the overall shift of $Z^\pm$, following
  the notation of ref.~\cite{moduli}.
  }
\centering
\renewcommand{\arraystretch}{1.25}
\setlength{\tabcolsep}{5pt}
\begin{tabular}{|c|cc|c|c|}
\hline
$\alpha(y^i,w^i),\,\alpha(\byy^i,\bw^i)$
  & $X_L^i$ & $X_R^i$
  & $Z^{\pm}\,\to$
  & Asymmetric action \\
\hline
$(0,0),\,(1,0)$
  & $X_L^i+\pi$ & $-X_R^i$
  & $Z^{\pm}+2\pi+2\pi i$
  & Asymmetric twist \\
$(0,0),\,(0,1)$
  & $X_L^i+\pi$ & $-X_R^i+\pi$
  & $Z^{\pm}+2\pi$
  & Asymmetric roto-translation \\
$(0,0),\,(1,1)$
  & $X_L^i+\pi$ & $X_R^i$
  & $Z^{\pm}+2\pi$
  & Asymmetric shift \\
$(1,1),\,(1,0)$
  & $X_L^i$ & $-X_R^i$
  & $Z^{\pm}+2\pi i$
  & Asymmetric twist \\
$(1,1),\,(0,1)$
  & $X_L^i$ & $-X_R^i+\pi$
  & $Z^{\pm}$
  & Asymmetric roto-translation \\
$(1,1),\,(0,0)$
  & $X_L^i$ & $X_R^i+\pi$
  & $Z^{\pm}+2\pi i$
  & Asymmetric shift \\
\hline
\end{tabular}
\end{table}

In terms of moduli, we can determine the invariant Thirring terms by
considering which of the six antiholomorphic currents
$J^j_R={:}\byy^j\bw^j{:}$ are invariant under $\Bga$.  Any asymmetric assignment $\alpha(\byy^j,\bw^j)=(1,0)$ or $(0,1)$
generates a twist action and projects two Thirring terms. Due to modular invariance consistency we must have an even number of asymmetric twist assignments, so any additional vector can project
$0\bmod 4$ Thirring terms. We observe that we can either freeze all
4 real (2 complex) moduli in one of the three tori, or
freeze half the moduli in two tori.  For the latter case, the residual
coset moduli-space factor is
\[
\frac{\mathrm{SO}(2,1)}{\mathrm{SO}(2)}
\;\cong\;
\frac{\mathrm{SU}(1,1)}{\mathrm{U}(1)},
\]
where the right-hand side corresponds to the complexified description.
The surviving modulus in this case is an $H$-type modulus.

In Table \ref{tb:alpha_geometry} the five types of asymmetric twist
$\Bga$ case are distinguished by their residual moduli spaces. We label them Classes 1a), 1b), 2a), 2b) and 3), where the integer $n$ in the class label encodes the number of internal tori \emph{fully
frozen} by the asymmetric twist, i.e.\ $n=3-N_M/4$ (with $N_M$ the total number of real untwisted moduli). The sub-label a) or b)
distinguishes geometrically inequivalent configurations that share the same value of $N_M$.  
In addition, the $\Bga$ case with no
moduli freezing is dubbed Class 0), with $N_M=12$, is not listed in Table \ref{tb:alpha_geometry} since there is no asymmetric twist; it
is classified in Section \ref{sc:Bga_class} when $\Bga$ vectors with asymmetric shifts are explored.

\begin{table}[!ht]
\caption{Summary of geometric moduli spaces and K\"{a}hler potentials for the five classes of asymmetric $\mathbb{Z}_2$ twist orbifold
  actions.  A representative $\Bga$ vector is given for each class.
  The K\"{a}hler potential $K$ is expressed using Poincar\'{e} disk
  coordinates as in eq.~(\ref{eq:Kahler_pot}); each surviving complex
  modulus contributes a factor $-\log(1-|M|^2)$.  Here $T_i$ and
  $U_i$ denote the K\"{a}hler and complex-structure moduli of the
  $i$-th torus when it is fully untwisted ($M_i=4$), while $H_j$
  denotes the single surviving complex modulus of $H$-type from the
  $j$-th half-frozen torus ($M_j=2$): $H_j$ is a linear combination
  of $T_j$ and $U_j$ fixed by the asymmetric projection, and spans
  $\mathrm{SU}(1,1)/\mathrm{U}(1)$ with the same disk-form K\"{a}hler
  potential \cite{Halyo:1994kn}. The class label $n$ (integer part) encodes the number of fully frozen tori, $n=3-N_M/4$; see text for the a)/b) distinction.\label{tb:alpha_geometry}}
\footnotesize
\centering
\vspace{0.5em}
\renewcommand{\arraystretch}{1.45}
\setlength{\tabcolsep}{4pt}
\newcommand{\BodyStrut}{\vrule width 0pt height 6.0ex depth 6.0ex}
\newcolumntype{C}{>{\centering\arraybackslash}X}
\newcolumntype{L}{>{\raggedright\arraybackslash}X}
\renewcommand{\tabularxcolumn}[1]{m{#1}}
\newcommand{\BgaRep}[2]{%
  $\Bga=\left\{\begin{array}{@{}l@{}}#1;\\#2\end{array}\right\}$}
\begin{tabularx}{\linewidth}{|c|c|C|C|L|}
\hline
\begin{tabular}{c}Class\\Label\end{tabular}
&
\begin{tabular}{c}Real moduli\\$(M_1,M_2,M_3)$\end{tabular}
&
Coset (complex) moduli space
&
K\"{a}hler potential
&
Representative $\Bga$ vector
\\
\hline

\BodyStrut 1a)
& $(4,4,0)$
& $\left(\dfrac{SU(1,1)}{U(1)} \otimes
         \dfrac{SU(1,1)}{U(1)}\right)^{\!2}$
& $\displaystyle
  -\!\sum_{i=1}^{2}\!\bigl[
      \log(1{-}|T_i|^2)
    + \log(1{-}|U_i|^2)\bigr]$
& \BgaRep{\byy^{56}}{\bgps^{1,2,3},\bget^{2,3},\bgf^{1,2}}
\\
\hline

\BodyStrut 1b)
& $(4,2,2)$
& $\left(\dfrac{SU(1,1)}{U(1)}\right)^{\!\!2}
  \!\times\!
  \left(\dfrac{SU(1,1)}{U(1)}\right)^{\!\!2}$
& $\begin{aligned}
    &-\log(1{-}|T_1|^2)\\
    &-\log(1{-}|U_1|^2)\\
    &-\log(1{-}|H_2|^2)\\
    &-\log(1{-}|H_3|^2)
  \end{aligned}$
& \BgaRep{\byy^{46}}{\bgps^{1,2,3},\bget^{2,3},\bgf^{1,2}}
\\
\hline

\BodyStrut 2a)
& $(4,0,0)$
& $\dfrac{SU(1,1)}{U(1)} \otimes \dfrac{SU(1,1)}{U(1)}$
& $-\log(1{-}|T_1|^2)-\log(1{-}|U_1|^2)$
& \BgaRep{\bw^{34},\byy^{56}}{\bgps^{1,2,3},\bget^{2},\bgf^{1,2}}
\\
\hline

\BodyStrut 2b)
& $(2,2,0)$
& $\left(\dfrac{SU(1,1)}{U(1)}\right)^{\!2}$
& $-\log(1{-}|H_1|^2)-\log(1{-}|H_2|^2)$
& \BgaRep{\bw^{24},\byy^{56}}{\bgps^{1,2,3},\bget^{2},\bgf^{1,2}}
\\
\hline

\BodyStrut 3)
& $(0,0,0)$
& $1$\newline(All geometric\newline moduli frozen)
& $0$\newline(no contribution)
& \BgaRep{\byy^1\bw^6,\bw^{24},\byy^{35}}{\bgps^{1,2,3},\bgf^{1,2}}
\\
\hline
\end{tabularx}
\end{table}

\subsection{Additional asymmetric shifts}\label{sc:add_asymm_shifts}

Within each Class 0)--3) there are possible additional asymmetric shift actions that modify the underlying geometric Narain lattice.  In the free fermionic formulation, this modification of the lattice can
be considered in terms of the standard set of symmetric lattice shift vectors $\bm{e}_{1,...,6}$, where we find that an asymmetrically shifted
or twisted direction $i$ renders the corresponding symmetric shift
vector $\bm{e}_i$ incompatible with modular invariance.

From the real moduli $(M_1,M_2,M_3)$ we can write the number of twisted (or roto-translated) directions in each internal torus as
\beq \label{eq:N_t}
(N_t^1,N_t^2,N_t^3)
= \Bigl(2-\tfrac{M_1}{2},\;2-\tfrac{M_2}{2},\;2-\tfrac{M_3}{2}\Bigr).
\eeq
It is then useful to define the following quantity
\beq
\mathsf{E}=(E_{12},E_{34},E_{56})=(E_1+E_2,\;E_3+E_4,\;E_5+E_6),
\eeq
which counts the number of symmetric shift vectors compatible with the
choice of $\Bga$ in each torus, such that
\begin{align}\label{eq:Eis}
E_i=\begin{cases}
  1 & \text{if } \Bga \cdot \bm{e}_{i}=0,\\
  0 & \text{else},
\end{cases}
\end{align}
for $i\in\{1,\ldots,6\}$. From these definitions, the number of asymmetric shifts in each torus for a particular $\Bga$ vector is
\beq \label{eq:N_as}
(N_{as}^1,N_{as}^2,N_{as}^3)
= \bigl(2-N_t^1-E_{12},\;2-N_t^2-E_{34},\;2-N_t^3-E_{56}\bigr).
\eeq
From these definitions, we expect to have up to 4 asymmetric shifts
for classes 1a) and 1b), up to 2 for classes 2a) and 2b), and no
asymmetric shifts for class 3).  Some numbers of asymmetric shifts will not be compatible with modular invariance, such as having a single asymmetrically shifted direction.  In Table \ref{tb:Bga_class_B_2}
of Section~\ref{sc:Bga_class} we present a classification of inequivalent $\Bga$ vectors with respect to their class 0)--3) and asymmetric-shift content.

\section{Classification of $\Bga$} \label{sc:Bga_class}

Having given details on the geometry, moduli spaces and compatible additional vectors resulting from asymmetric Pati--Salam breaking vector $\Bga$, a classification can now be performed for $\Bga$. The modular invariance constraints on $\Bga$ is considered in section \ref{sc:MI_Bga} and the additional key feature Doublet-Triplet splitting is explained in section \ref{sc:DTS}. Full classification tables for the PS-breaking vector $\Bga$ are then presented in section \ref{sc:Bga_Class_tables} with respect to its retained moduli, number of retained moduli and untwisted doublets/triplets. 

\subsection{Modular invariance conditions for $\Bga$}
\label{sc:MI_Bga}

To classify the basis vectors $\Bga$ we impose the modular invariance consistency
conditions summarised in eq.~(\ref{eq:ABKbvecs}).  

We isolate the left- and right-moving boundary-condition
sub-(bit)vectors associated with the internal fermions of $\Bv_{1,2,3}$:
\begin{align}\label{eq:AL_AR}
\bm{A}^{(1)}_L &\equiv A(y^{3,4,5,6}), &
\bm{A}^{(1)}_R &\equiv A(\byy^{3,4,5,6}), \\
\bm{A}^{(2)}_L &\equiv A(y^{1,2},w^{5,6}), &
\bm{A}^{(2)}_R &\equiv A(\byy^{1,2},\bw^{5,6}), \\
\bm{A}^{(3)}_L &\equiv A(w^{1,2,3,4}), &
\bm{A}^{(3)}_R &\equiv A(\bw^{1,2,3,4}),
\end{align}
each being a four-component binary vector.
We define the corresponding occupation numbers as their Hamming weights,
\begin{align}
\delta^j_L &\equiv \|\bm{A}^{(j)}_L\|_1, &
\delta^j_R &\equiv \|\bm{A}^{(j)}_R\|_1,
\end{align}
and their left–right difference
\begin{equation}\label{littledelta}
\delta^j \equiv \delta^j_L - \delta^j_R,
\qquad j=1,2,3.
\end{equation}
Additionally,
\beq
(k_1,k_2,k_3)\equiv(A(\bget^1),A(\bget^2),A(\bget^3)).
\eeq

In terms of these quantities, modular invariance implies
\begin{align}
\Bga \cdot \Bv_1 &\ \implies\ \delta^1 - 2k_1 = 2 \ \text{mod}\ 4 , \label{eq:MI_Ab1}\\
\Bga \cdot \Bv_2 &\ \implies\ \delta^2 - 2k_2 = 2 \ \text{mod}\ 4 , \label{eq:MI_Ab2}\\
\Bga \cdot \Bv_3 &\ \implies\ \delta^3 - 2k_3 = 2 \ \text{mod}\ 4 , \label{eq:MI_Ab3}\\ 
\Bga \cdot \Bga &\ \implies\ 
\delta^1+\delta^2+\delta^3 - 2(k_1+k_2+k_3) 
= 2 \ \text{mod}\ 8. 
\label{eq:MI_AA}
\end{align}

Once $\Bga$ is fixed, only certain additional vectors discussed in
section~\ref{sc:additional_vecs} remain compatible with modular invariance,
reflecting the underlying Narain lattice geometry of that choice.

\subsection{Doublet-triplet splitting}\label{sc:DTS}

Asymmetric boundary-condition assignments within $\Bga$ generate a
selection mechanism for triplets and doublets arising from the NS sector
in $\ztwo$-derived Pati–Salam models \cite{Faraggi:1994cv, dtsm}.  
The NS sector produces three vectorial $\mathbf{10}$ representations of
$SO(10)$,
\begin{align}
U^1&=\{\chi^{12}\}\{\bget^{1}\}\{\bgps^{a(*)}\}|0\rangle,\\
U^2&=\{\chi^{34}\}\{\bget^{2}\}\{\bgps^{a(*)}\}|0\rangle,\\
U^3&=\{\chi^{56}\}\{\bget^{3}\}\{\bgps^{a(*)}\}|0\rangle,
\end{align}
with $a=1,\dots,5$.

For $U^j$ ($j=1,2,3$), the $\Bga$ GGSO projection gives
\beq
\begin{cases}
k_j=1 \ \ \ \text{Triplets ($a=1,2,3$) retained},\\
k_j=0 \ \ \ \text{Doublets ($a=4,5$) retained}.
\end{cases}
\eeq
In terms of the Hamming weights $(\delta^j_L,\delta^j_R)$,
this corresponds to
\beq
\begin{cases}
\text{Triplets retained if} \ \ 
(\delta^j_L,\delta^j_R)\in \{(4,0),(0,4),(0,0),(1,1),(2,2),(3,3),(4,4)\}, \\ 
\text{Doublets retained if} \ \ 
(\delta^j_L,\delta^j_R)\in \{(4,2),(2,4),(3,1),(1,3),(2,0),(0,2)\}.
\end{cases}
\eeq
We note that at the level of the Pati--Salam models, triplets reside in the vectorial $(6,1,1)$ representation, whereas the doublets reside in the bi--doublet $(1,2,2)$ representation. 
It is interesting to note that considering these constraints and the definitions of the asymmetric actions in Table \ref{tb:Asymm_Bga_actions}, that the doublet-triplet splitting can arise from an asymmetric shift or an asymmetric twist. Therefore, even Class 0) (12 geometric moduli) models with asymmetric shifts can admit this doublet-triplet splitting within the three orbifold planes. 

The number of untwisted triplet pairs and bi--doublets can be defined as
\begin{align}\label{N_T_and_N_D}
N_T &= \sum_{j=1}^3 k_j,\\
N_D &= 3 - N_T. 
\end{align}
The surviving untwisted doublets are candidates for the Standard Model Higgs. Whilst the triplets are more problematic, and can lead to proton decay channels, they can be utilised in the missing partner mechanism to give mass to exotic or twisted sector triplets that may arise elsewhere in our model \cite{antoniadis1989supersymmetric}.

In our $\Bga$ classification, the case $(k_1,k_2,k_3)=(0,0,0)$ is of particular interest since all triplets are projected. It is interesting to note that this case does not allow for the addition of the $\Bx$ vector but is compatible with an $\tilde{\Bx}$ vector as explained in section \ref{sc:additional_vecs}. Within the free fermionic formulation the addition of the $\Bx$ vector is associated with the GUT enhancement $SO(10)\rightarrow E_6$. Therefore we find that only Pati--Salam models without an $E_6$ embedding can be used to project all three untwisted triplets.
\subsection{$\Bga$ classification results}\label{sc:Bga_Class_tables}

With the help of the SAT/SMT constraint solver Z3 \cite{10.1007/978-3-540-78800-3_24}, we enumerate all modular invariant choices of $\Bga$ and classify them according to the following characteristics\footnote{Our classification code to produce Table \ref{tb:Bga_class_B_2} is available at: \url{https://github.com/BPercivalMMU/AsymmPS}}: 
\begin{itemize}
\item The number of real geometric moduli in each torus, $(M_1,M_2,M_3)$, and total number of real moduli $N_M=M_1+M_2+M_3$.
\item Number of asymmetrically shifted directions, $(N^1_{as},N^2_{as},N^3_{as})$, 
as defined in~\eqref{eq:N_as}.
\item The number of untwisted doublets, $N_D$, as defined in eq. (\ref{N_T_and_N_D}). 
\end{itemize}

The results are given in Table \ref{tb:Bga_class_B_2}. 
For each case a representative $\Bga$ is given, along with the total count of $\Bga$'s with these properties. This frequency value also accounts for equivalent permutations across the three tori.

It is interesting to consider some example (quasi-)realistic asymmetric free fermionic models in the context of the classification scheme presented here. Firstly,
we identify the notable Pati–Salam models of refs \cite{alr,Leontaris:1999ce} as belonging to Class 1b) 
and having no asymmetric shifts, such that $M=(2,2,4)$ and $N_{\rm as}=(0,0,0)$. We can also apply this classification scheme to models with others subgroups such as the notable Standard-like model of ref. \cite{fny}, which is identified as Class 0) with $M=(4,4,4)$ and $N_{\rm as}=(1,1,1)$. The Standard-like model of \cite{Faraggi:1991jr} can be identified as class 3), such that $M=(0,0,0)$ and $N_{\rm as}=(0,0,0)$.

\section{ Defining a Basis Set for Obtaining Three Generations}
\label{sc:3genBasis}
\subsection{Additional basis vectors}
\label{sc:additional_vecs}
In addition to the basis set $\{\one,\Sv,\Bv_1,\Bv_2,\Bv_3,\Bga\}$, there are supplementary modular invariant basis vectors that may be included, regardless of the form of $\bm{\alpha}$, which preserve the Pati--Salam gauge group. We are free to include a hidden sector breaking vector
\beq 
\Bz_1=\{\bgf^{1,2,3,4}\}.
\eeq 
This vector can also be useful to project certain sectors, such as tachyons in the $\cN=0$ models explored in subsequent sections. We note the following other hidden sector combination
\beq 
\Bz_2=\Bz_1+\bm{H}=\Bz_1+\one+\Bv_1+\Bv_2+\Bv_3=\{\bgf^{5,6,7,8}\}.
\eeq 

Without the $\Bga$ vector our basis is: $\mathscr{B}=\{\one,\Sv,\Bv_1.\Bv_2,\Bv_3,\Bz_1\}$, from this we can consider what additional basis vectors can consistently be added to this set. 
All such additional vectors are `Narain-like', in the sense defined within ref. \cite{z2z25}. These vectors are symmetric and their effect is to modify the underlying Narain lattice. A generic form of such a vector that preserves the Pati--Salam gauge group may be written as
\begin{align}
\Bgx = \{ &m_1(y^1,w^1),...,m_6(y^6,w^6) \ | \ m_1(\byy^1,\bw^1),..., m_6(\byy^6,\bw^6); \nonumber \\ 
&\xi(\bgps^{1,2,3}),\xi(\bgps^{4,5}),\xi(\bget^1),\xi(\bget^2),\xi(\bget^3),\xi(\bgf^1),...,\xi(\bgf^8)\},
\end{align}
containing 19 free boundary condition assignments.
The additional vectors we find are given by
\begin{align}\label{eq:additional_vec_list}
\Bx &= \{\bgps^{1,2,3,4,5},\bget^{1,2,3}\} \\
\Bz_3 &= \{\bgf^{3,4,5,6} \} \\ 
\bm{e}_{i} &= \{y^i,w^i \ | \ \byy^i, \bw^i\}, \ \ \ i=1,...,6,\label{eq:eis}
\end{align}
along with their linear combinations.
It will prove useful to define the notation for the addition of the symmetric lattice shifts as
\beq
\bm{e}_{i_1\ldots i_p }= \bm{e}_{i_1}+\ldots + \bm{e}_{i_p}. 
\eeq 

The central motivation for including some additional basis vectors is that they can be used to reduce the degeneracy of the observable spinorial $\mathbf{16}/\overline{\mathbf{16}}$ producing sectors $\Bv_k$, $k=1,2,3$, and project unwanted states, so that three generation models can be obtained.  In the symmetric case, this role is filled by having all six symmetric shift vectors, $\bm{e}_i$. However, asymmetrically shifted or twisted directions from $\Bga$  mean the corresponding $\bm{e}_{i}$ cannot be included in the basis. In this case, vector combinations of the additional vectors (\ref{eq:additional_vec_list}) are the only options to reduce the degeneracies. 
One such combination that we make regular use of is
\beq 
\Bx_k = \Bx+\Bz_k 
\eeq 

where $\Bz_k$ is some combination of $\Bz_1$, $\Bz_2$ and $\Bz_3$, and indicates some set of four hidden fermions from $\bgf^{1,...,8}$.
The vector $\Bz_3$ from the list (\ref{eq:additional_vec_list}) is one choice that naturally satisfies the modular invariance relations with $\Bz_1,\Bz_2$ and $\Bga$. Since the $\Bz_k$ act only on hidden fermions, they can only act as projectors on our fermionic generations, and cannot reduce the degeneracy of the $\mathbf{16}/\overline{\mathbf{16}}$ spinorials. Note, however, that if we take an $\Bga$ with 
four hidden complex fermions, then the following combination emerges 
\beq 
\Bga'=\Bga+\Bz = \bm{A}+\{\bgps^{1,2,3}\},
\eeq 
where $\bm{z}$ is the combination $\bm{z}_i$'s which cancels the hidden sector of $\bm{\alpha}$. This additional observable sector that would necessitate careful case-by-case analysis to avoid exotic states in the spectrum. To avoid this possibility, we exclude, for example, $\Bz_3$ in the basis whenever $\Bga$ has hidden sector $\{\bgf^{1,2,5,6}\}$.


It is important to note that when $\Bx$ or $\Bx_k$ is present in the basis set, the modular invariance constrains (\ref{eq:ABKbvecs}) imposes
\beq \label{eq:x_eta_constraint}
A(\bget^1)=1 \ \text{and } \ A(\bget^2)=A(\bget^3),
\eeq 
without loss of generality.
In Section \ref{sc:DTS}, we show that this necessitates the retention of a Higgs triplet in the spectrum of the model. Should we wish to project all Higgs triplets, we would have to modify our $\bm{x}$-like vector such that
\begin{equation}
 \bm{B}\cdot \tilde{\bm{x}}  = 1 \; \text{mod} \; 2,
\end{equation}
where $\bm{B}$ is the hidden sector component of our vector $\bm{\alpha}$. A representative choice is then $\tilde{\bm{x}}=\Bx+\{\bgf^{1,3,5,7}\}$.

\subsection{Asymmetric Pairings and Three Generations} \label{sc:3gen_Bga}
With the NAHE basis set (\ref{eq:NAHE}), we recall the sectors $\{\Bv_1,\Bv_2,\Bv_3\}$ generate $3\cdot 16$ multiplets in the chiral $\mathbf{16}/\mathbf{\overline{16}}$ representation of $SO(10)$. 
This degeneracy comes from the sets of fermions $\{y^{3,4,5,6},\byy^{3,4,5,6},\bget^1\}$, $\{y^{1,2},w^{5,6},\byy^{1,2},\bw^{5,6},\bget^2\}$ and $\{w^{1,2,3,4},\bw^{1,2,3,4},\bget^3\}$ associated with the $SO(6)^3$ flavour and horizontal symmetry of NAHE-based models. The addition of vectors that break this symmetry can therefore reduce the degeneracy of the three spinorial sectors. For example, the addition of $\Bx$ would reduce the degeneracy, leaving $3\cdot 8$ $\mathbf{16}/\mathbf{\overline{16}}$ multiplets.

The vector $\Bga$ breaks the SO(10) gauge group into $SU(4)_C \times SU(2)_L \times SU(2)_R$, such that the sectors $\{\Bv_1,\Bv_2,\Bv_3\}$ contribute states with representations $(4,2,1)$, $(\bar{4},1,2)$, $(4,1,2)$ and $(\bar{4},2,1)$, the first two of which are the \textit{components} of a $\bm{16}$, and the second two being the \textit{components} of a $\overline{\bm{16}}$. Additionally, the vector $\Bga$ may act to fix helicities for the internal fermions, thus reducing the degeneracy. In our constructions this occurs as a result of the asymmetry in $\bm{\alpha}$. To quantify the role of $\Bga$ on the fermion generation degeneracy 
we define 
\beq
\bm{\Delta}=(\Delta^1,\Delta^2,\Delta^3)
\eeq
as a measure of the asymmetry in the three orbifold plane actions defined by $\Bv_{1,2,3}$
\begin{align}
 \ \begin{cases}
    \Delta^k=0 \ \ \text{if} \ \bm{A}_L^{(k)} = \bm{A}_R^{(k)}\\
    \Delta^k=1 \ \ \text{else} 
    \end{cases}
\end{align}
where $k=1,2,3$ and the $\bm{A}^{(k)}_{L,R}$ are defined in eq. \eqref{eq:AL_AR}. 
This notation has been employed, for example, in ref. \cite{towards}, and is sufficient for quantifying the degeneracy in eq. \ref{DEG}. However the mechanism for reducing the degeneracy is subtly different for each pairing $(\delta^i_L,\delta^i_R )$. For symmetric pairings, the degeneracy is effectively reduced through the models compatibility with all six $\bm{e}_{i}$ vectors, which fix the helicity of Ising, $f\bar{f}$-type, fermion pairs. As all helicities can be defined prior to acting with $\bm{\alpha}$, this vector now becomes a projector for one componant of the full generation.
When $(\delta^i_L,\delta^i_R ) = (2,0), (0,2), (2,4), (4,2)$, this will define the helicity of an $ff$ or $\bar{f}\bar{f}$ pair. Meanwhile,
$(\delta^i_L,\delta^i_R ) = (4,0), (0,4) $ will reduce the degeneracy by separating the left- from the right-moving fermions. All other combinations will reduce the degeneracy of the state by splitting the internal fermions into two groups with associated helicities.

Therefore, by considering the GGSO projection of $\Bga$ on the $\Bv_{k}$ sectors, we find that the degeneracy will be reduced by a factor of 2 whenever $\Delta^k=1$. This variable acts not only as a measure of the asymmetry in $\bm{\alpha}$, but will also indicate whether a sector, $\bm{b}_i$, will produce copies of a full $\bm{16}$ / $\overline{\bm{16}}$ generation, or copies of one component of the generation. When $\Delta^k=1$, full $\bm{16}$ / $\overline{\bm{16}}$ generations are produced, and when $\Delta^k=0$, just one component is produced.

We now consider the impact of adding compatible symmetric shift basis vectors into our classes of Pati--Salam models. For each $\Bga$ class, Table \ref{tab:AddBasVec} describes the simplest compatible vectors, $\Bgx_i$, built from $\bm{e}_{i}$'s and $\bm{z}_k$'s 
For each model we define
\beq 
\bm{N} = (N^1, N^2, N^3),
\eeq
where $N^k$ is the number of additional basis vectors, $\{\Bgx_1,...,\Bgx_{N^k}\}$ reducing the degeneracy of the corresponding spinorial sector $\Bv_k$, such that
\beq 
 (\Bgx_i \cap \Bv_k) \notin \{\emptyset,\Bv_k\}  .
\eeq
for $i=1,...,N^k$.
We can then define the degeneracy of each generation as
\beq\label{DEG}
D_k = \frac{16}{2^{\Delta^k + N^k}},
\eeq 
where we note that
\beq \label{DEGMIN}
D_{k_{min}} = 1.
\eeq
The necessary, but not sufficient, condition to produce three generation models is that at least one generation achieves $D_k = 1$.
We find that at least $N_k=3$ is required to reduce the degeneracy to one, thereby allowing for an odd number of generations. 
In Table \ref{tab:AddBasVec}, we give the fundamental basis set of compatible symmetric vectors for our representative $\bm{A}$ vectors. For brevity, we omit the additional $\bm{z}$ vectors which may also be necessary. We have also constructed the models such that $\bm{x}'$ is not in the basis but can be constructed from the basis vectors. This ensures that the degeneracy can only take integer values and eq. \ref{DEGMIN} holds. 

This list provides the simplest set of irreducible vectors and reflects the maximum number of symmetric shift vectors, $\bm{e}_{i_1...i_p}$, compatible with the model. It is important to note that the above additional vectors can introduce undesirable states such as tachyons and exotic states. For the most part, these states can be removed through GGSO projection, however, as we shall see in Sections \ref{ClassC} and \ref{ClassD}, this is not always the case. Instead, phenomenological models must be built with a basis set of \textit{heavier} vectors constructed from those appearing in Table \ref{tab:AddBasVec}. In this sense, the vectors listed can be viewed as the indivisible building blocks, from which more optimal basis vectors may be built.

{\footnotesize
\setlength{\tabcolsep}{3pt} 
\renewcommand{\arraystretch}{1.1} 
\setlength\LTleft{0pt}
\setlength\LTright{0pt}
\begin{longtable}{|c|c|p{4.8cm}|p{3.5cm}|c|c|}
\caption{Representative $\bm{A}$ vectors for each class label, together with the corresponding additional basis vectors. Note that, from these vectors  $\bm{x}$ or $\tilde{\bm{x}}$ vectors can be constructed, where models with $|k| = 1 \ \text{ mod } \ 2 $ are compatible with $\bm{x}$ and $|k| = 0 \ \text{ mod } \ 2 $ are compatible with $\tilde{\bm{x}}$. In line with this, in the above table we have used the  $\tilde{\bm{e}}$ notation to denote $\bm{e}+\bm{z}$. Writing the basis in this way ensures that $D_{k_{min}} = 1$.}\label{tab:AddBasVec}\\

\hline
Class label & $(N^1_{\rm as},N^2_{\rm as},N^3_{\rm as})$ & Representative $\bm{A}$ & \makecell{Additional internal\\symmetric shift\\vectors} & $\bm{D}$ & $\bm{\Delta}$ \\
\hline \hline
\endfirsthead

\hline
Class label & $(N^1_{\rm as},N^2_{\rm as},N^3_{\rm as})$ & Representative $\bm{A}$ & \makecell{Additional internal \\symmetric shift\\vectors} & $\bm{D}$ & $\bm{\Delta}$\\
\hline \hline
\endhead

\hline
\endfoot

\hline
\endlastfoot
0) & (2,0,0) & $\{\bar{y}^{12}, \bar{w}^{12} ; \bget^1 \}$ & $\{\bm{e}_3,\bm{e}_4,\bm{e}_5,\bm{e}_6,\bm{e}_{12}\}$ & $(1,1,1)$ & $(0,1,1)$ \\

\hline
0) & (1,1,1) & $\{ \byy^1\bw^5,\byy^{35},\bw^{13}\}$ & $\{\bm{e}_2,\bm{e}_4,\bm{e}_6,\tilde{\bm{e}}_{135}\}$ & $(1,1,1)$ & $(1,1,1)$\\

\hline
0) & $(2,2,0)$ & $\{y^1,w^1 \ | \  \byy^2, \byy^{34},\bw^{2,3,4} ;  \bget^2\}$ & $\{\bm{e}_5,\bm{e}_6,\bm{e}_{12},\bm{e}_{34}\}$ & $(1,1,2)$ & $(1,1,1)$\\

\hline
0) & (2,2,2) & $\{y^{1,3},w^{1,3} \ | \ \byy^{2,4,5,6},\bw^{2,4,5,6} ; \ \bget^3 \}$ & $\{\bm{e}_{12},\bm{e}_{34},\bm{e}_{56}\}$ & $(2,2,2)$ & $(1,1,1)$\\
\hline

1a) & (0,0,0) & $\{ \byy^{56}  ; \bget^{2,3}\}$ & $\{\bm{e}_1,\bm{e}_2,\bm{e}_3,\bm{e}_4,\tilde{\bm{e}}_{56}\}$ & $(1,1,1)$ & $(1,0,0)$\\
\hline
1a) & (2,0,0) & $\{\byy^{12},\bw^{12},\byy^{56}\}$ & $\{\bm{e}_3,\bm{e}_4,\bm{e}_{12},\tilde{\bm{e}}_{56}\}$ & $(1,2,1)$ & $(1,1,1)$\\
\hline
1a) & (1,1,0) & $\{ \byy^{46},\bw^{24},\byy^2\bw^5\}$ & $\{\bm{e}_1,\bm{e}_3,\tilde{\bm{e}}_{2456}\}$ & $(2,2,1)$ & $(1,1,1)$\\
\hline
1a) & (2,2,0) & $\{ y^2,y^{34}, w^{2,3,4} \ | \ \byy^1,\bw^{1,5,6}\}$ & $\{\tilde{\bm{e}}_{12},\bm{e}_{34},\tilde{\bm{e}}_{56}\}$ & $(2,2,2)$ & $(1,1,1)$\\

\hline
1b) & (0,0,0) & $\{\bar{y}^{46},\bget^{2,3}\}$ & $\{\bm{e}_1,\bm{e}_2,\bm{e}_3,\bm{e}_5,\tilde{\bm{e}}_{46}\}$ & $(1,1,1)$ & $(1,0,0)$\\

\hline
1b) & (0,1,0) & $\{\bar{y}^{46}, \bar{w}^{34}\bget^2\}$ & $\{\bm{e}_1,\bm{e}_2,\bm{e}_5,\bm{e}_{346}\}$ & $(2,1,1)$ & $(1,0,1)$\\

\hline
1b) & (1,0,0) & $\{\bar{y}^{1}\bar{w}^{5},\bar{w}^{1,3},\bget^1\}$ & $\{\bm{e}_2,\bm{e}_4,\bm{e}_6,\bm{e}_{135}\}$ & $(2,1,1)$ & $(0,1,1)$\\

\hline
1b) & (0,1,1) & $\{\bar{y}^{4,6},\bar{w}^{5,6},\bar{w}^{3,4}\}$ & $\{\bm{e}_1,\bm{e}_2,\tilde{\bm{e}}_{3456}\}$ & $(4,1,1)$ & $(1,1,1)$\\

\hline
1b) & (1,1,0) & $\{\byy^2\bw^6,\byy^{34},\bw^{23}\}$ & $\{\bm{e}_1,\bm{e}_5,\tilde{\bm{e}}_{2346}\}$ & $(2,1,2)$ & $(1,1,1)$\\

\hline
1b) & (2,0,0) & $\{y^{12},w^{12} \;|\; \byy^{46}\}$ & $\{\bm{e}_3,\bm{e}_5,\bm{e}_{12},\tilde{\bm{e}}_{46}\}$ & $(1,1,1)$ & $(1,1,1)$\\
\hline

1b) & (1,1,1) & $\{y^6,w^6 \;|\; \byy^{2,3,4,5},\bw^{24},\bw^{5,6}\}$ & $\{\bm{e}_1,\bm{e}_{36},\tilde{\bm{e}}_{245}\}$ & $(2,1,1)$ & $(1,1,1)$\\

\hline

1b) & (2,1,0) & $\{y^{1,2,4},w^{1,2,4} \ | \ \byy^{3,4,6},\bw^{3}\}$ & $\{\bm{e}_5,\bm{e}_{12},\tilde{\bm{e}}_{346}\}$ & $(2,1,2)$ & $(1,1,1)$\\

\hline

1b) & (2,1,1) & $\{ y^2,w^2 \ | \ \byy^1,\byy^{46},\bw^{1,3,4},\bw^{5,6}\}$ & $\{\tilde{\bm{e}}_{146},\tilde{\bm{e}}_{235}\}$ & $(2,2,2)$ & $(1,1,1)$\\

\hline

2a) & (0,0,0) & $\{\byy^{34},\bw^{56}, \bget^3 \}$ & $\{\bm{e}_1,\bm{e}_2,\tilde{\bm{e}}_{34},\tilde{\bm{e}}_{56}\}$ & $(2,1,1)$ & $(1,1,0)$\\

\hline
2a) & (1,0,0) & $\{ \byy^{1,4,6},\bw^{1,3,5}\}$ & $\{\bm{e}_2,\tilde{\bm{e}}_{46},\bm{e}_{135}\}$ & $(2,1,1)$ & $(1,1,1)$\\

\hline
2a) & (2,0,0) & $\{y^{1,2,6},w^{1,2,6} \;|\; \byy^{3,4,6}, \bw^{5}\}$ & $\{\bm{e}_{12},\tilde{\bm{e}}_{34},\bm{e}_{56}\}$ & $(2,2,2)$ & $(1,1,1)$\\

\hline
2b) & (0,0,0) & $\{ \bw^{24},\bw^{56}; \bget^1\}$ & $\{\bm{e}_1,\bm{e}_3,\tilde{\bm{e}}_{24},\tilde{\bm{e}}_{56}\}$ & $(2,1,1)$ & $(0,1,1)$\\

\hline
2b) & (1,0,0) & $\{ \byy^2\bw^5,\byy^{46},\bw^{12}\}$ & $\{\bm{e}_3,\tilde{\bm{e}}_{46},\bm{e}_{125}\}$ & $(1,2,1)$ & $(1,1,1)$ \\

\hline
2b) & (1,1,0) & $\{y^{2,3,4},w^{2,3,4} \ | \ \byy^1,\bw^4,\bw^{5,6}\}$ & $\{\bm{e}_{56},\bm{e}_{1234}\}$ & $(2,2,4)$ & $(1,1,1)$\\

\hline

3) & (0,0,0) & $\{\byy^1\bw^6,\bw^{24},\byy^{35}\}$ & $\{\tilde{\bm{e}}_{16},\tilde{\bm{e}}_{35},\tilde{\bm{e}}_{24}\}$ & $(1,1,1)$ & $(1,1,1)$\\

\hline

\end{longtable}
}

Additional vectors in the basis 
can increase the number of spinorial sectors in the model. For symmetric models, these sectors are constructed through adding asymmetric shift vectors to $\bm{b}_1, \bm{b}_2, \bm{b}_3$ and take the form, 
\begin{align}\label{generalgenerationsSYM}
\hspace{-1.5cm}
    \begin{split}
        \bm{F}^1_{pqrs} &= \bm{b_1}+ p\bm{e_3}+ q\bm{e_4}+ r\bm{e_5}+ s\bm{e_6}, \\ 
        \bm{F}^2_{pqrs} &= \bm{b_2}+ p\bm{e_1}+ q\bm{e_2}+ r\bm{e_5}+ s\bm{e_6}, \\ 
        \bm{F}^3_{pqrs} &= \bm{b_3}+p\bm{e_1}+q\bm{e_2}+r\bm{e_3}+s\bm{e_4}. 
    \end{split}
\end{align}
In our asymmetric construction, these equations become more complex, and will involve the addition of $\Bgx_i=\bm{e}_{i_1...i_p}$ vectors, when present in the model. To fully define all spinorial sectors in the model, we must also draw attention to the case when the $\bm{x}$--vector is present in the model, in which case it is possible to construct the $SO(12)$ lattice vector
\begin{align} 
\begin{split}
  \bm{G} = \bm{S} + \sum_{i=1,2,3}\bm{b}_{i} + \bm{x}= \{y^{1,...,6}, w^{1,...,6} \ | \ \bar{y}^{1,...,6}, \bar{w}^{1,...,6}\}, 
\end{split}
\end{align}
and the three combinations: 

\begin{align}
\begin{split}
  \bm{e_{3456}}=&\bm{G}+\bm{e_1}+\bm{e_2}=\{y^{3456},w^{3456} \ | \ \bar{y}^{3456},\bar{w}^{3456}\} \ \ \text{for } \ \mathsf{E}_{12}=2 \\
\bm{e_{1256}}=&\bm{G}+\bm{e_3}+\bm{e_4}=\{y^{1256},w^{1256} \ | \ \bar{y}^{1256},\bar{w}^{1256}\} \ \ \text{for } \ \mathsf{E}_{34}=2\\
\bm{e_{1234}}=&\bm{G}+\bm{e_5}+\bm{e_6}=\{y^{1234},w^{1234} \ | \ \bar{y}^{1234},\bar{w}^{1234}\} \ \ \text{for } \ \mathsf{E}_{56}=2.
\end{split}
\end{align}

Therefore we can give a more general form of our spinorial vectors to account for these.

\begin{align}\label{generalgenerations}
\hspace{-1.5cm}
    \begin{split}
        \bm{F}^1_{pqrstu} &= \bm{b_1}+E_3 p\bm{e_3}+E_4 q\bm{e_4}+E_5 r\bm{e_5}+E_6 s\bm{e_6} \\
        &+\sum_{t_{ij},i\neq j = 3,4,5,6} (1-E_i)(1-E_j)t_{ij} \bm{e}_{ij} +\prod_{i=3,5}(1-E_iE_{i+1}) u \bm{e_{3456}}\\ 
        \bm{F}^2_{pqrstu} &= \bm{b_2}+E_1 p\bm{e_1}+E_2 q\bm{e_2}+E_5 r\bm{e_5}+E_6 s\bm{e_6} \\ 
        &+\sum_{t_{ij},i\neq j = 1,2,5,6}(1-E_i)(1-E_j) t_{ij} \bm{e}_{ij}+\prod_{i=1,5}(1-E_iE_{i+1}) u \bm{e_{1256}}\\ 
        \bm{F}^3_{pqrstu} &= \bm{b_3}+E_1p\bm{e_1}+E_2q\bm{e_2}+E_3r\bm{e_3}+E_4s\bm{e_4} \\ 
        &+\sum_{t_{ij},i\neq j = 1,2,3,4}(1-E_i)(1-E_j) t_{ij} \bm{e}_{ij}+\prod_{i=1,3}(1-E_iE_{i+1}) u \bm{e_{1234}},
    \end{split}
\end{align}

 However it should be noted that $\bm{F}^i_{000001}$ contribute to both the $\bm{16}$ and $\overline{\bm{16}}$ equally and so will not give a net effect on the number of generations. This is also the case for $t_{i,j} = 1$, for $\bm{e}_{i,j} \neq \bm{e}_{12}\text{, } \bm{e}_{34}\text{, } \bm{e}_{56}$. The number of generations will be determined by the survival, chirality, and degeneracy of $\bm{F}^i_{pqrst0}$ states. In particular, we will see that retaining enough projectors in the basis plays a crucial role. The candidates for projectors are:
\begin{align}\label{FUpsilons}
    \begin{split}
        \Upsilon(\bm{F}^1_{pqrstu})&=\{\bm{z_1}, \bm{z_2}, \bm{z_3}, \bm{e_1}, \bm{e_2}, \bm{e}_{12}\}\\
        \Upsilon(\bm{F}^2_{pqrstu})&=\{\bm{z_1}, \bm{z_2}, \bm{z_3}, \bm{e_3}, \bm{e_4}, \bm{e}_{34}\}\\
        \Upsilon(\bm{F}^3_{pqrstu})&=\{\bm{z_1}, \bm{z_2}, \bm{z_3}, \bm{e_5}, \bm{e_6}, \bm{e}_{56}\}
    \end{split}
\end{align}
Projecting moduli causes $\bm{e}_{i}$ vectors to become incompatible with the model, and so reduces the number of projecting vectors. Because of this, we are more reliant on creating projecting vectors from the hidden sector. We can give the general survival equation as
\begin{align}
\mathbb{P}_{\bm{F}^1_{pqrstu}}&=\frac{1}{2^5}\prod_{a=1,2,3}\left(1- \CC{\bm{F}^{1}_{pqrstu}}{\bm{z}_{a}}\right)\prod_{i=1,2}\left(1 - \CC{\bm{F}^{1}_{pqrstu}}{\bm{e}_{i}}\right) \left(1 - \CC{\bm{F}^{1}_{pqrstu}}{\bm{e}_{12}}\right)\label{PFa}\\
\mathbb{P}_{\bm{F}^2_{pqrstu}}&=\frac{1}{2^5}\prod_{a=1,2,3}\left(1- \CC{\bm{F}^{2}_{pqrstu}}{\bm{z}_{a}}\right)\prod_{i=3,4}\left(1 - \CC{\bm{F}^{2}_{pqrstu}}{\bm{e}_{i}}\right) \left(1 - \CC{\bm{F}^{2}_{pqrstu}}{\bm{e}_{34}}\right)\label{PFb}\\
\mathbb{P}_{\bm{F}^3_{pqrstu}}&=\frac{1}{2^5}\prod_{a=1,2,3}\left(1- \CC{\bm{F}^{3}_{pqrstu}}{\bm{z}_{a}}\right)\prod_{i=5,6}\left(1 - \CC{\bm{F}^{3}_{pqrstu}}{\bm{e}_{i}}\right) \left(1 - \CC{\bm{F}^{3}_{pqrstu}}{\bm{e}_{56}}\right).\label{PFc}
\end{align}  

Defining the chirality of the states as,
\begin{align}\label{ChProjs}
\begin{split}
\bm{X}^1_{pqrst0} &= -ch(\psi^\mu)\CC{\bm{F}^1_{pqrst0}}{\bm{b_2}+r\bm{e_5}+s\bm{e_6}+t_{56} \bm{e}_{56}}^*\\
\bm{X}^2_{pqrst0} &= -ch(\psi^\mu)\CC{\bm{F}^2_{pqrst0}}{\bm{b_1}+r\bm{e_5}+s\bm{e_6}+ t_{56} \bm{e}_{56}}^*\\
\bm{X}^3_{pqrst0} &= -ch(\psi^\mu)\CC{\bm{F}^3_{pqrst0}}{\bm{b_1}+p\bm{e_3}+q\bm{e_4}+ t_{34} \bm{e}_{34}}^* , 
\end{split}
\end{align}
we can then give the formula that describe which states contribute to the components of the Pati--Salam reprentations. Without loss of generality, we can take $ch(\psi^\mu)=\ket{+}$ to specify the chirality, since the  CPT-conjugate states will then have the opposite choice.
\begin{align} \label{gens}
\begin{split}
    n_{4L}&=\sum_{\substack{k=1,2,3 \\ p,q,r,s,t=0,1}} \frac{1}{2^{2-\Delta^k}}D_k
\mathbb{P}_{\bm{F}^k_{pqrst0}}\left(1 + \bm{X}^k_{pqrst0}\right)\left(1 +(1-\Delta^k) \CC{\bm{F}^k_{pqrst0}}{\bm{\alpha}}\right) +\frac{D_k}{2}\mathbb{P}_{F^k_{000001}} \\
    n_{\bar{4}R}&=\sum_{\substack{k=1,2,3 \\ p,q,r,s,t=0,1}} \frac{1}{2^{2-\Delta^k}}D_k
\mathbb{P}_{\bm{F}^k_{pqrst0}}\left(1 + \bm{X}^k_{pqrst0}\right)\left(1 -(1-\Delta^k) \CC{\bm{F}^k_{pqrst0}}{\bm{\alpha}}\right)+\frac{D_k}{2}\mathbb{P}_{F^k_{000001}}\\
    n_{\overline{4}L}&=\sum_{\substack{k=1,2,3 \\ p,q,r,s,t=0,1}} \frac{1}{2^{2-\Delta^k}}D_k
\mathbb{P}_{\bm{F}^k_{pqrst0}}\left(1 - \bm{X}^k_{pqrst0}\right)\left(1 +(1-\Delta^k) \CC{\bm{F}^k_{pqrst0}}{\bm{\alpha}}\right)+\frac{D_k}{2}\mathbb{P}_{F^k_{000001}}\\
    n_{4R}&=\sum_{\substack{k=1,2,3 \\ p,q,r,s,t=0,1}} \frac{1}{2^{2-\Delta^k}}D_k
\mathbb{P}_{\bm{F}^k_{pqrst0}}\left(1 - \bm{X}^k_{pqrst0}\right)\left(1 -(1-\Delta^k) \CC{\bm{F}^k_{pqrst0}}{\bm{\alpha}}\right)+\frac{D_k}{2}\mathbb{P}_{F^k_{000001}},
\end{split}
\end{align}
where
\begin{align}
  n_{\bm{16}} &= n_{(4,2,1)} + n_{(\overline{4},1,2)} = n_{4L} + n_{\bar{4}R}, \\
  n_{\overline{\bm{16}}} &=  n_{(\bar{4},2,1)} + n_{(4,1,2)} = n_{\bar{4}L} + n_{4R}.
\end{align}

The representations of the states in these sectors, along with their associated charges, are given in Table \ref{charges}. In each state, the oscillators with fermion number $F=0$ are labelled with a `$+$', and those with $F=-1$ labelled with `$-$'.

\begin{table}[h]
\centering
\footnotesize
\renewcommand{\arraystretch}{1.2}
\caption{\label{charges} Charge assignments of states in $n_{\bar{4}R}$, $n_{4R}$, $n_{\bar{4}L}$ and $n_{4L}$ sectors.}
\begin{tabularx}{\textwidth}{|X| X| X| r| r|}
\hline
Representation & $\bar{\psi}^{1,2,3}$ & $\bar{\psi}^{4,5}$ & $Y$ & $Q_{\text{em}}$ \\
\hline

$(\overline{4},1,2)$ 
& $(+,+,+)$ & $(+,+)$ & $1$ & $1$ \\
& $(+,+,+)$ & $(-,-)$ & $0$ & $0$ \\
& $(+,-,-)$ & $(+,+)$ & $\frac{1}{3}$ & $\frac{1}{3}$ \\
& $(+,-,-)$ & $(-,-)$ & $-\frac{2}{3}$ & $-\frac{2}{3}$ \\ \hline

$(4,1,2)$
& $(-,-,-)$ & $(-,-)$ & $-1$ & $-1$ \\
& $(-,-,-)$ & $(+,+)$ & $0$ & $0$ \\
& $(+,+,-)$ & $(-,-)$ & $-\frac{1}{3}$ & $-\frac{1}{3}$ \\
& $(+,+,-)$ & $(+,+)$ & $\frac{2}{3}$ & $\frac{2}{3}$ \\ \hline

$(\overline{4},2,1)$
& $(+,+,+)$ & $(+,-)$ & $\frac{1}{2}$ & $1,0$ \\
& $(+,-,-)$ & $(+,-)$ & $-\frac{1}{6}$ & $\frac{1}{3},-\frac{2}{3}$ \\ \hline

$(4,2,1)$
& $(-,-,-)$ & $(+,-)$ & $-\frac{1}{2}$ & $-1,0$ \\
& $(+,+,-)$ & $(+,-)$ & $\frac{1}{6}$ & $-\frac{1}{3},\frac{2}{3}$ \\ \hline

\hline
\end{tabularx}
\end{table}

From this, we can define a \textit{Complete Generation} as one where
\begin{align}
\begin{split}
n_{4L} - n_{\overline{4}L} =& \; n_{\overline{4}R} - n_{4R} \\
n_{4L} - n_{\overline{4}L} >&  \;  0
\end{split}
\end{align}
and therefore define the number of generations as
\begin{equation}
    n_{g} = n_{4L} - n_{\overline{4}L}.
\end{equation}

\section{Phenomenological Features of Asymmetric Pati--Salam models}\label{Phenofeatures}
There are several phenomenological features common to all the Pati--Salam models discussed thus far. 
We note that we will be considering models with only one asymmetric vector, $\bm{\alpha}$, since 
our aim is to build the simplest extension to the NAHE set, with the number of retained moduli and the observable gauge group determined by a single Pati--Salam vector. 

\subsection{Supersymmetry}
Common across all models is the $\bm{S}$ vector, which is responsible for supersymmetry in the spectrum. Supersymmetry can be retained or removed through appropriate choice of GGSO phases. Given the general form of $\bm{\alpha}$, we can write the survival conditions for supersymmetry as:
\begin{equation}\label{eq:SUSY}
  \CC{\bm{S}}{\bm{e_{i}}} = \CC{\bm{S}}{\bm{z_{k}}} = \CC{\bm{S}}{\bm{x}'} = \CC{\bm{S}}{\bm{\alpha}} = -1
\end{equation}
where $i \subseteq 1,...,6 $, and $k \subseteq 1,2,3$, where these vectors are present in the spectrum. Here and moving forward we use $\bm{x}'$ to represent $\bm{x}$, $\bm{x}_k$ or $\tilde{\bm{x}}$ as appropriate.
If these conditions are satisfied, the chirality of the $\bm{S}$ vector is then given by
\begin{equation}
  \CC{\bm{1}}{\bm{S}} = \CC{\bm{S}}{\bm{b_{1}}}  \CC{\bm{S}}{\bm{b_{2}}}  \CC{\bm{S}}{\bm{b_{3}}}.
\end{equation}
Under these conditions, a gravitino is retained and produces a model with $\mathcal{N} =1$ spacetime supersymmetry.



\subsection{Heavy Higgs}\label{HeavyHiggs}
Heavy Higgs states in Pati--Salam models arise as $n_{\bar{4}R}+n_{4R}$ representations. These states can be constructed from the unpaired spinor sectors listed above, $(n_{\bar{4}R}+n_{4R})^F$ and in the non-supersymmetric case, from the would-be superpartners 
\begin{equation}\label{eq:HHiggs} 
\bm{B}^{k}_{pqrstu} =\bm{S}+\bm{F}^k_{pqrstu}, \ \ \ k=1,2,3.
\end{equation}

Any surviving sector, $\bm{B}^{k}_{pqrstu}$, that survives in a model with $\Delta^k = 1$ will contribute to both $n_{\bar{4}R}+n_{4L}$ and $n_{4R}+n_{\bar{4}L}$, and so generates at least one Heavy Higgs state. When $\Delta^k = 0$ and $ \bm{D} = 1$, each of these states will conribute either $n_{\bar{4}R}+n_{4R}$ or $n_{\bar{4}L}+n_{4L}$. In this case, we must ensure it is the former state that survives, through appropriate choice of GGSO phases. These states have quantum numbers $(4,1,2)$ and $(\bar{4},1,2)$ and will act to break the Pati--Salam gauge group to give Standard-Like Models.
We can express the number of Heavy Higgs states coming from $\bm{B}^{k}_{pqrs}$ as
\beq 
n_{4R}^{B}=\sum_{\substack{k=1,2,3 \\ p,q,r,s,t=0,1}} \mathbb{P}_{\bm{B}^k_{pqrst}},
\eeq 
where $\mathbb{P}_{\bm{B}^k_{pqrst}}$ has the same form as $\mathbb{P}_{\bm{F}^k_{pqrst}}$ in eq \ref{PFc}. Therefore, for non-supersymmetric models, 
\beq 
n_{4R}^H=n_{4R}^{B} + (n_{4R}+n_{\bar{4}R})^{F},
\eeq 
This equation can become more complex depending on the form of $\bm{\alpha}$ and the presence of other basis vectors. In particular, if the helicities of all internal fermions is fixed, we must ensure that the chiralities of $\bar{\psi^{4,5}}$ take the correct form, as is the case in Section \ref{ClassA}.

\subsection{Light Higgs}\label{TQMC}
In our models, states from the twisted sectors can form vectorial $\bm{10}$ representations and act as light Higgs states. These sectors take the form
\begin{equation}\label{HeavyBs} 
\bm{V}^{k}_{pqrstu} =\bm{S}+\bm{F}^k_{pqrstu} + \bm{x}, \ \ \ k=1,2,3.
\end{equation}
The Standard Model Higgs arises when the vacuum is acted on with the oscillator $ \bar{\psi}^a$, where $a=1,...,5$. When present, these states may give rise to the Top Quark Mass Coupling (TQMC), a detailed account of which can be found in \cite{cfr, RizosTop2014}.
The twisted type couplings has the general form
\begin{align}\label{TTQMCA}
    \begin{split}
        &\bm{F}^1\bm{F}^2 \bm{V}^3_{\{\bar{\psi}^a\}}, \ \ \ 
        \bm{F}^1\bm{V}^2_{\{\bar{\psi}^a\}}\bm{F}^3,\ \ \ \bm{V}^1_{\{\bar{\psi}^a\}}\bm{F}^2\bm{F}^3 .
    \end{split}
\end{align}

The NS sector also contributes a TQMC term of the form
 \begin{align}
    \begin{split}\label{UTqmc}
        \bm{F}^1\bm{F}^1{\bar h}_1, \ \ \ \bm{F}^2\bm{F}^2{\bar h}_2, \ \ \  
        \bm{F}^3\bm{F}^3{\bar h}_3,
    \end{split}
\end{align}
where ${\bar h}_k$, $k=1,2,3$, are the Higgs representations from the Neveu-Schwarz sector. In Table \ref{tb:Bga_class_B_2} we have chosen to give examples of $\bm{\alpha}$ vectors which maximise the number of untwisted Higgs doublets, so that there is always at least one present and TQMC will always be a feature of the models.

\subsection{Enhancements}\label{enhancements}
A complete description of gauge enhancements requires the full basis set specification. In general, the enhancements may arise from the following set of sectors.
\begin{equation}\label{enhancementsEx}
\footnotesize
E=\begin{Bmatrix}
\{\psi^{\mu}\}_{\frac{1}{2}} \{ \bar{\lambda}\}_{\frac{1}{2}}: & \ket{\bm{z}_1} & \ket{\bm{z}_2} & \ket{\bm{z}_{3}} \\
 &\ket{\bm{z}_{1} + \bm{z}_{3}} & \ket{\bm{z}_{2} + \bm{z}_{3}} & \ket{\bm{z}_{1} + \bm{z}_{2} + \bm{z}_{3}}\\
& & \\
\{\psi^{\mu}\}_{\frac{1}{2}}: &\ket{ \bm{z}_{1}+ \bm{z}_{2}} &\ket{ \bm{x}}
\end{Bmatrix}.
\end{equation}
The subset of these that are enhacements to the observable sector are:
\begin{equation}\label{ObenhacementsEx}
\footnotesize
E_{Ob}=\begin{Bmatrix}
\{\psi^{\mu}\}_{\frac{1}{2}} \{ \bar{\psi}^{1,...5}\}_{\frac{1}{2}}: & \ket{\bm{z}_1} & \ket{\bm{z}_2} & \ket{\bm{z}_{3}} \\
 &\ket{\bm{z}_{1} + \bm{z}_{3}} & \ket{\bm{z}_{2} + \bm{z}_{3}} & \ket{\bm{z}_{1} + \bm{z}_{2} + \bm{z}_{3}}\\
& & \\
\{\psi^{\mu}\}_{\frac{1}{2}}: &\ket{ \bm{x}}
\end{Bmatrix}.
\end{equation}

In addition to these there may be ``exotic enhancements'', which take the same form, but transform under the hidden and the observable gauge group factors.
These are listed for each Example Class below, and will be projected to ensure phenomenological viability. 

\subsection{Exotics}\label{exotics}
Exotic sectors are those with fractional charge with respect to the gauge group. As with the enhancements, it is difficult to describe the exact form of exotic sectors before defining an $\bm{\alpha}$ vector. However, in general these sectors are constructed in the following way:
\begin{equation}\label{exoticsEx}
  G^{k}_{pqrstu, a} \approx \bm{F}^k_{pqrstu} + \bm{\alpha} + \sum_i a_i \bm{z}_i,
\end{equation}
where $a_i$ are the coefficients describing contributions from the hidden sector. 
These sectors produce states with non-standard structures of $((2_L,1),(2,1)_H)$, $((2_L,1),(1,2)_H)$, $((1,2_R),(2,1)_H)$ or $((1,2_R),(1,2)_H)$ under $SU(2)_{L} \times SU(2)_{R} \times SO(4)_{Hidden} $. We will define $n_{2L}$ and $n_{2R}$ as the number of these states that survive, where
\begin{align}
    n_{2L} &: \pm \tfrac{1}{2} \text{ leptons,} \\
    n_{2R} &: \pm \tfrac{1}{2} \text{ singlets.}
\end{align}

The addition of the $\bm{x}$ vector can produce $(4,1,1)$ and $(\bar{4},1,1)$ states under $SU(4) \times SU(2)_{L} \times SU(2)_{R} $. We will define $n_{4}$ and $n_{\bar{4}}$ as the number of these states in the model, where 
\begin{equation}
    n_{4} + n_{\bar{4}} : \pm \tfrac{1}{6} \text{ exotic coloured particles and singlets.}
\end{equation}
These cases will be highlighted in more detail for each example model. We can sometimes remove these fractionally charged states from the massless spectrum by pairing them into vector-like representations, which can subsequently acquire a mass. A model is therefore defined to be free from chiral exotic states if,
\begin{align} 
n_{2L} &= 0 \; \text{mod } 2, \\
n_{2R} &= 0 \; \text{mod } 2, \\  
n_{4} &= n_{\bar{4}},
\end{align}
which is slightly less restrictive than the condition to produce a model containing no massless fractionally charged fermionic states. It is preferable to eliminate all exotic sectors, as vector-like exotics can induce phenomenologically undesirable couplings, and their decoupling via mass generation is not guaranteed. Furthermore, this makes the statistics more comparable to the symmetric case (e.g. \cite{PStclass}).

\subsection{Tachyons}\label{tachyons}
In the absence of supersymmetry, level-matched tachyonic states tend to proliferate and must be GGSO projected.
 These models tend to have a large number of tachyons due to the three hidden sector breaking vectors $\bm{z}_{1,2,3}$, which give additional permutations of hidden sector fermions that produce level-matched states. The number of tachyons is reduced when symmetric shift vectors $\bm{e}_i$ become incompatible with the model due to an asymmetric internal pairing, however projection of tachyons remains a computationally expensive process, and can limit how much of the landscape is sampled. 
We list the common level matched tachyons across the classes in Table \ref{table:tachyons_combined}, omitting the tachyons which arise from the NS sector. We remark that tachyons arising from the NS--sector are always projected out by the $\Sv$--vector.
\begin{table}[!ht]
\centering

\begin{tabular}{|c|c|c|c|}
\hline
$(-\frac{1}{2},-\frac{1}{2})$ 
& $(-\frac{1}{4},-\frac{1}{4})$ 
& $(-\frac{3}{8},-\frac{3}{8})$ 
& $(-\frac{1}{8},-\frac{1}{8})$ \\
\hline

$|\bm{z}_{1}\rangle$ 
& $|\bm{e}_{i}+\bm{e}_{j}+\bm{z}_{1}\rangle$
& $|\bm{e}_{i}+\bm{z}_{1}\rangle$
& $|\bm{e}_{i}+\bm{e}_{j}+\bm{e}_{k}+\bm{z}_{1}\rangle$ \\

$|\bm{z}_{2}\rangle$ 
& $|\bm{e}_{i}+\bm{e}_{j}+\bm{z}_{2}\rangle$
& $|\bm{e}_{i}+\bm{z}_{2}\rangle$
& $|\bm{e}_{i}+\bm{e}_{j}+\bm{e}_{k}+\bm{z}_{2}\rangle$ \\

$|\bm{z}_{3}\rangle$ 
& $|\bm{e}_{i}+\bm{e}_{j}+\bm{z}_{3}\rangle$
& $|\bm{e}_{i}+\bm{z}_{3}\rangle$
& $|\bm{e}_{i}+\bm{e}_{j}+\bm{e}_{k}+\bm{z}_{3}\rangle$ \\

$|\bm{z}_{1}+\bm{z}_{3}\rangle$ 
& $|\bm{e}_{i}+\bm{e}_{j}+\bm{z}_{1}+\bm{z}_{3}\rangle$
& $|\bm{e}_{i}+\bm{z}_{1}+\bm{z}_{3}\rangle$
& $|\bm{e}_{i}+\bm{e}_{j}+\bm{e}_{k}+\bm{z}_{1}+\bm{z}_{3}\rangle$ \\

$|\bm{z}_{2}+\bm{z}_{3}\rangle$ 
& $|\bm{e}_{i}+\bm{e}_{j}+\bm{z}_{2}+\bm{z}_{3}\rangle$
& $|\bm{e}_{i}+\bm{z}_{2}+\bm{z}_{3}\rangle$
& $|\bm{e}_{i}+\bm{e}_{j}+\bm{e}_{k}+\bm{z}_{2}+\bm{z}_{3}\rangle$ \\

$|\bm{z}_{1}+\bm{z}_{2}+\bm{z}_{3}\rangle$ 
& $|\bm{e}_{i}+\bm{e}_{j}+\bm{z}_{1}+\bm{z}_{2}+\bm{z}_{3}\rangle$
& $|\bm{e}_{i}+\bm{z}_{1}+\bm{z}_{2}+\bm{z}_{3}\rangle$
& $|\bm{e}_{i}+\bm{e}_{j}+\bm{e}_{k}+\bm{z}_{1}+\bm{z}_{2}\rangle$ \\

& $\bar{\lambda}|\bm{e}_{i}+\bm{e}_{j}\rangle$
& $\bar{\lambda}|\bm{e}_{i}\rangle$
& $\bar{\lambda}|\bm{e}_{i}+\bm{e}_{j}+\bm{e}_{k}\rangle$ \\

\hline
\end{tabular}

\caption{Level-matched tachyons — Mass levels and states. 
Here $i,j,k = 1,\dots,6$ with $i \neq j \neq k$, and $\bar{\lambda}$ is a right-moving oscillator.}
\label{table:tachyons_combined}

\end{table}

\subsection{Vacuum Energy}

At the free fermionic point, the partition function is expressed as
\begin{equation}\label{PartFunc}
  Z = \sum_{\bm{\alpha},\bm{\beta}} \CC{\bm{\alpha}}{\bm{\beta}} \prod_{f} Z
  \sqbinom{\bm{\alpha}(f)}{\bm{\beta}(f)},
\end{equation}
where  $Z\smb{a}{b}=\sqrt{\vartheta\smb{a}{b}}$ and the Jacobi theta functions take their standard form
\begin{equation}\label{ThetaQ}
\vartheta\smb{a}{b} = \sum_{n\in\mathbb{Z}} q^{(n+a/2)^{2} / 2} e^{ 2 \pi i(n+a/2)b/2},
\end{equation}
with $a,b\in\{0,1\}$ for real boundary conditons. 
To gain further insight into the effect of the asymmetric action 
$\bm{\alpha}$, we express the partition function in a more general form. 
Since our models are build from the NAHE set, and the number of additional vectors is undetermined, we consider the partition function of the NAHE set with $n$ symmetric shift vectors,

\begin{align}\label{PFnahe}
Z=&\frac{1}{\eta^{10}\bar{\eta}^{22}}\;\frac{1}{2^3}\sum_{\substack{a,k,r\\b,l,s,}} \;\frac{1}{2^n} \sum_{\substack{H_i\\G_i}} \;\frac{1}{2^2} \sum_{\substack{h_1,h_2\\g_1,g_2}} (-1)^{\Phi\left[\begin{smallmatrix}
a&k&r&H_i&h_1&h_2\\b&l&s&G_i&g_1&g_2\end{smallmatrix} \right]}\nonumber\\[0.1cm]
&\times \vth_{\psi^{\mu}}\smb{a}{b} \;\, \vth_{\chi^{12}}\smb{a+h_1}{b+g_1} \vth_{\chi^{34}}\smb{a+h_2}{b+g_2} \vth_{\chi^{56}}\smb{a-h_1-h_2}{b-g_1-g_2} \\[0.2cm]
&\times \Gamma_{(6,6)}\left[\begin{smallmatrix}
r&H_i&h_1&h_2\\s&G_i&g_1&g_2
\end{smallmatrix} \right] \nonumber\\[0.2cm]
&\times \vthb_{\bar{\psi}^{1,...,5}}\smb{k}{l}^5 \vthb_{\bar{\eta}^{1}}\smb{k+h_1}{l+g_1} \vthb_{\bar{\eta}^{2}}\smb{k+h_2}{l+g_2} \vthb_{\bar{\eta}^{3}}\smb{k-h_1-h_2}{l-g_1-g_2} \vthb_{\bar{\phi}^{1,...,8}}\smb{r}{s}^8, \nonumber
\end{align}
where, in the symmetric case with six symmetric shift vectors, the internal lattice $\Gamma_{(6,6)}$ would be
\begin{align}\label{BaseLattice}
\Gamma_{(6,6)}\left[\begin{smallmatrix}r&H_i&h_1&h_2\\s&G_i&g_1&g_2\end{smallmatrix}\right] =& 
\;\; \Big| \; \vth_{y\bar{y}^1}\smb{r+h_1+H_1}{s+g_1+G_1} \vth_{y\bar{y}^2}\smb{r+h_1+H_2}{s+g_1+G_2} \vth_{y\bar{y}^3}\smb{r+h_2+H_3}{s+g_2+G_3}  \nonumber\\
&\times  \vth_{y\bar{y}^4}\smb{r+h_2+H_4}{s+g_2+G_4} \vth_{y\bar{y}^5}\smb{r+h_2+H_5}{s+g_2+G_5} \vth_{y\bar{y}^6}\smb{r+h_2+H_6}{s+g_2+G_6}  \nonumber\\[0.1cm]
&\times  \vth_{w\bar{w}^1}\smb{r-h_1-h_2+H_1}{s-g_1-g_2+G_1} \vth_{w\bar{w}^2}\smb{r-h_1-h_2+H_2}{s-g_1-g_2+G_2} \vth_{w\bar{w}^3}\smb{r-h_1-h_2+H_3}{s-g_1-g_2+G_3}  \\ 
&\times  \vth_{w\bar{w}^4}\smb{r-h_1-h_2+H_4}{s-g_1-g_2+G_4} \vth_{w\bar{w}^5}\smb{r+h_1+H_5}{s+g_1+G_5} \vth_{w\bar{w}^6}\smb{r+h_1+H_6}{s+g_1+G_6}\Big|, \nonumber
\end{align}
with $\left|\vth\smb{a}{b}\right|=\sqrt{\vth\smb{a}{b}\vthb\smb{a}{b}}$, and $\bm{\Phi}$ being a polynomial which ensures modular invarince. We have also labelled the theta functions with their corresponding worldsheet fermions.
The addtion of $\bm{\alpha}$ has the effect of breaking the gauge group and ensuring asymmetric shift vectors $\bm{e}_i$ are not present in the model, and this is reflected in the partition function through the addition of the variable $H'$.

\begin{align}\label{PF}
Z=&\frac{1}{\eta^{10}\bar{\eta}^{22}}\;\frac{1}{2^3}\sum_{\substack{a,k,r\\b,l,s}} \;\frac{1}{2^n} \sum_{\substack{H_i\\G_i}} \;\frac{1}{2^2} \sum_{\substack{h_1,h_2\\g_1,g_2}} (-1)^{\Phi\left[\begin{smallmatrix}
a&k&r&H_i&h_1&h_2\\b&l&s&G_i&g_1&g_2\end{smallmatrix} \right]}\nonumber\\[0.1cm]
&\times \vth_{\psi^{\mu}}\smb{a}{b} \;\, \vth_{\chi^{12}}\smb{a+h_1}{b+g_1} \vth_{\chi^{34}}\smb{a+h_2}{b+g_2} \vth_{\chi^{56}}\smb{a-h_1-h_2}{b-g_1-g_2} \\[0.2cm]
&\times \Gamma^{\alpha}_{(6,6)}\left[\begin{smallmatrix}
r&H_i&h_1&h_2&H'\\s&G_i&g_1&g_2&G'
\end{smallmatrix} \right] \nonumber\\[0.2cm]
&\times \vthb_{\bar{\psi}^{1,2,3}}\smb{k + H'}{l + G'}^3 \vthb_{\bar{\psi}^{4,5}}\smb{k}{l}^2 \vthb_{\bar{\eta}^{1}}\smb{k+h_1+ \bm{k}_{1}H'}{l+g_1+ \bm{k}_{1}G'} \vthb_{\bar{\eta}^{2}}\smb{k+h_2+ \bm{k}_{2}H'}{l+g_2+ \bm{k}_{2}G'} \vthb_{\bar{\eta}^{3}} \smb{k-h_1-h_2+ \bm{k}_{3}H'}{l-g_1-g_2+ \bm{k}_{3}G'} \nonumber\\
&\times \vthb_{\bar{\phi}^{12}}\smb{r+H'}{s+G'}^2 \vthb_{\bar{\phi}^{34}}\smb{r}{s}^2 \vthb_{\bar{\phi}^{56}}\smb{r}{s}^2 \vthb_{\bar{\phi}^{78}}\smb{r}{s}^2, \nonumber
\end{align}
where it should be noted that $\bm{k}_{1,2,3} = (A(\bar{\eta}^1),A(\bar{\eta}^2),A(\bar{\eta}^3))$ and are not a variable to be summed over. We write them in bold to avoid confusion.

The effect of an asymmetric twist on the partion function is to break the left-right symmetric (Ising) pairing of the internal fermions such that $(y^i\bar{y}^i)$ and $(y^j\bar{y}^j)$ become  $(y^iy^j)(\bar{y}^i\bar{y}^j)$. Given that each plane is associated with four left moving and four right moving fermions, this breaking of symmetry can happen up to two times on the plane, For example, 
\begin{align}
 \Gamma_1^{\alpha} =& \;\; \vth_{y^3}\smb{r+h_2+H_3}{s+g_2+G_3}^{1/2} 
 \vth_{y^4}\smb{r+h_2+H_4}{s+g_2+G_4}^{1/2} \vth_{y^{5,6}}\smb{r+h_2+H'}{s+g_2+G'}\nonumber\\ 
 &\times\vth_{\bar{y}^3}\smb{r+h_2+H_3}{s+g_2+G_3}^{1/2} \vthb_{\bar{y}^4}\smb{r+h_2+H_4}{s+g_2+G_4}^{1/2} \vthb_{\bar{y}^{5,6}}\smb{r+h_2}{s+g_2}.
\end{align}
describes one asymmetric pair  and the two complementary symmetric shift vectors being present in the spectrum, and 
\begin{equation}
 \Gamma_1^{\alpha} = \;\;  \vth_{y^{3,4,5,6}}\smb{r+h_2+H'}{s+g_2+G'}\times \vthb_{\bar{y}^{3,4,5,6}}\smb{r+h_2}{s+g_2}
\end{equation}
describes the case when no symmetric shift vectors are compatible. Defining $\bm{e}_{34}$ would split this more explicitly into two asymmetric pairings.


Expanding the $\vartheta$ functions allows us to express the partition function as a polynomial
\begin{equation}\label{QExpPF}
    Z = \sum_{n,m} a_{mn} q^m \bar{q}^n, 
\end{equation}
from which we can read off the net number of states at each mass level.

From this position we can calculate the one-loop partition function by integrating the partition function over the fundamental domain
\begin{equation}
    \Lambda = \int_\mathcal{F}\frac{d^2\tau}{\tau_2^2}\, Z_B Z_F = \int_\mathcal{F}\frac{d^2\tau}{\tau_2^3}\, \sum_{n.m} a_{mn} q^m \bar{q}^n,
    \label{CoC}
\end{equation}
where $Z_B$ is the contribution from the bosonic degrees of freedom given by
\begin{equation}
  Z_B = \frac{1}{\tau_2} \frac{1}{\eta^2\bar{\eta}^2}.
  \label{Z_B}
\end{equation}

\section{Example Model - Class 0)}\label{ClassA}
The first model we will explore is a Class 0) model as defined in section \ref{sc:Non_Geom_Pic}, such that it has no asymmetric twist action, retaining the 12 real moduli of the $\ztwo$ orbifold. Therefore we define the following basis,

\begin{align}\label{basisA}
\bm{\mathds{1}}&=\{\psi^\mu,
\chi^{1,\dots,6},y^{1,\dots,6}, w^{1,\dots,6}\ | \ \bar{y}^{1,\dots,6},\bar{w}^{1,\dots,6};
\bar{\psi}^{1,\dots,5},\bar{\eta}^{1,2,3},\bar{\phi}^{1,\dots,8}\},\nonumber\\
\bm{S}&=\{{\psi^\mu},\chi^{1,\dots,6} \},\nonumber\\
\bm{e_3}&=\{y^{3},w^{3}\; | \; \bar{y}^{3},\bar{w}^{3}\}, 
\nonumber\\
\bm{e_4}&=\{y^{4},w^{4}\; | \; \bar{y}^{4},\bar{w}^{4}\}, 
\nonumber\\
\bm{e_5}&=\{y^{5},w^{5}\; | \; \bar{y}^{5},\bar{w}^{5}\}, 
\nonumber\\
\bm{e_6}&=\{y^{6},w^{6}\; | \; \bar{y}^{6},\bar{w}^{6}\}, 
\nonumber\\
\bm{b_1}&=\{\psi^\mu,\chi^{12},y^{3,4,5,6}\; | \; \bar{y}^{3,4,5,6};\bar{\psi}^{1,\dots,5},\bar{\eta}^1\},\nonumber\\
\bm{b_2}&=\{\psi^\mu,\chi^{34},y^{1,2},w^{5,6}\; | \; \bar{y}^{1,2},
\bar{w}^{5,6};\bar{\psi}^{1,\dots,5},\bar{\eta}^2\},\\
\bm{b_3}&=\{\psi^\mu,\chi^{56},w^{1,2,3,4}\; | \; \bar{w}^{1,2,3,4};\bar{\psi}^{1,\dots,5},
\bar{\eta}^3\}, \nonumber\\
\bm{z_1}&=\{\bar{\phi}^{1,\dots,4}\},\nonumber\\
\bm{\alpha} &= \{y^{12}, w^{12} \; | \;  \bar{\psi}^{1,2,3}, \bar{\eta}^{1}, \bar{\phi}^{3,4} \}\nonumber.
\end{align}
This model has the following defining characteristics:
\begin{align}
    \begin{split}
        \bm{M} &= (4,4,4), \\
        \bm{\Delta}&=(0,1,1),\\
        \bm{D}&=(1,2,2),\\
        \bm{N}_{as} &= (2,0,0) ,
    \end{split}
\end{align}
which suggests three generations are achievable as eq. (\ref{DEGMIN}) is satisfied.

\subsection{Phenomenological Characteristics}\label{PhenoA}
\subsubsection{Observable Spinorial Repressentations}

Because this model contains four asymmetric shift vectors, eq \ref{generalgenerations}, is reduced to a much simpler form and gives the following spinorial sectors, which give rise to fermionic generations,
\begin{align}
   \bm{F}^1_{pqrs}&=\bm{b_1}+p\bm{e_3}+q\bm{e_4}+r\bm{e_5}+s\bm{e_6},\\ 
   \bm{F}^2_{rs}&=\bm{b_2}+r\bm{e_5}+s\bm{e_6},\\
   \bm{F}^3_{rs}&=\bm{b_3}+r\bm{e_3}+s\bm{e_4},
\end{align}
and the number of surviving sectors is defined through
\begin{align}
\mathbb{P}_{\bm{F}^1_{pqrs}}&=\frac{1}{2^2}
\prod_{a=1,2}\left(1- \CC{\bm{F}^{(1)}_{pqrs}}{\bm{z_a}}\right),\label{PF1}\\
        \mathbb{P}_{\bm{F}^2_{rs}}&=\frac{1}{2^4}\prod_{i=3,4}\left(1 - \CC{\bm{F}^{(2)}_{rs}}{\bm{e_i}}\right)
\prod_{a=1,2}\left(1- \CC{\bm{F}^{(2)}_{rs}}{\bm{z_a}}\right),\label{PF2}\\
\mathbb{P}_{\bm{F}^3_{rs}}&=\frac{1}{2^4}\prod_{i=5,6}\left(1 - \CC{\bm{F}^{(3)}_{rs}}{\bm{e_i}}\right)
\prod_{a=1,2}\left(1- \CC{\bm{F}^{(3)}_{rs}}{\bm{z_a}}\right).\label{PF3}
\end{align}

Applying eq. (\ref{ChProjs}) to get the chirality phases,
\begin{align}
\begin{split}
\bm{X}^1_{pqrs} &= -\CC{\bm{F}^1_{pqrs}}{\bm{F}^2_{rs}}^*, \nonumber\\
\bm{X}^2_{rs} &= -\CC{\bm{F}^2_{rs}}{\bm{F}^1_{00rs}}^*,\\
\bm{X}^3_{rs} &= -\CC{\bm{F}^3_{rs}}{\bm{F}^1_{pq00}}^*,\nonumber
\end{split}
\end{align}
where $ch(\psi^\mu)=+1$ for the spacetime fermion chirality.  We can write the multiplicities of the subsequent Pati--Salam representations as
\begin{align} 
\begin{split}
    n_{4L}=&\sum_{\substack{ p,q,r,s=0,1}}\frac{1}{4}
\mathbb{P}_{\bm{F}^1_{pqrs}}\left(1+ \bm{X}^1_{pqrs}\right)\left(1+\CC{\bm{F}^1_{pqrs}}{\bm{\alpha}+\bm{x}'}\right)+\sum_{\substack{ r,s=0,1}}2
\mathbb{P}_{\bm{F}^2_{rs}}\frac{1}{2}\left(1 + \bm{X}^2_{rs}\right)\\
&+\sum_{\substack{ r,s =0,1}} 2
\mathbb{P}_{\bm{F}^3_{rs}}\frac{1}{2}\left(1 + \bm{X}^3_{rs}\right), \\
    n_{\overline{4}R}=&\sum_{\substack{ pqrs=0,1}} \frac{1}{4}
\mathbb{P}_{\bm{F}^1_{pqrs}}\left(1 + \bm{X}^1_{pqrs}\right)\left(1-\CC{\bm{F}^1_{pqrs}}{\bm{\alpha}+\bm{x}'}\right)+\sum_{\substack{ r,s=0,1}} 2
\mathbb{P}_{\bm{F}^2_{rs}}\frac{1}{2}\left(1 + \bm{X}^2_{rs}\right)\\\
&+\sum_{\substack{ r, s=0,1}} 2
\mathbb{P}_{\bm{F}^3_{rs}}\frac{1}{2}\left(1 + \bm{X}^3_{rs}\right),\\
    n_{\overline{4}L}=&\sum_{\substack{ pqrs=0,1}}\frac{1}{4}
\mathbb{P}_{\bm{F}^1_{pqrs}}\left(1-\bm{X}^1_{pqrs}\right)\left(1+\CC{\bm{F}^1_{pqrs}}{\bm{\alpha}+\bm{x}'}\right)+\sum_{\substack{ r,s=0,1}} 2
\mathbb{P}_{\bm{F}^2_{rs}}\frac{1}{2}\left(1 - \bm{X}^2_{prs}\right)\\
&+\sum_{\substack{ r, s=0,1}} 2
\mathbb{P}_{\bm{F}^3_{rs}}\frac{1}{2}\left(1 - \bm{X}^3_{rs}\right),\\
    n_{4R}=&\sum_{\substack{ pqrs=0,1}}\frac{1}{4}
\mathbb{P}_{\bm{F}^1_{pqrs}}\left(1 - \bm{X}^1_{pqrs}\right)\left(1-\CC{\bm{F}^1_{pqrs}}{\bm{\alpha}+\bm{x}'}\right)+\sum_{\substack{ r,s=0,1}} 2
\mathbb{P}_{\bm{F}^2_{rs}}\frac{1}{2}\left(1 - \bm{X}^2_{rs}\right)\\
&+\sum_{\substack{r,s =0,1}} 2
\mathbb{P}_{\bm{F}^3_{rs}}\frac{1}{2}\left(1 - \bm{X}^3_{rs}\right),
\end{split}
\end{align}

where
\begin{align}\label{XBAR-A}
 \bm{x}' &= \bm{S} + \sum \bm{b}_i + \sum_{i=3}^6 \bm{e}_i \\ 
 &= \{y^{12}, w^{12} \; | \; \bar{y}^{1,2}, \bar{w}^{1,2} ; \  \bar{\psi}^{1,2,3,4,5},\bar{\eta}^{1,2,3}\}, \nonumber
 \end{align}

and is added to $\bm{\alpha}$ to isolate the helicity of $\bar{\psi}^{4,5}$.
Imposing the condition for complete generations 
results in the condition
\begin{align}
\begin{split} \label{completegens}
&\sum_{\substack{ p,q,r,s=0,1}}\frac{1}{4}
\mathbb{P}_{\bm{F}^1_{pqrs}}\left(2 \bm{X}^1_{pqrs}\right)\left(1+\CC{\bm{F}^1_{pqrs}}{\bm{\alpha}+\bm{x}'}\right) \\ 
&= \sum_{\substack{ p,q,r,s=0,1}}\frac{1}{4}
\mathbb{P}_{\bm{F}^1_{pqrs}}\left(2 \bm{X}^1_{pqrs}\right)\left(1-\CC{\bm{F}^1_{pqrs}}{\bm{\alpha}+\bm{x}'}\right),
\end{split}
\end{align}

\beq \label{completegens2}
\sum_{\substack{ p,q,r,s=0,1}}
\mathbb{P}_{\bm{F}^1_{pqrs}} \bm{X}^1_{pqrs}\CC{\bm{F}^1_{pqrs}}{\bm{\alpha}+\bm{x}'}= 0 .
\eeq 

Plugging this into the $n_{4L}-n_{\overline{4}L}=3$ equation for three generations gives
\beq \label{3genFs}
3=\sum_{\substack{ p,q,r,s=0,1}} \frac{1}{2}
\mathbb{P}_{\bm{F}^1_{pqrs}}\bm{X}^1_{pqrs} +\sum_{r,s} 2 \mathbb{P}_{F^2_{rs}}\bm{X}^2_{rs}+\sum_{p,q} 2 \mathbb{P}_{\bm{F}^3_{pq}}\bm{X}^3_{pq},
\eeq 
which is only possible if
\beq 
\sum_{p,q,r,s}\mathbb{P}_{\bm{F}^{1}_{pqrs}} \bm{X}^1_{pqrs} = 2 \text{ mod } 4.
\eeq 
However, without more projecting vectors, we have
\beq 
\sum_{p,q,r,s}\mathbb{P}_{\bm{F}^{1}_{pqrs}} \bm{X}^1_{pqrs} = 0 \text{ mod } 4\text{ ,}
\eeq 
Therefore another basis vector is required to act as a projector and ensure three generation models are possible. In fact there is some freedom in what form this additional vector takes, so for simplicity, we choose not to enhance the gauge group with the addition of $\bm{x}$ (or equivalently $\bm{e}_{12}$) 
and choose
\beq 
\bm{z}_{3} = \{ \bar{\phi}^{3,4,5,6}\}.
\eeq 
This forces the hidden sector to break to $SO(4)^4$.


\subsubsection*{Heavy Higgs} 
From the spinorial $\mathbf{16}/\mathbf{\overline{16}}$ states the additional Heavy Higgs producing sectors can be built:
\begin{align}
    \begin{split}
    \bm{B}^1_{pqrs}&= \bm{F}^1_{pqrs} + \bm{S} \\
   &= \bm{S} + \bm{b_1}+p\bm{e_3}+q\bm{e_4}+r\bm{e_5}+s\bm{e_6},  \\
    \bm{B}^2_{rs}&= \bm{F}^2_{rs} + \bm{S} \\
   &= \bm{S} + \bm{b_2}+r\bm{e_5}+s\bm{e_6},  \\
   \bm{B}^3_{rs}&= \bm{F}^3_{rs} + \bm{S}\\
   &=\bm{S} + \bm{b_3}+r\bm{e_3}+s\bm{e_4}, 
   \end{split}
\end{align}
as defined in (\ref{eq:HHiggs}). It is important to note that whilst surviving $\bm{B}^{2,3}_{r,s}$ will always contribute an $n_{4R}^B$, the $n_{4R}^B$ states from $\bm{B}^{1}_{pqrs}$ may be incompatible with $\bm{\alpha}$ and $\bar{\bm{x}}$, and so require additional projection critria to be considered.
\subsubsection*{Light Higgs} 

Without the $\bm{x}$ vector in the basis set, the twisted sector vectorial Higgs states cannot be constructed across all three tori. However, using the same definition of $\bm{x}'$ as in equation \ref{XBAR-A}, 
the following sectors in the model
\begin{align}
   \bm{\bar{V}}^2_{rs}&= \bm{S} + \bm{b_2}+r\bm{e_5}+s\bm{e_6} + \bm{x}',\\
   \bm{\bar{V}}^3_{rs}&=\bm{S} + \bm{b_3}+r\bm{e_3}+s\bm{e_4} + \bm{x}' .
\end{align}
These states act as the light Higgs states when accompanied by the right moving oscillator $ \bar{\psi}^{1,...,5}$. To ensure a TQMC is present, we should account for the Higgs states from the NS sector as well as potential couplings from these vectorial states, as defined in equations (\ref{UTqmc}) and (\ref{TTQMCA}). These sectors contain Higgs triplets which are vector-like and can be removed from the massless spectrum.

\subsubsection*{Tachyonic Sector}\label{tachyA}
Since in the vast majority of cases, the models in this class will violate the SUSY condition on the GGSO phases of eq. (\ref{eq:SUSY}), tachyonic states must be projected through appropriate combinations of GGSO phases. With the addition of the $\bm{z}_{3}$ vector, the number of tachyon producing sectors increases substantially to 125.  We list all class dependent tachyonic sectors in Table \ref{table:tachyonsA1}, in addition to those given in Table \ref{table:tachyons_combined}.

\begin{table}[ht]
\centering

\begin{tabular}{|c|c|}
\hline
 $(-\frac{1}{4} ,-\frac{1}{4})$ &$(-\frac{1}{8} ,-\frac{1}{8})$ \\
\hline
 $|\bm{\alpha} >$ & $ | \bm{e}_{i} + \bm{\alpha}> $ \\
 $|\bm{\alpha} + \bm{z}_{1} >$ & $ | \bm{e}_{i} + \bm{\alpha} + \bm{z}_{1}> $ \\
 $|\bm{\alpha} + \bm{z}_{3} >$ &  $ | \bm{e}_{i} + \bm{\alpha} + \bm{z}_{3}> $\\
  $|\bm{\alpha} + \bm{z}_{2} + \bm{z}_{3} >$ & $ | \bm{e}_{i} + \bm{\alpha} + \bm{z}_{2} + \bm{z}_{3}> $\\
 
\hline
\end{tabular}
\caption{Class specific level-matched tachyons — Mass levels and states, where $i = 3,4,5,6$}
\label{table:tachyonsA1}
\end{table}
Each of these states must be projected individually, which increases the computational time required. This is the limiting factor when considering how much of the landscape to scan.

\subsubsection*{Enhancements}
We will first consider the enhancements to the observable sector. Again, due to the additional $\bm{z}_{3}$ vector there are a comparitively large number of enhancements to consider. 

\begin{equation}\label{Enhancements1A}
\footnotesize
\noindent E_{ob}=\begin{Bmatrix}
\{\psi^{\mu}\} \{ \bar{\psi}^{1,...,5}\}: & \ket{\bm{z}_1} & \ket{\bm{z}_{1} + \bm{z}_{3}} \\
& \ket{\bm{z}_2} &  \ket{\bm{z}_{2} + \bm{z}_{3}}\\
& \ket{\bm{z}_{3}}& \ket{\bm{z}_{1} + \bm{z}_{2} + \bm{z}_{3}} \\
&&\\
\{\psi^{\mu}\}: &\ket{ \bm{S}+ \sum_{i}\bm{e}_{i}+ \sum_{i}\bm{b}_{i} + \bm{\alpha}}&  \ket{ \bm{S}+ \sum_{i}\bm{e}_{i}+ \sum_{i}\bm{b}_{i}  + \bm{\alpha} + \bm{z}_{1}}\\
 &\ket{ \bm{S}+ \sum_{i}\bm{e}_{i}+ \sum_{i}\bm{b}_{i}  + \bm{\alpha} + \bm{z}_{3}} &\ket{ \bm{S}+ \sum_{i}\bm{e}_{i}+ \sum_{i}\bm{b}_{i}  + \bm{\alpha} + \bm{z}_{2} + \bm{z}_{3}}
\end{Bmatrix}.
\end{equation}

Furthermore, we will remove all enhancements to the hidden sector, given below, through appropriate GGSO phases.

\begin{equation}\label{Enhancements2A}
\footnotesize
E_H=\begin{Bmatrix}
\{\psi^{\mu}\}_{\frac{1}{2}} \{ \bar{\lambda}\}_{\frac{1}{2}}: & \ket{\bm{z}_1} & \ket{\bm{z}_2} & \ket{\bm{z}_{3}} \\
 &\ket{\bm{z}_{1} + \bm{z}_{3}} & \ket{\bm{z}_{2} + \bm{z}_{3}} & \ket{\bm{z}_{1} + \bm{z}_{2} + \bm{z}_{3}}\\
& & \\
\{\psi^{\mu}\}_{\frac{1}{2}}: &\ket{ \bm{z}_{1}+ \bm{z}_{2}}
\end{Bmatrix}
\end{equation}

\subsubsection*{Exotics}
The first 32 fermionic exotic sectors to consider take the following form:

\begin{align}\label{exoticsA}
    \begin{split}
        \bm{F}^4_{rs, abc} =& \bm{b_2}+r\bm{e_5}+s\bm{e_6} + \bm{\alpha} + a \bm{z_1} + b \bm{z_3} + c(\bm{z_2} + \bm{z_3}) \\ 
        =&\{\psi^{\mu} ,\chi^{3,4}, w^{12}, \bar{y}^{12}, (1-r)w^{5}\bar{w}^{5}, ry^{5}\bar{y}^{5}, (1-s)w^{6}\bar{w}^{6}, sy^{6}\bar{y}^{6}, \bar{\eta}^{1,2}, \bar{\psi}^{4,5}, \\
         & a \bar{\phi}^{1,2}, (1-a)(1-b)(1-c)\bar{\phi}^{3,4}, b \bar{\phi}^{5,6}, c \bar{\phi}^{7,8}  \}, \\
        \bm{F}^5_{rs, abc} =& \bm{b_3}+r\bm{e_3}+s\bm{e_4}+ \bm{\alpha} + a \bm{z_1} + b \bm{z_3} + c(\bm{z_2} + \bm{z_3}), \\
    \end{split}
\end{align}
where $p,q,r,s,t\in \{0,1\}$, $a,b,c\in \{0,1\}$ and $a+b+c \in \{0,1\}$. These states have repressentations $((2_L,1),(2,1)_H)$, $((2_L,1),(1,2)_H)$, $((1,2_R),(2,1)_H)$ or $((1,2_R),(1,2)_H)$ under $SU(2)_{L} \times SU(2)_{R} \times SO(4)_{(1/2/3/4)} $, where the $SO(4)$ contribution can come from one of the four pairs of hidden sector fermions given in (\ref{exoticsA}). The degeneracy of these states is two, and as such they can be combined into vector-like pairs and lifted out of the massless spectrum. Therefore, these states contribute no chiral exotics. 
The second set of 32 fermionic exotic sectors can be constructed from the previous 32:

\begin{align}\label{exoticsA2}
    \begin{split}
        \bm{F}^6_{rs, abc} =& \bm{F}^4_{rs, abc} + \bm{x}' \\ 
        =&\{\psi^{\mu} ,\chi^{3,4}, y^{12}, \bar{w}^{12}, (1-r)w^{5}\bar{w}^{5}, ry^{5}\bar{y}^{5}, (1-s)w^{6}\bar{w}^{6}, sy^{6}\bar{y}^{6}, \bar{\eta}^{3}, \bar{\psi}^{1,2,3}, \\ 
         & a \bar{\phi}^{1,2}, (1-a)(1-b)(1-c)\bar{\phi}^{3,4}, b \bar{\phi}^{5,6}, c \bar{\phi}^{7,8}  \}, \\
        \bm{F}^7_{rs, abc} =& \bm{F}^5_{rs, abc} + \bm{x}' .\\
    \end{split}
\end{align}

These states have representations of $(4_{Obs},(2,1)_H)$, $(4_{Obs},(1,2)_H)$, $(\bar{4}_{Obs},(2,1)_H)$ and $(\bar{4}_{Obs},(1,2)_H)$, under $SU(4)_{Obs} \times SO(4)_{(1,2,3,4)}$, 
where again the doublets depend on the structure of the hidden sector. Each surviving sector contributes to either $n_{4}$ or $n_{\bar{4}}$, with the chirality phases defined by 

\begin{align}
    \bm{X}^{6} &= -\CC{\bm{F}_{rs}^{6}}{\bm{b}_1+r\bm{e}_5+s\bm{e}_6}^*, \\
    \bm{X}^{7} &= -\CC{\bm{F}_{rs}^{7}}{\bm{b}_1+r\bm{e}_3+s\bm{e}_4}^*.
\end{align}

Finally, there are fermionic exotics that do not come from the observable spinorial sectors and take the form:
\begin{align}\label{exoticsA3}
\begin{split}
    \bm{F}^8_{abc} =& \bm{S} + \bm{x}' + \bm{\alpha} + a \bm{z_1} + b \bm{z_3} + c(\bm{z_2} + \bm{z_3}) \\
                =& \{\psi^{\mu}, \chi^{1,...,6}, \bar{y}^{12}, \bar{w}^{12}, \bar{\eta}^{2,3}, \bar{\psi}^{4,5}, \\
         & a \bar{\phi}^{1,2}, (1-a)(1-b)(1-c)\bar{\phi}^{3,4}, b \bar{\phi}^{5,6}, c \bar{\phi}^{7,8}  \}. \\
         \end{split}
\end{align}
These states have a degeneracy of two and so will not contribute to the net number of chiral exotics. To conform to the stricter condition of having no fermionic exotics in the spectrum, all of the above must be projected, which can be achieved by suitable choice of GGSO phases with the $\bm{z}_i$ vectors. In the SUSY case, this ensures the model is exophobic.

\subsection{Phenomenological Classification}
To statistically analyse this construction, we take a random sample of $10^{9}$ models and classify them according to the above criteria. In particular, we define the following characteristics for phenomenologically viable models.

\begin{align*}\label{ClassConstraints}
    \begin{split}
        &(1) \ \text{No On-Shell Tachyons as discussed in Section \ref{tachyons} and Table \ref{table:tachyonsA1}} \\
        &(2) \ \text{No Observable Enhancements as given in Section \ref{enhancements} and  eqs (\ref{Enhancements1A}, \ref{Enhancements2A})} \\ 
        &(3) \ \text{No Hidden Enhancements }\\ 
        &(4) \ \text{Complete Generations: } \ n_g\neq 0 \text{ and }  n_{4L}-n_{\overline{4}L}=n_{\bar{4}R}-n_{4R} \ \ \\
        &(5) \ \text{Three singly degenerate generations: }\ n_g=3: \ \ \\
        &(6) \ \text{Presence of Heavy Higgs: } \ n_{4R}^H \geq 1 \ \ \\ 
         &(7) \ \text{TQMC, although this is a feature of all models, as discussed in \ref{TQMC}} \\
        &(8) \ \text{No Chiral Exotics, given by $n_{2L/2R}=0$ mod $2$ and $n_4=n_{\bar{4}}$ according to Section \ref{exotics}}  \\ 
        &(9) \ \text{No Fermionic Exotics}  \\ 
    \end{split}
\end{align*}

Below we give the results of our classification in Table \ref{StatstableA}. 
\begin{table}[H]
\small
\centering
\caption{\label{StatstableA} \emph{Phenomenological statistics from sample of $10^9$ Class 0) models.}}
\begin{tabular}{|c|l|r|c|c|c|r|}
\hline
 & \multicolumn{5}{|l|}{Total models in sample: $10^9$}   \\ \hline
  & SUSY or Non-SUSY: & $\mathcal{N}=1$ &Probability& $\mathcal{N}=0$& Probability  \\ \hline
&{ Total} & 3905613 &$3.91\times 10^{-3}$ &996096718& 0.996  \\  \hline
(1)&{+ Tachyon-Free} & \cellcolor{gray!25} &\cellcolor{gray!25}&2073173& $2.07\times 10^{-3}$   \\  \hline
(2)& {+ No Observable Enhancements} & 3498607 &$3.50\times 10^{-3}$&1968976&$1.97\times 10^{-3}$  \\ \hline
(3)& {+ No Enhancements} & 2204351 &$2.20\times 10^{-3}$&1223035&$1.22\times 10^{-3}$  \\ \hline
(4)&{+ Complete Generations} &285093  &$2.85\times 10^{-4}$&164856& $1.65\times10^{-4}$ \\  \hline
(5)&{+ Three Generations} & 26817 &$2.69\times 10^{-5}$&15941& $1.59\times 10^{-5}$  \\  \hline
(6)&{+ Heavy Higgs}& 166 &$1.66\times 10^{-7}$&15841& $1.59\times 10^{-5}$  \\  \hline
(7)&{+ TQMC}& 166 &$1.66\times 10^{-7}$&15941& $1.59\times 10^{-5}$  \\  \hline
(8)&{+ No Chiral Exotics}& 87 &$8.7\times 10^{-8}$&6235& $6.24\times 10^{-6}$  \\  \hline
(9)&{+ No Fermionic Exotics}& 41 &$4.1 \times 10^{-8}$&644&$6.44 \times 10^{-7}$  \\ \hline
\end{tabular}

\end{table}
Table \ref{StatstableA} shows that models that satisfy all phenomenological conditions can be found for both $N=0$ and $N=1$ cases. 
Because we generate the GGSO matrices randomly for our sample, we must take care to check for double counting, and ensure that the program isn't landing on the same GGSO matrix repeatedly. Accounting for this, we found that 38 SUSY and 586 Non-SUSY phenomenologically viable models had unique GGSO matrices.
In the SUSY case, this indicates 38 exophobic models were found.
In the non-SUSY case further investigation indicated that 138 of our phenomenologically viable states were also exophobic.

\subsection{Vacuum Energy Analysis}

\begin{figure}[h]
    \centering
    \includegraphics[width=1\textwidth]{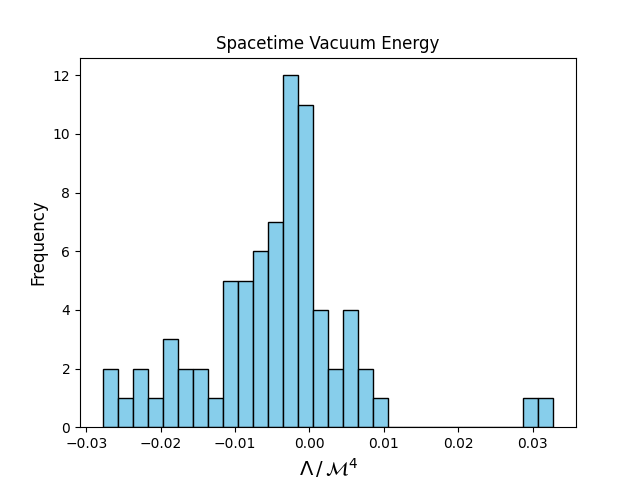}
    \caption{Distribution of spacetime vacuum energy at the free fermionic point for exophobic, non-supersymmetric, phenomenologically viable Class 0) models, displaying 75 distinct values.}
    \label{fig:HistA}
    
\end{figure}

With our phenomenologically viable model gathered, the distribution of the vacuum energy is now examined for the 138 exophobic non-supersymmetric models, as the supersymmetic models have vanishing potential. This calculation is done at the free fermionic point, and the distribution is shown in Figure \ref{fig:HistA}, which shows that negative values are more likely, however positive values are possible. The graph also demonstrates a moderate level of degeneracy of these vacuum energies, as only 75 unique values were found. This degeneracy of vacuum energies in asymmetric models becomes more apparent in Sections \ref{ClassB} and \ref{ClassD}.

\subsection{Spectrum Analysis}
Taking one of our models that fulfils all classification criteria as an example, we find the GGSO matrix has the form given in (\ref{GGSOA}).

\begin{equation}\label{GGSOA}
\small
\CC{\bm{v_i}}{\bm{v_j}}= 
\begin{blockarray}{cccccccccccccc}
&\mathbf{1}& \bm{S} & \bm{e_3}& \bm{e_4}&\bm{e_5}&\bm{e_6}& \bm{b_1}&\bm{b_2}&\bm{b_3}&\bm{z_1}& \bm{z_3}&\bm{\alpha}& \\
\begin{block}{c(rrrrrrrrrrrrr)}
\mathbf{1}&  -1 & -1 & 1 & -1 & 1 & 1 & 1 & -1 & 1 & 1 & -1 & 1 & \\ 
\bm{S}&   -1 & -1 & -1 & -1 & -1 & -1 & 1 & -1 & 1 & -1 & -1 & 1 & \\ 
\bm{e_3}& 1 & -1 & -1 & 1 & -1 & -1 & -1 & 1 & 1 & -1 & 1 & -1 & \\ 
\bm{e_4}& -1 & -1 & 1 & 1 & -1 & -1 & -1 & 1 & -1 & 1 & 1 & 1 & \\ 
\bm{e_5}& 1 & -1 & -1 & -1 & -1 & -1 & -1 & -1 & -1 & 1 & -1 & -1 & \\ 
\bm{e_6}&  1 & -1 & -1 & -1 & -1 & -1 & -1 & -1 & -1 & 1 & -1 & 1  & \\ 
\bm{b_1}& 1 & -1 & -1 & -1 & -1 & -1 & 1 & 1 & -1 & 1 & 1 & 1 & \\ 
\bm{b_2}&   -1 & 1 & 1 & 1 & -1 & -1 & 1 & -1 & -1 & 1 & -1 & 1 & \\ 
\bm{b_3}&   1 & -1 & 1 & -1 & -1 & -1 & -1 & -1 & 1 & 1 & -1 & -1 & \\  
\bm{z_1}&    1 & -1 & -1 & 1 & 1 & 1 & 1 & 1 & 1 & 1 & -1 & -1 & \\ 
\bm{z_3} &   -1 & -1 & 1 & 1 & -1 & -1 & 1 & -1 & -1 & 1 & -1 & 1 & \\ 
\bm{\alpha}&   1 & 1 & -1 & 1 & -1 & 1 & 1 & -1 & 1 & 1 & -1 & 1  & \\ 
\end{block}
\end{blockarray}
\end{equation}

This model is not supersymmetric and 
gives the following partition function and worldsheet 
vacuum energy
\begin{align}
    Z =& \, 2\,q^{0}\bar{q}^{-1} +40\,q\bar{q}^{-1} -8\,q^{\frac{1}{4}}\bar{q}^{-\frac{3}{4}} +16 \,q^{\frac{1}{2}}\bar{q}^{-\frac{1}{2}} +32\,q^{\frac{5}{8}}\bar{q}^{-\frac{3}{8}} +512\,q^{\frac{3}{4}}\bar{q}^{-\frac{1}{4}}\\
    & -1408\,q^{\frac{7}{8}}\bar{q}^{-\frac{1}{8}} + 72 -2208\,q +96\,q^{\frac{1}{8}}\bar{q}^{\frac{1}{8}} +272\,q^{\frac{1}{4}}\bar{q}^{\frac{1}{4}}\nonumber \\
    & +2048\,q^{\frac{3}{8}}\bar{q}^{\frac{3}{8}} -5376\,q^{\frac{1}{2}}\bar{q}^{\frac{1}{2}} +576\,q^{-\frac{3}{8}}\bar{q}^{\frac{5}{8}} -62496\,q^{\frac{5}{8}}\bar{q}^{\frac{5}{8}} \nonumber \\
    & +3904\,q^{-\frac{1}{4}}\bar{q}^{\frac{3}{4}} -138368\,q^{\frac{3}{4}}\bar{q}^{\frac{3}{4}} +18816\,q^{-\frac{1}{8}}\bar{q}^{\frac{7}{8}} -138240\,q^{\frac{7}{8}}\bar{q}^{\frac{7}{8}} \nonumber \\
    & +71520\,\bar{q} +343168\,q\bar{q} \nonumber \\
    \Lambda_\text{WS} =& \,0.770 \\
    \Lambda_\text{ST} =& \, -2.47 \times 10^{-4} \mathcal{M}^4
\end{align}
This model has a complete $\bm{16}$ generation being constructed from $\bm{b}_{1}+\bm{e}_{3}+\bm{e}_{4}+\bm{e}_{5}$, which gives a single $(\bar{4},1,2)$,  and $\bm{b}_{1}+\bm{e}_{3}+\bm{e}_{4}+\bm{e}_{6}$, which gives a single $(4,2,1)$ There is a further 2 $\bm{16}$s coming from $\bm{b}_{3}+\bm{e}_{3}+\bm{e}_{4}$. The model is free from level matched tachyons, and this is confirmed by the form of the partition function and the finite potential. The Heavy Higgs states come from $\bm{S} + \bm{b}_{3}+\bm{e}_{3}+\bm{e}_{4}$. The TQMC terms come from the untwisted Higgs sector and $\bar{\bm{V}}^{3}_{10}$ sector, which survives. This model is also exophobic.

\section{Example Model 1b)}\label{ClassB}
The second class of models to consider is an example of Class 1b), in which $M=(2,2,4)$ real moduli are retained. 
We choose the following basis set

 \begin{align}\label{basisB}
\bm{\mathds{1}}&=\{\psi^\mu,
\chi^{1,\dots,6},y^{1,\dots,6}, w^{1,\dots,6}\ | \ \bar{y}^{1,\dots,6},\bar{w}^{1,\dots,6};
\bar{\psi}^{1,\dots,5},\bar{\eta}^{1,2,3},\bar{\phi}^{1,\dots,8}\},\nonumber\\
\bm{S}&=\{{\psi^\mu},\chi^{1,\dots,6} \},\nonumber\\
\bm{e_2}&=\{y^{2},w^{2}\; | \; \bar{y}^{2},\bar{w}^{2}\}, \ \ \  \nonumber\\
\bm{e_6}&=\{y^{6},w^{6}\; | \; \bar{y}^{6},\bar{w}^{6}\}, \ \ \  \nonumber\\
\bm{b_1}&=\{\psi^\mu,\chi^{12},y^{3,4,5,6}\; | \; \bar{y}^{3,4,5,6}; \bar{\psi}^{1,\dots,5}, \bar{\eta}^1\},\\
\bm{b_2}&=\{\psi^\mu,\chi^{34},y^{1,2},w^{5,6}\; | \; \bar{y}^{1,2},
\bar{w}^{5,6}; \bar{\psi}^{1,\dots,5},\bar{\eta}^2\},\nonumber\\
\bm{b_3}&=\{\psi^\mu,\chi^{56},w^{1,2,3,4}\; | \; \bar{w}^{1,2,3,4}; \bar{\psi}^{1,\dots,5},\bar{\eta}^3\}, \nonumber\\
\bm{z_1}&=\{\bar{\phi}^{1,\dots,4}\},\nonumber\\
\bm{x}'&=\{ \bar{\psi}^{1,\dots,5},\bar{\eta}^{1,2,3},\bar{\phi}^{1,2,7,8} \},\nonumber\\
\bm{\alpha}&=\{y^{1,5}, w^{1,5} \;|\; \bar{y}^{3},\bar{w}^{1,3,4};  \bar{\psi}^{1,2,3},\bar{\eta}^{1}, \bar{\phi}^{3,4,5,6} \},\nonumber 
\end{align}
where in this case the $\bm{x}'$ vector is equivalent to including $\bm{e}_{1345}$ in reducing the degeneracy of our spinorial sectors.

This model is defined by the following characteristics:
\begin{align}
    \begin{split}
    \bm{M} &= (2,2,4)\\
        \bm{\Delta}&=(1,1,1),\\
        \bm{D}&=(2,1,2),\\
        \bm{N}_{\rm as} &= (0,1,1),
    \end{split}
\end{align}
where the projected moduli are $h_{11},h_{21},h_{34}$ and $h_{44}$ via twists in the first and second torus.
The gauge symmetry group of these models is
\begin{align} \label{GGB}
\text{Observable: } \ &SO(6)\times SO(4) \times U(1)_{k=1,2,3} \times U(1)_{l=4,5}  \\
\text{Hidden: } \ & SO(4)^{4},
\end{align} 
where $U(1)_{k=1,2,3}$ correspond to antiholomorphic currents $\bar{\eta}^k\bar{\eta}^{k*}$ and the $U(1)_{l}$ are horizontal symmetries arising 
from the asymmetric pairings $\bar{y}^1\bar{w}^5$ and $\bar{w}^3\bar{w}^4$. \\

\subsection{Phenomenological Features}\label{CPheno}
\subsubsection*{Observable Spinorial Representations}
As more of our asymmetric shift vectors become incompatible with the model, the number of spinorial sectors is also reduced, Applying \ref{generalgenerations}, the sectors producing fermionic generations are:
\begin{align}
    \begin{split} 
    \bm{F}^1_{s}&=\bm{b_1}+s\bm{e_6}, \\
    \bm{F}^2_{qs}&=\bm{b_2}+q\bm{e_2} +s\bm{e_6}, \\
    \bm{F}^3_{q}&=\bm{b_3}+q\bm{e_2},
    \end{split}
\end{align}
where each surviving $\bm{F}^2_{qs}$ states will produce one complete $\bm{16} / \overline{\bm{16}}$ generation, and $\bm{F}^1_{s}$ and $\bm{F}^3_{q}$ will produce two copies of a $\bm{16} / \overline{\bm{16}}$. 

 The projectors for these sectors,
\begin{align}
\mathbb{P}_{\bm{F}^1_{s}}&=\frac{1}{2^3}\left(1 - \CC{\bm{F}^{1}_{s}}{\bm{e}_{2}}\right)
\prod_{a=1,2}\left(1- \CC{\bm{F}^1_{s}}{\bm{z_a}}\right),\label{PF1}\\
\mathbb{P}_{\bm{F}^2_{qs}}&=\frac{1}{2^2}
\prod_{a=1,2}\left(1- \CC{\bm{F}^2_{qs}}{\bm{z_a}}\right),\label{PF2}\\
\mathbb{P}_{\bm{F}^3_{q}}&=\frac{1}{2^3}\left(1 - \CC{\bm{F}^{3}_{q}}{\bm{e_6}}\right)
\prod_{a=1,2}\left(1- \CC{\bm{F}^3_{q}}{\bm{z_a}}\right),\label{PF3}
\end{align}

and we apply eq. (\ref{ChProjs}) to define the chirality phases of these states.
\begin{align}
\begin{split}
\bm{X}^1_{s} &= -\CC{\bm{F}^1_{s}}{\bm{b_2}+s \bm{e_6}}^* \nonumber\\
\bm{X}^2_{qs} &= -\CC{\bm{F}^2_{qs}}{\bm{b_1}+s \bm{e_6}}^*\\
\bm{X}^3_{q} &= -\CC{\bm{F}^3_{q}}{\bm{b_1}}^*\nonumber
\end{split}
\end{align}

\subsubsection*{Heavy Higgs}
In the case that five or more of the spinoral states listed above survive with the correct chirality, the additional $n_{4R}$ and $n_{\bar{4}R}$ states can produce the Heavy Higgs state. However we note that we would still need an odd number of surviving states, each producing a complete generation. Since $\bm{F}^{1,3}$ have degeneracy two, this would require one or three $\bm{F}^{2}_{qs}$ states. Using a similar analysis of phases as in Section \ref{ClassA}, we can conclude that three surviving $\bm{F}^{2}_{qs}$ states is not possible. Therefore, we will require a single state from $\bm{F}^{2}_{qs}$ and for three out of the four $\bm{F}^{1,3}$ states to survive. Again, it can be shown that this is not possible due to the number of available GGSO phases we have, and can't be resolved through the addition of a $\bm{z}_3$ projecting vector. Therefore, for SUSY model, a Heavy Higgs state can't be found. This problem appears to be a consequence of each of our surviving spinorial sectors producing a full $\bm{16}$ or $\overline{\bm{16}}$ generation, and could be mitigated by selecting models without this feature. More complex basis sets have been constructed to achieve this goal for this class of model, found in \cite{alr} and \cite{Leontaris:1999ce}, and we will explore the idea of more complex basis sets further in Example Class 3).

For the non-SUSY models, the Heavy Higgs states can also be found in the sectors
\begin{align}
    \begin{split}
    \bm{B}^1_{s}&=\bm{S}+\bm{b_1}+s\bm{e_6}, \\
    \bm{B}^2_{qs}&=\bm{S}+\bm{b_2}+q\bm{e_2}+s\bm{e_6},\\
    \bm{B}^3_{q}&=\bm{S}+\bm{b_3}+q\bm{e_2}. 
    \end{split}
\end{align}
The additional options for Heavy Higgs states negates the problem described above, and can allow us to produce phenomenologically viable models. Applying the rules set out in Section \ref{HeavyHiggs}, we note that $\bm{B}^{1}_{s}$ and $\bm{B}^{3}_{q}$ will both contribute two copies of a $n_{4R}^B$ states while $\bm{B}^{2}_{qs}$ each contribute a single $n_{4R}^B$.

\subsubsection*{Tachyonic Sector}
In addition to the NS sector, there are only 17 level matched tachyons in this model, which are 
\begin{equation}\label{eq:Tachyons1B}
\footnotesize
T_B=\begin{Bmatrix}
 \{ \bar{\lambda}\}_{\frac{1}{2}}: & \ket{\bm{e}_2} & \ket{\bm{e}_6} & \ket{\bm{e}_{2}+\bm{e}_{6}} \\
 &\ket{ \bm{z}_a }&\ket{ \bm{e}_i + \bm{z}_a} & \ket{ \bm{e}_2 + \bm{e}_6 + \bm{z}_a}\\
 & \ket{\bm{1} + \bm{S}+ \bm{e}_{i} + \bm{\alpha} + \bm{x}'} & &\ket{\bm{1} + \bm{S}+ \bm{e}_{2}+ \bm{e}_{6} + \bm{\alpha} + \bm{x}'} \\
 &\ket{ \bm{z}_1 + \bm{z}_2 + \bm{x}'+\bm{\alpha}} & & \ket{ \bm{z}_1 + \bm{z}_2 + \bm{x}'+\bm{\alpha}+ \bm{e}_{i}}
\end{Bmatrix}.
\end{equation}
Each of these 
can be removed from the spectrum by GGSO projections. 
\subsubsection*{Enhancements}\label{enB}
The number of enhancements in this model is reduced in comparison to the previous model, largely due to $\bm{z}_3$ being absent. 
\begin{equation}\label{Enhs1C}
E_1 = \begin{Bmatrix}
\psi^\mu \{\bar{\lambda}\}_{\frac{1}{2}}: &\ket{\bm{z}_{1}} & \ket{\bm{z}_{2}} \\

\\
\psi^\mu: & \ket{\bm{z}_{1}+\bm{z}_{2}} &
\end{Bmatrix},
\end{equation}
where the observable enhancements, which come from the $\bm{z}_{1}$ and $\bm{z}_{2}$ sectors with $\bar{\lambda} = \bar{\psi}^{1,...,5}$, will be removed as well as the hidden enhancements.

\subsubsection*{Exotics}
In this class the following fermionic exotics arise, 
\begin{align}\label{exoticsB}
\bm{F}^4_{s, ab} =& \bm{1} + \bm{S} + \bm{b_1}+\bm{e_2} + \bm{\alpha} + \bm{x}' + s\bm{e_6} + a \bm{z_1} + b \bm{z_2}, \\ 
\bm{F}^5_{qs} =& \bm{b_2}+ \bm{\alpha} + \bm{x}' + \bm{z_1} + \bm{z_2} + q\bm{e_{2}}+s\bm{e_6}, \\
\bm{F}^6_{s} =& \bm{1} + \bm{S} + \bm{b_3}+\bm{e_6}+ \bm{\alpha} + \bm{x}' + q\bm{e_2},
\end{align}
where $q,s\in \{0,1\}$, $a,b\in \{0,1\}$. The sectors $\bm{F}^4_{s, ab}$ have degeneracy one, and so we must take care to ensure no unpaired chiral exotics arise from these sectors. However $\bm{F}^5_{qs}$ will contribute to both $n_{4}$ and $n_{\bar{4}}$, and $\bm{F}^6_{s}$ have degeneracy four, and therefore these sectors will not produce chiral exotics.  

\subsubsection*{Light Higgs}
Since we have constructed our $\bm{\alpha}$ vector in a way which ensures $(k_1, k_2, k_3) = (1,0,0)$, we have ensured two Higgs bi--doublets and one Higgs triplets pair are retained in our model. All TQMC come from the untwisted type coupling as there are no twisted light Higgs states in this model.

\subsection{Phenomenological Classification}
Taking a random sample of $10^9$ models in this class, we classify them according to the same criteria as in Class 0). The results of this classification are presented in Table \ref{StatstableB}. 
\begin{table}[H]
\small
\centering
\caption{\label{StatstableB} \emph{Phenomenological statistics from sample of $10^9$ Class 1b) models. }}
\begin{tabular}{|c|l|r|c|c|c|r|}
\hline
 & \multicolumn{5}{|l|}{Total models in sample: $10^9$}   \\ \hline
  & SUSY or Non-SUSY: & $\mathcal{N}=1$ &Probability& $\mathcal{N}=0$& Probability  \\ \hline
&{ Total} & 15623650 &$1.56\times 10^{-2}$ &984376350& 0.984  \\  \hline
(1)&{+ Tachyon-Free} & \cellcolor{gray!25} &\cellcolor{gray!25}&190953165& $0.191$   \\  \hline
(2)& {+ No Observable Enhancements} & 14678547 &$1.47\times 10^{-2}$&187601088 &$0.188$  \\ \hline
(3)& {+ No Hidden Enhancements} & 8117340 &$8.12 \times 10^{-3}$&129620167&$0.130$  \\ \hline
(4)&{+ Complete Generations} & 8117340 &$8.12 \times 10^{-3}$&129620167&$0.130$  \\ \hline
(5)&{+ Three Generations} & 244573 &$2.45\times 10^{-4}$&5201786& $5.20\times 10^{-3}$  \\  \hline
(6)&{+ Heavy Higgs}& 0 &0&4262795& $4.28\times 10^{-3}$  \\  \hline
(7)&{+ TQMC}& \cellcolor{gray!25} &\cellcolor{gray!25}&4262795& $4.28\times 10^{-3}$  \\  \hline
(8)&{+ No Chiral Exotics}& \cellcolor{gray!25} &\cellcolor{gray!25} &1926868& $1.93\times 10^{-3}$  \\  \hline
(9)&{+ No Fermionic Exotics }& \cellcolor{gray!25} & \cellcolor{gray!25} &26147& $2.61 \times 10^{-5}$  \\ \hline
\end{tabular}

\end{table}
\noindent The table shows that $\cN=0$ type phenomenological models can be found, but $\cN=1$ models were not found. The abundance of models is largly a consequence of the small number of tachyonic and 
exotic sectors to be projected. We note that, whilst models free from Fermionic exotics were found, no exophobic models were identified.

\subsection{Vacuum Energy Analysis}
As in Section \ref{ClassA}, we analyse the vacuum energy of our phenomenologically viable $\cN=0$ models. 
We analysed a sample of phenomenologically viable $10^4$ GGSO phase configurations and Table \ref{freqB} gives the frequency of vacuum energies we found. This table shows that we have only found nine distinct partition functions and vacuum energies. We note that a matching of $q$-expansions does not mean two models are the same, which would require further analysis of all states, for example of their respective $U(1)$ charges. However, these results strongly suggest that the space of distinct asymmetric models satisfying all phenomological constraints collapses down to only a few models. This is not too surprising given the highly constrained space of asymmetrically twisted models in particular. We note that this constrained space arises both from reducing the moduli space, and concomitant lack of internal shift freedom, as well as the requirement that the twisted sectors produce three chiral generations. 

\newpage

\renewcommand{\arraystretch}{1.2}
\small

\captionof{table}{Distribution of vacuum energy values of $10^4$ phenomenological $\cN = 0$ Class 1b) models, the associated partition function, and a representative GGSO matrix.}
\label{freqB}
\small
\begin{longtable}{|>{\centering\arraybackslash}p{6.5cm}|
>{\centering\arraybackslash}m{5cm}|
>{\centering\arraybackslash}m{1.5cm}|
>{\centering\arraybackslash}m{1.5cm}|}

\hline
Representitive GGSO Matrix & Partition Function & Vacuum Energy / $\mathcal{M}^{4}$ & Frequency \\
\hline
\endfirsthead

\hline
Representitive GGSO Matrix & Partition Function & Vacuum Energy / $\mathcal{M}^{4}$ & Frequency \\
\hline
\endhead

\hline
\endfoot

\hline
\endlastfoot
{\scriptsize
$\begin{array}{c|cccccccccc}
 & \mathbf{1} & \bm{S} & \bm{e}_2 & \bm{e}_6 & \bm{b_1} & \bm{b_2} & \bm{b_3}
 & \bm{z}_1 & \bm{x}' & \bm{\alpha}  \\ \hline
\mathbf{1}  & -1 & 1 & -1 & -1 & -1 & 1 & -1 & 1 & -1 & 1  \\
\bm{S}      &1 & 1 & -1 & -1 & -1 & -1 & 1 & 1 & 1 & 1  \\
\bm{e}_2    &-1 & -1 & 1 & 1 & -1 & -1 & 1 & 1 & -1 & 1  \\
\bm{e}_6    &-1 & -1 & 1 & 1 & -1 & 1 & -1 & 1 & -1 & 1  \\
\bm{b_1}    &-1 & 1 & -1 & -1 & -1 & 1 & -1 & -1 & -1 & -1  \\
\bm{b_2} &1 & 1 & -1 & 1 & 1 & 1 & -1 & -1 & -1 & -1  \\
\bm{b_3}  &-1 & -1 & 1 & -1 & -1 & -1 & -1 & -1 & 1 & -1  \\
\bm{z}_1 &1 & 1 & 1 & 1 & -1 & -1 & -1 & 1 & 1 & 1  \\
\bm{x}' &-1 & 1 & -1 & -1 & 1 & 1 & -1 & -1 & -1 & 1  \\
\bm{\alpha} & 1 & 1 & 1 & 1 & -1 & 1 & -1 & -1 & 1 & -1  \\
\end{array}$

}

 &  $Z = \,  24 +2\,\bar{q}^{-1} +48\,q\bar{q}^{-1} +8\,q^{\frac{1}{4}}\bar{q}^{-\frac{3}{4}}+40\,q^{\frac{1}{2}}\bar{q}^{-\frac{1}{2}} +112\,q^{\frac{5}{8}}\bar{q}^{-\frac{3}{8}} +768 \,q^{\frac{3}{4}}\bar{q}^{-\frac{1}{4}}-1024\,q^{\frac{7}{8}}\bar{q}^{-\frac{1}{8}}  -960\,q +336\,q^{\frac{1}{8}}\bar{q}^{\frac{1}{8}} -1520\,q^{\frac{1}{4}}\bar{q}^{\frac{1}{4}}-7168\,q^{\frac{3}{8}}\bar{q}^{\frac{3}{8}} +192\,q^{-\frac{1}{2}}\bar{q}^{\frac{1}{2}} -13440\,q^{\frac{1}{2}}\bar{q}^{\frac{1}{2}} +928\,q^{-\frac{3}{8}}\bar{q}^{\frac{5}{8}}-12336\,q^{\frac{5}{8}}\bar{q}^{\frac{5}{8}} +5888\,q^{-\frac{1}{4}}\bar{q}^{\frac{3}{4}} -181248\,q^{\frac{3}{4}}\bar{q}^{\frac{3}{4}} +22528\,q^{-\frac{1}{8}}\bar{q}^{\frac{7}{8}}+15360\,q^{\frac{7}{8}}\bar{q}^{\frac{7}{8}} +80544\,\bar{q} +1758080\,q\bar{q}$  & 0.0222 & 2,226\\ \hline 

{\scriptsize
$\begin{array}{c|cccccccccc}
 & \mathbf{1} & \bm{S} & \bm{e}_2 & \bm{e}_6 & \bm{b_1} & \bm{b_2} & \bm{b_3}
 & \bm{z}_1 & \bm{x}' & \bm{\alpha}  \\ \hline
\mathbf{1}  & -1 & -1 & -1 & 1 & -1 & 1 & -1 & -1 & 1 & 1  \\
\bm{S}      &-1 & -1 & -1 & -1 & -1 & 1 & 1 & 1 & 1 & 1 \\
\bm{e}_2    &-1 & -1 & 1 & 1 & -1 & -1 & 1 & 1 & 1 & -1  \\
\bm{e}_6    &1 & -1 & 1 & -1 & -1 & -1 & -1 & 1 & -1 & 1  \\
\bm{b_1}    &-1 & 1 & -1 & -1 & -1 & 1 & -1 & -1 & -1 & 1  \\
\bm{b_2} & 1 & -1 & -1 & -1 & 1 & 1 & -1 & -1 & -1 & -1  \\
\bm{b_3}  & -1 & -1 & 1 & -1 & -1 & -1 & -1 & -1 & -1 & -1  \\
\bm{z}_1 &-1 & 1 & 1 & 1 & -1 & -1 & -1 & -1 & -1 & 1  \\
\bm{x}' &  1 & 1 & 1 & -1 & 1 & 1 & 1 & 1 & 1 & 1  \\
\bm{\alpha} &  1 & 1 & -1 & 1 & 1 & 1 & -1 & -1 & 1 & -1 \\
\end{array}$

}

 &  $Z = \,- 40 + 2\,\bar{q}^{-1} +48\,q\bar{q}^{-1} +40\,q^{\frac{1}{2}}\bar{q}^{-\frac{1}{2}} +176\,q^{\frac{5}{8}}\bar{q}^{-\frac{3}{8}} -512 \,q^{\frac{3}{4}}\bar{q}^{-\frac{1}{4}}  -5824\,q +336\,q^{\frac{1}{8}}\bar{q}^{\frac{1}{8}} -512\,q^{\frac{1}{4}}\bar{q}^{\frac{1}{4}}-7168\,q^{\frac{3}{8}}\bar{q}^{\frac{3}{8}} +128\,q^{-\frac{1}{2}}\bar{q}^{\frac{1}{2}} +1920\,q^{\frac{1}{2}}\bar{q}^{\frac{1}{2}} +864\,q^{-\frac{3}{8}}\bar{q}^{\frac{5}{8}}-12336\,q^{\frac{5}{8}}\bar{q}^{\frac{5}{8}} +4352\,q^{-\frac{1}{4}}\bar{q}^{\frac{3}{4}} -243200\,q^{\frac{3}{4}}\bar{q}^{\frac{3}{4}} +21504\,q^{-\frac{1}{8}}\bar{q}^{\frac{7}{8}}+15360\,q^{\frac{7}{8}}\bar{q}^{\frac{7}{8}} +70560\,\bar{q} +753536\,q\bar{q}$  & 0.0255 & 2,220\\ \hline 

 {\scriptsize
$\begin{array}{c|cccccccccc}
 & \mathbf{1} & \bm{S} & \bm{e}_2 & \bm{e}_6 & \bm{b_1} & \bm{b_2} & \bm{b_3}
 & \bm{z}_1 & \bm{x}' & \bm{\alpha}  \\ \hline
\mathbf{1}  & -1 & -1 & 1 & 1 & 1 & 1 & -1 & 1 & -1 & 1 \\
\bm{S}      &-1 & -1 & -1 & -1 & 1 & 1 & -1 & 1 & 1 & -1 \\
\bm{e}_2    &1 & -1 & -1 & 1 & -1 & 1 & 1 & 1 & -1 & 1 \\
\bm{e}_6    &1 & -1 & 1 & -1 & 1 & 1 & -1 & 1 & 1 & -1 \\
\bm{b_1}    & 1 & -1 & -1 & 1 & 1 & 1 & -1 & -1 & 1 & 1 \\
\bm{b_2} & 1 & -1 & 1 & 1 & 1 & 1 & -1 & -1 & 1 & 1 \\
\bm{b_3}  &  -1 & 1 & 1 & -1 & -1 & -1 & -1 & -1 & 1 & -1 \\
\bm{z}_1 & 1 & 1 & 1 & 1 & -1 & -1 & -1 & 1 & -1 & -1 \\
\bm{x}' &  -1 & 1 & -1 & 1 & -1 & -1 & -1 & 1 & -1 & -1 \\
\bm{\alpha} &  1 & -1 & 1 & -1 & 1 & -1 & -1 & 1 & -1 & -1 \\
\end{array}$

}

 &  $Z = \, 88 + 2\,\bar{q}^{-1} +48\,q\bar{q}^{-1} +40\,q^{\frac{1}{2}}\bar{q}^{-\frac{1}{2}} +112\,q^{\frac{5}{8}}\bar{q}^{-\frac{3}{8}} -576 \,q^{\frac{3}{4}}\bar{q}^{-\frac{1}{4}} -4288\,q +16\,q^{\frac{1}{8}}\bar{q}^{\frac{1}{8}} +640\,q^{\frac{1}{4}}\bar{q}^{\frac{1}{4}}-10240\,q^{\frac{3}{8}}\bar{q}^{\frac{3}{8}} +128\,q^{-\frac{1}{2}}\bar{q}^{\frac{1}{2}} -3968\,q^{\frac{1}{2}}\bar{q}^{\frac{1}{2}} +864\,q^{-\frac{3}{8}}\bar{q}^{\frac{5}{8}}-15216\,q^{\frac{5}{8}}\bar{q}^{\frac{5}{8}} +3904\,q^{-\frac{1}{4}}\bar{q}^{\frac{3}{4}} -242944\,q^{\frac{3}{4}}\bar{q}^{\frac{3}{4}} +22528\,q^{-\frac{1}{8}}\bar{q}^{\frac{7}{8}}+34816\,q^{\frac{7}{8}}\bar{q}^{\frac{7}{8}} +72096\,\bar{q} +771968\,q\bar{q}$  & 0.00711 & 1,162\\ \hline 

{\scriptsize
$\begin{array}{c|cccccccccc}
 & \mathbf{1} & \bm{S} & \bm{e}_2 & \bm{e}_6 & \bm{b_1} & \bm{b_2} & \bm{b_3}
 & \bm{z}_1 & \bm{x}' & \bm{\alpha}  \\ \hline
\mathbf{1}  & -1 & -1 & 1 & 1 & 1 & 1 & 1 & -1 & -1 & -1  \\
\bm{S}      & -1 & -1 & 1 & 1 & -1 & -1 & -1 & 1 & 1 & 1  \\
\bm{e}_2    &1 & 1 & -1 & 1 & -1 & -1 & -1 & 1 & -1 & -1  \\
\bm{e}_6    &1 & 1 & 1 & -1 & 1 & 1 & -1 & 1 & -1 & -1  \\
\bm{b_1}    &1 & 1 & -1 & 1 & 1 & 1 & -1 & -1 & -1 & -1  \\
\bm{b_2} & 1 & 1 & -1 & 1 & 1 & 1 & -1 & -1 & -1 & -1  \\
\bm{b_3}  &1 & 1 & -1 & -1 & -1 & -1 & 1 & -1 & -1 & -1  \\
\bm{z}_1 &-1 & 1 & 1 & 1 & -1 & -1 & -1 & -1 & -1 & 1  \\
\bm{x}' &-1 & 1 & -1 & -1 & 1 & 1 & 1 & 1 & -1 & 1  \\
\bm{\alpha} & -1 & 1 & -1 & -1 & -1 & 1 & -1 & -1 & 1 & 1  \\
\end{array}$

}

 &  $Z = \, 88 +2\,\bar{q}^{-1} +48\,q\bar{q}^{-1} +8\,q^{\frac{1}{4}}\bar{q}^{-\frac{3}{4}} +40\,q^{\frac{1}{2}}\bar{q}^{-\frac{1}{2}} +176\,q^{\frac{5}{8}}\bar{q}^{-\frac{3}{8}} +768 \,q^{\frac{3}{4}}\bar{q}^{-\frac{1}{4}}-1536\,q^{\frac{7}{8}}\bar{q}^{-\frac{1}{8}} -704\,q +144\,q^{\frac{1}{8}}\bar{q}^{\frac{1}{8}} -2032\,q^{\frac{1}{4}}\bar{q}^{\frac{1}{4}}-5120\,q^{\frac{3}{8}}\bar{q}^{\frac{3}{8}} +192\,q^{-\frac{1}{2}}\bar{q}^{\frac{1}{2}} -13440\,q^{\frac{1}{2}}\bar{q}^{\frac{1}{2}} +928\,q^{-\frac{3}{8}}\bar{q}^{\frac{5}{8}}-11504\,q^{\frac{5}{8}}\bar{q}^{\frac{5}{8}} +6400\,q^{-\frac{1}{4}}\bar{q}^{\frac{3}{4}} -178176\,q^{\frac{3}{4}}\bar{q}^{\frac{3}{4}} +20992\,q^{-\frac{1}{8}}\bar{q}^{\frac{7}{8}}-15360\,q^{\frac{7}{8}}\bar{q}^{\frac{7}{8}} +80800\,\bar{q} +1759104\,q\bar{q}$  & 0.0155 & 1,126\\ \hline 

{\scriptsize
$\begin{array}{c|cccccccccc}
 & \mathbf{1} & \bm{S} & \bm{e}_2 & \bm{e}_6 & \bm{b_1} & \bm{b_2} & \bm{b_3}
 & \bm{z}_1 & \bm{x}' & \bm{\alpha}  \\ \hline
\mathbf{1}  & -1 & 1 & 1 & 1 & 1 & -1 & 1 & 1 & -1 & -1  \\
\bm{S}      & 1 & 1 & -1 & -1 & -1 & 1 & 1 & 1 & 1 & -1  \\
\bm{e}_2    &1 & -1 & -1 & 1 & -1 & 1 & 1 & 1 & -1 & 1  \\
\bm{e}_6    &1 & -1 & 1 & -1 & -1 & -1 & -1 & 1 & -1 & 1  \\
\bm{b_1}    &1 & 1 & -1 & -1 & 1 & 1 & -1 & -1 & 1 & -1  \\
\bm{b_2} & -1 & -1 & 1 & -1 & 1 & -1 & -1 & -1 & 1 & 1  \\
\bm{b_3}  & 1 & -1 & 1 & -1 & -1 & -1 & 1 & -1 & -1 & -1  \\
\bm{z}_1 &1 & 1 & 1 & 1 & -1 & -1 & -1 & 1 & -1 & -1  \\
\bm{x}' & -1 & 1 & -1 & -1 & -1 & -1 & 1 & 1 & -1 & -1  \\
\bm{\alpha} & -1 & -1 & 1 & 1 & -1 & -1 & -1 & 1 & -1 & 1  \\
\end{array}$

}

 &  $Z = \, 152 + 2\,\bar{q}^{-1} +48\,q\bar{q}^{-1} +8\,q^{\frac{1}{4}}\bar{q}^{-\frac{3}{4}}+40\,q^{\frac{1}{2}}\bar{q}^{-\frac{1}{2}} +112\,q^{\frac{5}{8}}\bar{q}^{-\frac{3}{8}} +320 \,q^{\frac{3}{4}}\bar{q}^{-\frac{1}{4}}  +576\,q +16\,q^{\frac{1}{8}}\bar{q}^{\frac{1}{8}} -368\,q^{\frac{1}{4}}\bar{q}^{\frac{1}{4}}-10240\,q^{\frac{3}{8}}\bar{q}^{\frac{3}{8}} +192\,q^{-\frac{1}{2}}\bar{q}^{\frac{1}{2}} -11392\,q^{\frac{1}{2}}\bar{q}^{\frac{1}{2}} +864\,q^{-\frac{3}{8}}\bar{q}^{\frac{5}{8}}-15216\,q^{\frac{5}{8}}\bar{q}^{\frac{5}{8}} +5952\,q^{-\frac{1}{4}}\bar{q}^{\frac{3}{4}} -180992\,q^{\frac{3}{4}}\bar{q}^{\frac{3}{4}} +22528\,q^{-\frac{1}{8}}\bar{q}^{\frac{7}{8}}+34816\,q^{\frac{7}{8}}\bar{q}^{\frac{7}{8}} +82080\,\bar{q} +1776512\,q\bar{q}$  & 0.00383 & 1,113\\ \hline 

{\scriptsize
$\begin{array}{c|cccccccccc}
 & \mathbf{1} & \bm{S} & \bm{e}_2 & \bm{e}_6 & \bm{b_1} & \bm{b_2} & \bm{b_3}
 & \bm{z}_1 & \bm{x}' & \bm{\alpha}  \\ \hline
\mathbf{1}  &-1 & 1 & 1 & -1 & 1 & -1 & 1 & -1 & 1 & -1 \\
\bm{S}      & 1 & 1 & -1 & -1 & -1 & 1 & -1 & 1 & -1 & 1 \\
\bm{e}_2    & 1 & -1 & -1 & 1 & -1 & 1 & 1 & -1 & 1 & -1 \\ 
\bm{e}_6    &-1 & -1 & 1 & 1 & -1 & 1 & -1 & -1 & -1 & 1 \\ 
\bm{b_1}    & 1 & 1 & -1 & -1 & 1 & 1 & -1 & 1 & -1 & -1 \\ 
\bm{b_2} & -1 & -1 & 1 & 1 & 1 & -1 & 1 & -1 & -1 & -1 \\ 
\bm{b_3}  &  1 & 1 & 1 & -1 & -1 & 1 & 1 & 1 & -1 & -1 \\
\bm{z}_1 & -1 & 1 & -1 & -1 & 1 & -1 & 1 & -1 & -1 & -1 \\
\bm{x}' &  1 & -1 & 1 & -1 & 1 & 1 & 1 & 1 & 1 & 1 \\
\bm{\alpha} &  -1 & 1 & -1 & 1 & -1 & 1 & -1 & 1 & 1 & 1 \\
\end{array}$

}

 &  $Z = \,  24 +2\,\bar{q}^{-1} +48\,q\bar{q}^{-1} +40\,q^{\frac{1}{2}}\bar{q}^{-\frac{1}{2}} +176\,q^{\frac{5}{8}}\bar{q}^{-\frac{3}{8}} +1024 \,q^{\frac{3}{4}}\bar{q}^{-\frac{1}{4}}-1536\,q^{\frac{7}{8}}\bar{q}^{-\frac{1}{8}}  -5568\,q +144\,q^{\frac{1}{8}}\bar{q}^{\frac{1}{8}} -1024\,q^{\frac{1}{4}}\bar{q}^{\frac{1}{4}}-5120\,q^{\frac{3}{8}}\bar{q}^{\frac{3}{8}} +128\,q^{-\frac{1}{2}}\bar{q}^{\frac{1}{2}} +1920\,q^{\frac{1}{2}}\bar{q}^{\frac{1}{2}} +928\,q^{-\frac{3}{8}}\bar{q}^{\frac{5}{8}}-11504\,q^{\frac{5}{8}}\bar{q}^{\frac{5}{8}} +4352\,q^{-\frac{1}{4}}\bar{q}^{\frac{3}{4}} -240128\,q^{\frac{3}{4}}\bar{q}^{\frac{3}{4}} +20992\,q^{-\frac{1}{8}}\bar{q}^{\frac{7}{8}}-15360\,q^{\frac{7}{8}}\bar{q}^{\frac{7}{8}} +70816\,\bar{q} +754560\,q\bar{q}$  & 0.0187 & 1,105\\ \hline 

{\scriptsize
$\begin{array}{c|cccccccccc}
 & \mathbf{1} & \bm{S} & \bm{e}_2 & \bm{e}_6 & \bm{b_1} & \bm{b_2} & \bm{b_3}
 & \bm{z}_1 & \bm{x}' & \bm{\alpha}  \\ \hline
\mathbf{1}  & -1& 1& 1& 1& 1& 1& -1& -1& -1& -1 \\
\bm{S}      &  1& 1& -1& 1& 1& 1& 1& 1& -1& -1 \\
\bm{e}_2    &  1& -1& -1& -1& -1& 1& 1& -1& 1& 1  \\
\bm{e}_6    & 1& 1& -1& -1& -1& -1& 1& 1& -1& -1   \\
\bm{b_1}    &  1& -1& -1& -1& 1& -1& -1& -1& -1& 1  \\
\bm{b_2} &  1& -1& 1& -1& -1& 1& -1& -1& -1& -1 \\
\bm{b_3}  &  -1& -1& 1& 1& -1& -1& -1& 1& -1& -1 \\
\bm{z}_1 &  -1& 1& -1& 1& -1& -1& 1& -1& -1& 1 \\
\bm{x}' &  -1& -1& 1& -1& 1& 1& 1& 1& -1& 1 \\
\bm{\alpha} &  -1& -1& 1& -1& 1& 1& -1& -1& 1& 1 \\
\end{array}$
}

 &  $Z = \, 496 + 2\,\bar{q}^{-1} +48\,q\bar{q}^{-1} +152\,q^{\frac{1}{2}}\bar{q}^{-\frac{1}{2}} -80\,q^{\frac{5}{8}}\bar{q}^{-\frac{3}{8}} -576 \,q^{\frac{3}{4}}\bar{q}^{-\frac{1}{4}}-2048\,q^{\frac{7}{8}}\bar{q}^{-\frac{1}{8}}  +18144\,q +320\,q^{\frac{1}{8}}\bar{q}^{\frac{1}{8}} -1920\,q^{\frac{1}{4}}\bar{q}^{\frac{1}{4}}-9216\,q^{\frac{3}{8}}\bar{q}^{\frac{3}{8}} +200\,q^{-\frac{1}{2}}\bar{q}^{\frac{1}{2}} -30592\,q^{\frac{1}{2}}\bar{q}^{\frac{1}{2}} +656\,q^{-\frac{3}{8}}\bar{q}^{\frac{5}{8}}+31040\,q^{\frac{5}{8}}\bar{q}^{\frac{5}{8}} +4032\,q^{-\frac{1}{4}}\bar{q}^{\frac{3}{4}} -41728\,q^{\frac{3}{4}}\bar{q}^{\frac{3}{4}} +21504\,q^{-\frac{1}{8}}\bar{q}^{\frac{7}{8}}-236544\,q^{\frac{7}{8}}\bar{q}^{\frac{7}{8}} +81088\,\bar{q} +1064704\,q\bar{q}$  & -0.0606 & 357\\ \hline 

{\scriptsize
$\begin{array}{c|cccccccccc}
 & \mathbf{1} & \bm{S} & \bm{e}_2 & \bm{e}_6 & \bm{b_1} & \bm{b_2} & \bm{b_3}
 & \bm{z}_1 & \bm{x}' & \bm{\alpha}  \\ \hline
\mathbf{1}  & -1& -1& -1& 1& -1& 1& 1& 1& 1& -1 \\
\bm{S}      &  -1& -1& 1& -1& 1& -1& 1& 1& -1& -1 \\
\bm{e}_2    &  -1& 1& 1& 1& 1& -1& -1& -1& -1& -1  \\
\bm{e}_6    & 1& -1& 1& -1& -1& 1& -1& 1& -1& -1   \\
\bm{b_1}    &  -1& -1& 1& -1& -1& -1& -1& 1& 1& 1  \\
\bm{b_2} &  1& 1& -1& 1& -1& 1& -1& -1& -1& 1 \\
\bm{b_3}  &  1& -1& -1& -1& -1& -1& 1& -1& -1& -1 \\
\bm{z}_1 &  1& 1& -1& 1& 1& -1& -1& 1& -1& 1 \\
\bm{x}' &  1& -1& -1& -1& -1& 1& 1& 1& 1& -1 \\
\bm{\alpha} &  -1& -1& -1& -1& 1& -1& -1& -1& -1& 1 \\
\end{array}$
}
 &  $ Z = \, 312 +2\,\bar{q}^{-1} +48\,q\bar{q}^{-1} +108\,q^{\frac{1}{2}}\bar{q}^{-\frac{1}{2}} +112 \,q^{\frac{5}{8}}\bar{q}^{-\frac{3}{8}} -64\,q^{\frac{3}{4}}\bar{q}^{-\frac{1}{4}} -2048 \,q^{\frac{7}{8}}\bar{q}^{-\frac{1}{8}} +12480\,q -96\,q^{\frac{1}{8}}\bar{q}^{\frac{1}{8}} -384\,q^{\frac{1}{4}}\bar{q}^{\frac{1}{4}}-14336\,q^{\frac{3}{8}}\bar{q}^{\frac{3}{8}} +156\,q^{-\frac{1}{2}}\bar{q}^{\frac{1}{2}} +18784\,q^{\frac{1}{2}}\bar{q}^{\frac{1}{2}} +944\,q^{-\frac{3}{8}}\bar{q}^{\frac{5}{8}}-71264\,q^{\frac{5}{8}}\bar{q}^{\frac{5}{8}} +4032\,q^{-\frac{1}{4}}\bar{q}^{\frac{3}{4}} -14080\,q^{\frac{3}{4}}\bar{q}^{\frac{3}{4}} +2048\,q^{-\frac{1}{8}}\bar{q}^{\frac{7}{8}}-288768\,q^{\frac{7}{8}}\bar{q}^{\frac{7}{8}} +83104\,\bar{q} +1532800\,q\bar{q}$ & -0.0246 & 353 \\ 
 \hline 

{\scriptsize
$\begin{array}{c|cccccccccc}
 & \mathbf{1} & \bm{S} & \bm{e}_2 & \bm{e}_6 & \bm{b_1} & \bm{b_2} & \bm{b_3}
 & \bm{z}_1 & \bm{x}' & \bm{\alpha}  \\ \hline
\mathbf{1}  & -1& -1& -1& -1& 1& 1& -1& 1& 1& 1 \\
\bm{S}      &  -1& -1& 1& -1& -1& 1& 1& 1& -1& 1 \\
\bm{e}_2    &  -1& 1& 1& 1& 1& -1& -1& 1& 1& 1 \\
\bm{e}_6    & -1& -1& 1& 1& -1& -1& -1& 1& 1& 1  \\
\bm{b_1}    &  1& 1& 1& -1& 1& 1& 1& 1& 1& 1  \\
\bm{b_2} &  1& -1& -1& -1& 1& 1& -1& -1& 1& -1 \\
\bm{b_3}  &  -1& -1& -1& -1& 1& -1& -1& -1& 1& -1 \\
\bm{z}_1 &  1& 1& 1& 1& 1& -1& -1& 1& 1& -1 \\
\bm{x}' &  1& -1& 1& 1& -1& -1& -1& -1& 1& 1 \\
\bm{\alpha} &  1& 1& 1& 1& 1& 1& -1& 1& 1& -1 \\
\end{array}$
}
&
   $ Z = \,376+ 2\,\bar{q}^{-1} +48\,q\bar{q}^{-1} +4\,q^{\frac{1}{8}}\bar{q}^{-\frac{7}{8}} +108\,q^{\frac{1}{2}}\bar{q}^{-\frac{1}{2}} -64\,q^{\frac{3}{4}}\bar{q}^{-\frac{1}{4}} -2048\,q^{\frac{7}{8}}\bar{q}^{-\frac{1}{8}}  +14784 \,q +164\,q^{\frac{1}{8}}\bar{q}^{\frac{1}{8}} -896\,q^{\frac{1}{4}}\bar{q}^{\frac{1}{4}} -17408\,q^{\frac{3}{8}}\bar{q}^{\frac{3}{8}} +220\,q^{-\frac{1}{2}}\bar{q}^{\frac{1}{2}}+5984\,q^{\frac{1}{2}}\bar{q}^{\frac{1}{2}} +1200\,q^{-\frac{3}{8}}\bar{q}^{\frac{5}{8}} -127952\,q^{\frac{5}{8}}\bar{q}^{\frac{5}{8}}+6592\,q^{-\frac{1}{4}}\bar{q}^{\frac{3}{4}} +1280\,q^{\frac{3}{4}}\bar{q}^{\frac{3}{4}} +29696\,q^{-\frac{1}{8}}\bar{q}^{\frac{7}{8}} -113664\,q^{\frac{7}{8}}\bar{q}^{\frac{7}{8}} +101792\,\bar{q} +2205568\,q\bar{q}$  & -0.0282 & 338\\ 
   \hline 


\end{longtable}

\subsection{Spectrum Analysis}
To explore the spectrum further, we will 
take a model from Table \ref{freqB} and see which sectors contribute to the spectrum. In this case we take
\begin{equation}
\small
\CC{\bm{v_i}}{\bm{v_j}}= 
\begin{blockarray}{cccccccccccc}
&\mathbf{1}& \bm{S} & \bm{e_2}& \bm{e_6}& \bm{b_1}&\bm{b_2}&\bm{b_3}&\bm{z_1}&\bm{x}' & \bm{\alpha}& \\
\begin{block}{c(rrrrrrrrrrr)}
\mathbf{1}&-1 & 1 & 1 & 1 &-1 &-1& -1 & 1 & 1 &-1&\ \\
\bm{S}   & 1 & 1 &-1 & 1 & 1 & 1 & 1 & 1 & 1 & 1&\ \\
\bm{e_2}& 1 &-1 &-1& -1 &-1 & 1 & 1 & 1 &-1 &-1&\ \\
\bm{e_6}&  1 & 1 &-1 &-1 &-1 & 1 &-1 & 1& -1 &-1&\ \\
\bm{b_1}&-1& -1 &-1& -1 &-1 &-1& -1& -1& -1 & 1&\ \\
\bm{b_2}& -1 &-1 & 1 & 1 &-1 &-1 & 1 &-1 & 1 & 1&\ \\
\bm{b_3}&  -1 &-1 & 1 &-1 &-1 & 1 &-1 & 1 & 1 &-1& \ \\  
\bm{z_1}& 1 & 1 & 1 & 1 &-1 &-1 & 1 & 1 & 1 &-1& \  \\
\bm{x}'& 1 & 1 &-1 &-1 & 1 &-1 &-1 &-1 & 1 & 1& \ \\ 
\bm{\alpha}& -1 & 1 &-1 &-1 & 1 &-1 &-1 & 1 & 1 & 1& \ \\
\end{block}
\end{blockarray} \; .
\end{equation}



This model has three complete $\bm{16}$ generations: two $\bm{16}$s coming from $\bm{b_1}$, and a $\bm{16}$ coming from $\bm{b}_{2}+\bm{e}_{2}$.

The model is free from level matched tachyons, and this is confirmed by the form of the partition function and the finite 
vacuum energy.
The Heavy Higgs states come from $\bm{S} + \bm{b}_{2}$ and $\bm{S} + \bm{b}_{3} + \bm{e}_{2}$, as this is a non-supersymmetric model.

\section{Example Model 2a)}\label{ClassC}
In this model we 
project eight of our 12 real moduli to give a Class 2a) model with $M=(0,0,4)$. In addition, we introduce the constraint that all untwisted Higgs triplets are to be projected, giving us more control over the spectrum. 
The basis for this class of models is
\begin{align}\label{basisC}
\bm{\mathds{1}}&=\{\psi^\mu,
\chi^{1,\dots,6},y^{1,\dots,6}, w^{1,\dots,6}\ | \ \bar{y}^{1,\dots,6},\bar{w}^{1,\dots,6}; 
\bar{\psi}^{1,\dots,5},\bar{\eta}^{1,2,3},\bar{\phi}^{1,\dots,8}\},\nonumber\\
\bm{S}&=\{{\psi^\mu},\chi^{1,\dots,6} \},\nonumber\\
\bm{e_6}&=\{y^{6},w^{6}\; | \; \bar{y}^{6},\bar{w}^{6}\}, \ \ \  \nonumber\\
\bm{e_{135}}&=\{y^{135},w^{135}\; | \; \bar{y}^{135},\bar{w}^{135}\}, \ \ \  \nonumber\\
\bm{b_1}&=\{\psi^\mu,\chi^{12},y^{3,4,5,6}\; | \; \bar{y}^{3,4,5,6}: \bar{\psi}^{1,\dots,5}, \bar{\eta}^1\},\\
\bm{b_2}&=\{\psi^\mu,\chi^{34},y^{1,2},w^{5,6}\; | \; \bar{y}^{1,2},
\bar{w}^{5,6}; \bar{\psi}^{1,\dots,5}, \bar{\eta}^2\},\nonumber\\
\bm{b_3}&=\{\psi^\mu,\chi^{56},w^{1,2,3,4}\; | \; 
\bar{w}^{1,2,3,4}; \bar{\psi}^{1,\dots,5}, \bar{\eta}^3\},\nonumber\\
\bm{x}'&=\{ \bar{\psi}^{1,\dots,5},\bar{\eta}^{1,2,3},\bar{\phi}^{2,3,7,8} \},\nonumber\\
\bm{z_1}&=\{\bar{\phi}^{1,\dots,4}\},\nonumber\\
\bm{\alpha}&=\{y^{1,3,5}, w^{1,3,5} \;|\; \bar{w}^{1,2,3,4}; \bar{\psi}^{1,2,3},\bar{\phi}^{3,4} \}.\nonumber \\
\end{align}
Note that we have used $\bm{x}'$ instead of $\tilde{\bm{e}}_{24}$, as suggested in Table \ref{tab:AddBasVec}, however the spectrum remains the same. 

The overlap of hidden sector fermions between $\bm{x}'$ and $\bm{\alpha}$ introduces additional possible chiral exotics to the model. The model characteristics are 
\begin{align}
    \begin{split}
    \bm{M} &= (0,0,4)\\
        \bm{\Delta}&=(1,1,1)\\
        \bm{D}&=(1,1,2)\\
        \bm{N}_{as} &= (0,0,1)
    \end{split}
\end{align}
and the gauge symmetry group 
is
\begin{align} \label{GGC}
\text{Observable: } \ &SO(6)\times SO(4) \times U(1)_{k=1,2,3} \times U(1)_{l=4,5,6}  \\
\text{Hidden: } \ &SO(2)^{4} \times SO(4)^{2}.
\end{align} 
where $U(1)_{k=1,2,3}$ correspond to antiholomorphic currents $\bar{\eta}^k\bar{\eta}^{k*}$ and the $U(1)_{l=4,5,6}$ are horizontal symmetries arising from the asymmetric pairings: $\bar{y}^3\bar{y}^5,\bar{y}^1\bar{w}^5$ and $\bar{w}^2\bar{w}^4$. 

\subsection{Class 2a) Phenomenological Features}\label{CPheno}
\subsubsection*{Observable Spinorial Representations}
The sectors producing fermionic generations are
\begin{align}
    \begin{split} 
    \bm{F}^1_{s}&=\bm{b_1}+s\bm{e_6} \\
    \bm{F}^2_{s}&=\bm{b_2}+s\bm{e_6} \\
    \bm{F}^3 &=\bm{b_3}.
    \end{split}
\end{align}
Due to the fact $\bm{F}^3$ has a degeneracy of two, we know that we need either one or three of the other four states to survive. The six phases that determine this are: 
\beq 
\CC{\bm{b_1}}{\bm{z_1}}, \ \CC{\bm{b_1}}{\bm{z_2}}, \CC{\bm{b_2}}{\bm{z_1}}, \ \CC{\bm{b_2}}{\bm{z_2}}, \ \CC{\bm{e_6}}{\bm{z_1}},\CC{\bm{e_6}}{\bm{z_2}}
\eeq 
Taking the case where $\bm{F}^3$ is projected, we can assume without loss of generality that both $\bm{F}^1_{0}$ and $\bm{F}^1_{1}$ survive, which fixes
\begin{align}
    \CC{\bm{b_1}}{\bm{z_1}} &= \CC{\bm{b_1}}{\bm{z_2}} = -1\\ 
    \CC{\bm{e_6}}{\bm{z_1}} &= \CC{\bm{e_6}}{\bm{z_2}} = 1 ,\nonumber
\end{align}
and so we would require one state from $\bm{F}^2_{s}$. However, for either of the states to survive we would fix
\beq 
\CC{\bm{b_2}}{\bm{z_1}} = \CC{\bm{b_2}}{\bm{z_2}} = -1,
\eeq
which guarentees the survival of both states and so rules out three generation models. Therefore we know that three generation models require the survival of $\bm{F}^3$ and just one other state. This substantially constrains the models and has an unwanted effect on the exotic states, as well shall discuss later.

\subsubsection*{Heavy Higgs}
In this class, for SUSY models, the model only has five candidates to produce a minimum of four $n_{\bar{4}R}$ states and one $n_{4R}$ state required for the presence of three complete generations and a Heavy Higgs state. However we have already shown that $\bm{F}^3$ must survive along with only one other state to produce odd numbers of generations and so it is not possible to produce the four full $\bm{16}$'s and a single $\overline{\bm{16}}$ required. Similar to Section \ref{ClassB}, we suggest that this is due to the sectors all producing complete generations. In order to build a model which doesn't produce full generations would require a model with at least one $\Delta_i = 0$. If this is the case, eq. \ref{littledelta}, tells us $\delta^j = 0$, and by the modular invarience rules of eq. \ref{eq:MI_Ab1}, this requires $k_j = 1$, and the retention of a Higgs Triplet. That implies that the projection of all untwisted Higgs Triplets is incompatible with the presence of a Heavy Higgs state in the spectrum for this class of models. We discuss this point in more detail in Section \ref{heavyhiggs}. 

In the non-SUSY case, the Heavy Higgs producing sectors can also be constructed through the addition of the $\bm{S}$ vector to our spinorial sectors:
\begin{align}
    \begin{split}
    \bm{B}^1_{s}&=\bm{S}+\bm{F}^1_{s} \\
    \bm{B}^2_{s}&=\bm{S}+\bm{F}^2_{s} \\
    \bm{B}^3 &=\bm{S}+\bm{F}^3, 
    \end{split}
\end{align}
 All $\bm{B}^i$ sectors produce $(\mathbf{16}+\overline{\mathbf{16}})$ pairs and so will contribute to $n_{4R}^{B}$. 
\subsubsection*{Light Higgs}
This model preserves all untwisted Higgs doublets since $(k_1, k_2, k_3)=(0,0,0)$. However, we note that additional twisted vectorial $\bm{10}$ arise generating light Higgs fields from the sector
\begin{align}
    \bm{V}^3 &= \bm{b}_1 + \bm{b}_2 +\bm{e}_6 + \bm{e}_{135} +  \bm{S} \\ 
    &=\{\psi^\mu,\chi^{56}, y^{2,4}, w^{1,3} \ | \ \byy^{2,4},\bw^{1,3} ; \ \bget^{1,2}\}. \nonumber
\end{align}

\subsubsection*{Tachyonic Sectors}
There are 18 on-shell tachyonic sectors to GGSO project
\begin{equation}\label{Enhancements1C}
\footnotesize
T_B=\begin{Bmatrix}
 \{ \bar{\lambda}\}_{\frac{1}{2}}: & \ket{\bm{e}_6} & & \ket{\bm{e}_{135}}  \\
 \\
  &\ket{ \bm{e}_6 + \bm{z}_a} & & \ket{ \bm{e}_{135} + \bm{z}_a}\\

 &\ket{ \bm{z}_a } & & \ket{ \bm{\alpha} + \bm{z}_{a}} \\
 &\ket{\bm{1} + \bm{S}+ \bm{e}_{135}+ \bm{x}'} & &\ket{\bm{1} + \bm{S}+\bm{e}_{6} + \bm{e}_{135}+ \bm{x}'} \\

  & \ket{\bm{1} + \bm{S}+ \bm{e}_{135}+ \bm{x}'+ \bm{z}_a} &  & \ket{\bm{1} + \bm{S}+\bm{e}_{6} + \bm{e}_{135}+ \bm{x}'+ \bm{z}_a} \\ 

    & \ket{\bm{1} + \bm{S}+ \bm{e}_{135}+ \bm{x}'+ \bm{z}_1 + \bm{z}_{2}} &  & \ket{\bm{1} + \bm{S}+\bm{e}_{6} + \bm{e}_{135}+ \bm{x}'+ \bm{z}_1 + \bm{z}_{2}} \\
\end{Bmatrix}.
\end{equation}

\subsubsection*{Enhancements}\label{enC}
The following enhancement sectors arise in this class of models, which will be projected from the spectrum. 
\begin{equation}\label{Enhs1C}
\begin{Bmatrix}
\psi^\mu \{\bar{\lambda}\}_{\frac{1}{2}}: &\ket{\bm{z_1}} & \ket{\bm{z_2}} \\

\psi^\mu: & \ket{\bm{z_1}+\bm{z_2}} &\ket{\bm{e_{135}}+\bm{\alpha}}& \ket{\bm{e_{135}}+\bm{\alpha} +\bm{z_1}}
\end{Bmatrix}.
\end{equation}

\subsubsection*{Exotic Sectors}\label{exoticsC}
The exotics take the form
\begin{align}
\bm{F}^{4}_{s,a} =& \bm{b}_{1} +\bm{e}_{135} + \bm{\alpha} + s \bm{e}_{6} + a \bm{z}_{1}  \\
 =&\{\psi^{\mu} ,\chi^{1,2}, y^{3,4,5},\bar{y}^{1,4}, \bar{w}^{2,4,5}, (1-s)y^{6}\bar{y}^{6}, sw^{6}\bar{w}^{6};\nonumber \\
         & \bar{\psi}^{4,5}, \bar{\eta}^{1},  (1-a) \bar{\phi}^{3,4}, a \bar{\phi}^{1,2}\}, \nonumber \\
\bm{F}^{5}_{s,a} =& \bm{b}_{2} +\bm{e}_{135} + \bm{\alpha} + s \bm{e}_{6} + a \bm{z}_{1}, \\ 
\bm{F}^{6}_{a} =& \bm{b}_{3} +\bm{e}_{135} + \bm{\alpha} + a \bm{z}_{1}, \\ 
 \bm{F}^{7}_{a} =& \bm{S} +\bm{e}_{135} + \bm{\alpha} + a \bm{z}_{1}, 
\end{align}
where $s\in \{0,1\}$, $a \in \{0,1\}$. We note that the first 10 of these exotic states have degeneracy of two and only have one projecting vector: $\bm{z}_2$. The final two states will contribute to both $n_{4}$ and $n_{\bar{4}}$ equally. Therefore we can conclude this model will fulfil the criteria of being free of chiral exotics regardless of the GGSO phases. 

However it can be shown that three generation models necessarily come with exotic states in this class of models. 
We have already proven that for a model to have three generations, we require the survival of $\bm{F}^3$, which fixes the phases
 \beq 
\CC{\bm{b_3}}{\bm{z_1}} = \CC{\bm{b_3}}{\bm{z_2}} = \CC{\bm{b_3}}{\bm{e_6}} = -1.
\eeq
To project state $\bm{F}^{6}_{0}$, we are therefore forced to fix
 \beq 
\CC{\bm{e}_{135}+\bm{\alpha}}{\bm{z}_{2}} = -1.
\eeq
No matter which of our other $\bm{F}^{1,2}_{s}$ sectors we choose to produce our third fermionic generation, we are forced to project at least one of these sectors by $\bm{z}_{2}$. We will define $\bm{f}$ to be the state that must be projected by $\bm{z}_{2}$. Therefore,
 \begin{align}
    \CC{\bm{z_2}}{\bm{f}} &= 1 \\ 
    \CC{\bm{z_2}}{\bm{f}+\bm{e}_{135}+\bm{\alpha}} &=\CC{\bm{z_2}}{\bm{f}}\CC{\bm{z}_2}{\bm{e}_{135}+\bm{\alpha}} \\
    &= -1
 \end{align}
 and so the exotic state $\bm{f}+\bm{e}_{135}+\bm{\alpha}$ will survive. Therefore three generation models free from exotics can not be found in this class of models. 
 This type of effect was common across other classes of models with all Higgs triplets projected and eight moduli projected, pointing to the fact that models constructed by the addition of the simplest compatible vectors as described in Table \ref{tab:AddBasVec} are not optimal. 
 We also note that this problem cannot be solved by breaking the hidden sector further with some vector, {\it e.g.}
 \begin{equation}
\bm{z}_{4} = \{ \bar{\phi}^{1,2,6,7}\}.
 \end{equation}
A similar analysis to the above can be done to show this.


\subsection{Phenomenological Classification}
To aid comparison with Examples 0) and 1b), we also take a random sample of $10^9$ models in this class and classify them according to the same criteria. The results of this classification are presented in Table \ref{StatstableC}. 

\begin{table}[H]
\small
\centering
\caption{\label{StatstableC} \emph{Phenomenological statistics from sample of $10^9$ Class 2a) models. }}
\begin{tabular}{|c|l|r|c|c|c|r|}
\hline
 & \multicolumn{5}{|l|}{Total models in sample: $10^9$}   \\ \hline
  & SUSY or Non-SUSY: & $\mathcal{N}=1$ &Probability& $\mathcal{N}=0$& Probability  \\ \hline
&{ Total} & 15619225 &$1.56\times 10^{-2}$ &984380775& 0.984  \\  \hline
(1)&{+ Tachyon-Free} & \cellcolor{gray!25} &\cellcolor{gray!25}&67876211& $6.79\times 10^{-2}$   \\  \hline
(2)& {+ No Observable Enhancements} & 14673058 &$1.47\times 10^{-2}$&64412994&$6.44 \times 10^{-2}$  \\ \hline
(3)& {+ No Hidden Enhancements} & 9355407 &$9.36\times 10^{-3}$&44002382&$4.40\times 10^{-2}$  \\ \hline
(4)&{+ Complete Generations} & 9355407 &$9.36\times 10^{-3}$&44002382&$4.40\times 10^{-2}$  \\ \hline
(5)&{+ Three Generations} & 108534 &$1.09\times 10^{-4}$&604041& $6.04\times 10^{-4}$  \\  \hline
(6)&{+ Heavy Higgs}& 0 &0&421737& $4.22\times 10^{-4}$  \\  \hline
(7)&{+ TQMC}& \cellcolor{gray!25} &\cellcolor{gray!25}&421737& $4.22\times 10^{-4}$  \\  \hline
(8)&{+ No Chiral Exotics}& \cellcolor{gray!25} &\cellcolor{gray!25} &421737&$4.22\times 10^{-4}$  \\ \hline
(9)&{+ No Fermionic Exotics}& \cellcolor{gray!25} &\cellcolor{gray!25}&0& 0 \\ \hline
\end{tabular}

\end{table}

As in Class 1b) this class of models initially shows a large number of stable, three generation models. However, as described in Section \ref{exoticsC}, there is an incompatibility between retaining three fermionic generations and projecting all exotic states.

The absence of models free from fermionic exotics appears to be a result of the specific representative basis set chosen for the class of models. In this respect, for issues with exotics and more involved phenomenological criteria, the additional vectors of Table \ref{tab:AddBasVec} should be utilised in more complex constructions, in particular combining them into heavier basis vectors. This idea is explored in the next section.



\section{Example Model 3)}\label{ClassD}
Finally, in the following section we give an example of a 
Class 3) model with no geometric moduli $M=(0,0,0)$. 
Our model could be built, according to Table \ref{tab:AddBasVec}, as 
\begin{align}\label{basisD}
\bm{\mathds{1}}&=\{\psi^\mu,
\chi^{1,\dots,6},y^{1,\dots,6}, w^{1,\dots,6}\ | \ \bar{y}^{1,\dots,6},\bar{w}^{1,\dots,6};
\bar{\psi}^{1,\dots,5},\bar{\eta}^{1,2,3},\bar{\phi}^{1,\dots,8}\},\nonumber\\
\bm{S}&=\{{\psi^\mu},\chi^{1,\dots,6} \},\nonumber\\
\tilde{\bm{e}}_{24}&=\{y^{2,4},w^{2,4}\; | \; \bar{y}^{2,4},\bar{w}^{2,4}; \bar{\phi}^{4,5,6,7}\}, \ \ \  \nonumber\\
\tilde{\bm{e}}_{36}&=\{y^{3,6},w^{3,6}\; | \; \bar{y}^{3,6},\bar{w}^{3,6}; \bar{\phi}^{4,5,6,7}\}, \ \ \  \nonumber\\
\tilde{\bm{e}}_{15}&=\{y^{1,5},w^{1,5}\; | \; \bar{y}^{1,5},\bar{w}^{1,5}; \bar{\phi}^{4,5,6,7}\}, \ \ \  \nonumber\\
\bm{b_1}&=\{\psi^\mu,\chi^{12},y^{3,4,5,6}\; | \; \bar{y}^{3,4,5,6}; \bar{\psi}^{1,\dots,5}, \bar{\eta}^1\},\\
\bm{b_2}&=\{\psi^\mu,\chi^{34},y^{1,2},w^{5,6}\; | \; \bar{y}^{1,2},
\bar{w}^{5,6}; \bar{\psi}^{1,\dots,5},\bar{\eta}^2\},\nonumber\\
\bm{b_3}&=\{\psi^\mu,\chi^{56},w^{1,2,3,4}\; | \; 
\bar{w}^{1,2,3,4}; \bar{\psi}^{1,\dots,5}, \bar{\eta}^3\},\nonumber\\
\bm{z_1}&=\{\bar{\phi}^{3,4,7,8}\},\nonumber\\
\bm{\alpha} &= \{y^{2,4}, w^{2,4} \;|\; \bar{y}^{1,2,3,4,6},\bar{w}^{5}; \bar{\psi}^{1,2,3},\bar{\phi}^{1,2,3,4} \},\nonumber 
\end{align}
to give the characteristics
\begin{align}
    \begin{split}
    \bm{M} &= (0,0,0),\\
        \bm{\Delta}&=(1,1,1),\\
        \bm{D}&=(1,1,1),\\
        \bm{N}_{as} &= (0,0,0).
    \end{split}
\end{align}
However, similar to Example Model 2a), 
we observe 
that exotic sectors such as $\bm{\alpha}+ \bm{b}_1 + \bm{e}_{24}+ \bm{z}_2$, can only projected by $\bm{z}_2 = \{\bar{\phi}^{1,2,5,6}\}$. 

To construct our model in such a way that no exotic sectors are permitted regardless of the GGSO matrix, 
we modify our basis by considering
\begin{align}\label{basisD1}
\bm{e}_{1245}&=\{y^{1,2,4,5},w^{1,2,4,5}\; | \; \bar{y}^{1,2,4,5},\bar{w}^{1,2,4,5}\}, \ \ \  \nonumber\\
\tilde{\bm{e}}_{1356}&=\{y^{1,3,5,6},w^{1,3,5,6}\; | \; \bar{y}^{1,3,5,6},\bar{w}^{1,3,5,6}, \bar{\phi}^{3,4,7,8}\}, \ \ \  \nonumber\\
\bm{e}_{2346}&=\{y^{2,3,4,6},w^{2,3,4,6}\; | \; \bar{y}^{2,3,4,6},\bar{w}^{2,3,4,6}\}, \ \ \  
\end{align} 

which are simply the pairwise addtion of the simplest symmetric vectors presented in \ref{basisD}. We then redefine the basis vectors as
\begin{align}\label{basisD2}
\bm{\beta}&= \bm{\alpha}+ \bm{e}_{1245}, \nonumber\\
\bm{\gamma}&= \bm{\alpha}+ \bm{e}_{2346}, \nonumber\\
\bm{\delta}&= \bm{\alpha}+ \tilde{\bm{e}}_{1356},
\end{align} 

to construct the modified basis set 
\begin{align}\label{basisD3}
\bm{\mathds{1}}&=\{\psi^\mu,
\chi^{1,\dots,6},y^{1,\dots,6}, w^{1,\dots,6}\ | \ \bar{y}^{1,\dots,6},\bar{w}^{1,\dots,6};
\bar{\psi}^{1,\dots,5},\bar{\eta}^{1,2,3},\bar{\phi}^{1,\dots,8}\},\nonumber\\
\bm{S}&=\{{\psi^\mu},\chi^{1,\dots,6} \},\nonumber\\
\bm{b_1}&=\{\psi^\mu,\chi^{12},y^{3,4,5,6}\; | \; \bar{y}^{3,4,5,6}; \bar{\psi}^{1,\dots,5}, \bar{\eta}^1\},\\
\bm{b_2}&=\{\psi^\mu,\chi^{34},y^{1,2},w^{5,6}\; | \; \bar{y}^{1,2},
\bar{w}^{5,6}; \bar{\psi}^{1,\dots,5},\bar{\eta}^2\},\nonumber\\
\bm{b_3}&=\{\psi^\mu,\chi^{56},w^{1,2,3,4}\; | \; 
\bar{w}^{1,2,3,4}; \bar{\psi}^{1,\dots,5},\bar{\eta}^3\},\nonumber\\
\bm{\alpha} &= \{y^{2,4}, w^{2,4} \;|\; \bar{y}^{1,2,3,4,6},\bar{w}^{5}; \bar{\psi}^{1,2,3}, \bar{\phi}^{1,2,3,4} \}, \nonumber \\
\bm{\beta} &= \{y^{1,5}, w^{1,5} \;|\; \bar{y}^{3,5,6}, \bar{w}^{1,2,4}; \bar{\psi}^{1,2,3}, \bar{\phi}^{1,2,3,4} \},\nonumber \\
\bm{\gamma} &= \{y^{3,6}, w^{3,6} \;|\; \bar{y}^{1},\bar{w}^{2,3,4,5,6}; \bar{\psi}^{1,2,3},\bar{\phi}^{1,2,3,4} \},\nonumber \\
\bm{\delta} &= \{y^{1,2,3,4,5,6}, w^{1,2,3,4,5,6} \;|\; \bar{y}^{2,4,5},\bar{w}^{1,3,6}; \bar{\psi}^{1,2,3},\bar{\phi}^{1,2,7,8} \},\nonumber 
\end{align}

The gauge group of these models is
\begin{align} \label{GGD}
\text{Observable: } \ &SO(6)\times SO(4) \times U(1)_{k=1,2,3} \times U(1)_{l=4,5,6}  \\
\text{Hidden: } \ &SO(4)^{4}.
\end{align} 
where $U(1)_{k=1,2,3}$ correspond to antiholomorphic currents $\bar{\eta}^k\bar{\eta}^{k*}$ and the $U(1)_{l=4,5,6}$ are horizontal symmetries arising from the asymmetric pairings: $\bar{y}^3\bar{y}^6,\bar{y}^1\bar{w}^5$ and $\bar{w}^2\bar{w}^4$. \\

\subsubsection*{Observable Spinorial Repressentations}

This model will have a single complete generation coming from each of $\bm{b}_1$, $\bm{b}_2$ and $\bm{b}_3$, which, for consistency, we will label
\begin{align}
    \begin{split} 
    \bm{F}^1&=\bm{b_1}, \\
    \bm{F}^2&=\bm{b_2},\\
    \bm{F}^3 &=\bm{b_3}.
    \end{split}
\end{align}

As we have no compatible symmetric shift vectors, we rely on all three of these sectors producing a chiral generaion with the correct chirality, which makes this class of models more constrained than the previous models considered.

\subsubsection*{Heavy Higgs}
The disadvantage of projecting all moduli, and therefore removing all symmetric shift vectors, is that for $\cN=1$ models, there are no longer enough sectors to produce three generations and Heavy Higgs states. Since we did not find Heavy Higgs states for the previous two classes, this model is phenomenologically comparable to the others. In the non-SUSY case, we will take our Heavy Higgs state to come from one of the three following states,

\begin{align}
    \begin{split} 
    \bm{B}^1&=\bm{S}+\bm{F}^1, \\
    \bm{B}^2&=\bm{S}+\bm{F}^2, \\
    \bm{B}^3 &=\bm{S}+\bm{F}^3,
    \end{split}
\end{align}
where all three contribute a full $(\bm{16}+ \overline{\bm{16}})$.

\subsubsection*{Light Higgs}
With the untwisted sector only producing Higgs doublets, there are no Higgs triplets inherent to this class. However, triplets arising from the twisted vectorial $\bm{10}$ sectors,
\begin{align}
\bm{V}_1 &= \bm{b}_2 +\bm{b}_3+\bm{\alpha}+\bm{\beta}, \\ 
\bm{V}_2 &= \bm{b}_1+\bm{b}_3+\bm{\alpha}+\bm{\gamma},  \nonumber \\
\bm{V}_{3} &= \bm{b}_{1}+\bm{b}_{2}+\bm{\beta}+\bm{\gamma}, \nonumber
\end{align}
can be generated for some choices of GGSO phases. As we have removed all untwisted Higgs Triplets, we will also classify our models by the survival of these states, in particular the associated triplets $\bgps^{1,2,3} \ket{ \bm{V}_i }$, as they can play a role in the missing 
partner mechanism.
\subsubsection*{Tachyonic Sectors}
There are only two tachyonic sectors in this class, 
\begin{align}
    \bm{z}_1 &=\bm{\alpha} + \bm{\beta} +\bm{\gamma} +\bm{\delta}, \\ 
    &= \{ \bar{\phi}^{3,4,7,8}  \}, \nonumber \\ 
    \bm{z}_2 &= \bm{z}_1 +\bm{H},
\end{align}
which have projectors,

\begin{equation}\label{ProjD}
\footnotesize
\Upsilon=\begin{Bmatrix}
 \bm{S}, & \bm{b}_i\\
\bm{\alpha}+\bm{\beta}, & \bm{\alpha}+\bm{\gamma}\\ 
\end{Bmatrix} ,
\end{equation}
as well as acting as projectors for each other. 

\subsubsection*{Enhancements}
The only enhancements to this model come from $\bm{z}_1$, $\bm{z}_2$ and $\bm{\zeta}$. The small number of tachyonic and gauge enhancing sectors suggests that this model will be very fertile, with the largest constraint being the reduced number of spinorial sectors. 

\subsubsection*{Exotics}
This class of models was constructed to be free of all exotics and so there are no constraints on the GGSO matrix coming from the projection of exotic states. All models produced in from this construction will be exophobic.


\subsection{Phenomenological Classification}
This model can be analysed and classified through the same process and the results are shown in Table \ref{StatstableD}. 
\begin{table}[H]
\small
\centering
\caption{\label{StatstableD} \emph{Phenomenological statistics from sample of $10^9$ Class 3) models, where we note that all models satisfy condition 9 an account of being exophobic. }}
\begin{tabular}{|c|l|r|c|c|c|r|}
\hline
 & \multicolumn{5}{|l|}{Total models in sample: $10^9$}   \\ \hline
  & SUSY or Non-SUSY: & $\mathcal{N}=1$ &Probability& $\mathcal{N}=0$& Probability  \\ \hline
&{ Total} & 31244781 &$3.12\times 10^{-2}$ &968755219& 0.969  \\  \hline
(1)&{+ Tachyon-Free} & \cellcolor{gray!25} &\cellcolor{gray!25}&851557918& $0.852$   \\  \hline
(2)& {+ No Observable Enhancements} & 24404805&$2.44\times 10^{-2}$&756848649 &$0.757$  \\ \hline
(3)& {+ No Hidden Enhancements} & 13733548 &$1.37 \times 10^{-2}$&590267212&$0.590$  \\ \hline
(4)&{+ Complete Generations} & 13733548 &$1.37 \times 10^{-2}$&590267212&$0.590$  \\ \hline
(5)&{+ Three Generations} & 275372 &$2.75\times 10^{-4}$&11683005& $1.17 \times 10^{-2}$  \\  \hline
(6)&{+ Heavy Higgs}& 0 &0&1923772& $1.92 \times 10^{-3}$  \\  \hline
(7)&{+ TQMC}& \cellcolor{gray!25} &\cellcolor{gray!25}&1923772& $1.92 \times 10^{-3}$  \\  \hline
(8)&{+ Survival of $\bm{V}_i$ }& \cellcolor{gray!25} &\cellcolor{gray!25}&643232& $6.43 \times 10^{-4}$  \\  \hline
(9)&{+ No Fermionic Exotics }& \cellcolor{gray!25} &\cellcolor{gray!25} &643232& $6.43 \times 10^{-4}$  \\  \hline
\end{tabular}

\end{table}
This table shows that the non-supersymmetric Class 3 example produces many models that adhere to the phenomenological constraints, including being exophobic. 
 
\subsection{Partition function and Vacuum Energy}
Taking the first $10^{4}$ randomly generated, phenomenologically viable, non-SUSY models from our scan and calculating their vacuum energy shows that only five unique values can be found. These results are recorded in Table \ref{freqD}.

\begin{table}[htbp]
\centering
\small
\caption{\label{freqD} Distribution of the vacuum energy values of $10^4$ phenomenological $\cN=0$ Class 3) models, the associated partition function, and a representitive GGSO matrix.}
\renewcommand{\arraystretch}{1.2}
\begin{tabular}{|>{\centering\arraybackslash}p{6cm}|
>{\centering\arraybackslash}m{5cm}|
>{\centering\arraybackslash}m{1.5cm}|
>{\centering\arraybackslash}m{1.5cm}|}
\hline
Representative GGSO Matrix & Partition Function & Vacuum Energy / $\mathcal{M}^{4}$ & Frequency\\
\hline

{\scriptsize
$\begin{array}{c|ccccccccc}
 & \mathbf{1} & \bm{S} & \bm{b_1} & \bm{b_2} & \bm{b_3}
 & \bm{\alpha} & \bm{\beta} & \bm{\gamma} & \bm{\delta} \\ \hline
\mathbf{1}  & -1 &1 &-1 &-1 &1 &1 &-1 &1 &-1 \\
\bm{S}      &  1& 1 &-1 &-1 &1 &1 &1 &-1& -1 \\
\bm{b_1}    &  -1 &1 &-1 &-1 &-1 &-1 &1 &1 &-1 \\
\bm{b_2}    &  -1 &1 &-1 &-1& -1 &-1 &-1 &1 &1 \\
\bm{b_3}    &  1 &-1 &-1 &-1 &1 &1 &-1& -1 &1 \\
\bm{\alpha} &  1 &1 &-1& -1& -1& -1 &-1& 1& 1 \\
\bm{\beta}  &  -1& 1 &1 &1 &-1 &-1& 1& 1& 1 \\
\bm{\gamma} &  1 &-1 &-1& 1 &-1& 1& 1& -1& -1 \\
\bm{\delta} &  -1 &-1& 1 &-1& -1& 1& 1& -1& -1
\end{array}$
}& $ Z =  296 +2\bar q^{-1} +24q\bar q^{-1} -32q^{\frac12}\bar q^{-\frac12}  +3552q -45568q^{\frac12}\bar q^{\frac12} +10240q^{-\frac14}\bar q^{\frac34} -20480q^{\frac34}\bar q^{\frac34} +183168\bar q +2198016q\bar q$ 
& $-0.0280$ & 3,337 \\
\hline

{\scriptsize
$\begin{array}{c|ccccccccc}
 & \mathbf{1} & \bm{S} & \bm{b_1} & \bm{b_2} & \bm{b_3}
 & \bm{\alpha} & \bm{\beta} & \bm{\gamma} & \bm{\delta} \\ \hline
\mathbf{1}  & -1 & 1 & 1 & 1 & 1 & 1 & 1 & 1 & 1 \\
\bm{S}      &  1 & 1 & 1 & -1 & -1 & -1 & -1 & 1 & 1 \\
\bm{b_1}    &  1 & -1 & 1 & -1 & -1 & 1 & -1 & 1 & -1 \\
\bm{b_2}    &  1 & 1 & -1 & 1 & -1 & -1 & 1 & -1 & 1 \\
\bm{b_3}    &  1 & 1 & -1 & -1 & 1 & -1 & -1 & 1 & 1 \\
\bm{\alpha} &  1 & -1 & 1 & -1 & 1 & -1 & -1 & -1 & -1 \\
\bm{\beta}  &  1 & -1 & -1 & -1 & -1 & -1 & -1 & -1 & -1 \\
\bm{\gamma} &  1 & 1 & -1 & -1 & 1 & -1 & -1 & -1 & 1 \\
\bm{\delta} &  1 & 1 & 1 & -1 & -1 & -1 & -1 & 1 & 1
\end{array}$
}
& {\centering
$Z =  296 +2\bar q^{-1} +56q\bar q^{-1} -64q^{\frac12}\bar q^{-\frac12}  +3168q + 2048q^{\frac14}\bar q^{\frac14} +128q^{-\frac12}\bar q^{\frac12} -36864q^{\frac12}\bar q^{\frac12} +7168q^{-\frac14}\bar q^{\frac34} -407552q^{\frac34}\bar q^{\frac34} +152448\bar q +1667584q\bar q$ \arraybackslash
}
& $-0.0372$ & 3,241 \\
\hline

{\scriptsize
$\begin{array}{c|ccccccccc}
 & \mathbf{1} & \bm{S} & \bm{b_1} & \bm{b_2} & \bm{b_3}
 & \bm{\alpha} & \bm{\beta} & \bm{\gamma} & \bm{\delta} \\ \hline
\mathbf{1}  & -1& 1& 1 &-1& -1& 1& 1& 1& 1 \\
\bm{S}      &  1& 1& -1&1 &-1& 1& 1 &1 &1 \\
\bm{b_1}    &  1& 1& 1 &-1& -1& 1 &1 &-1& -1 \\
\bm{b_2}    & -1 &-1 &-1 &-1 &-1& 1 &-1 &-1& 1  \\
\bm{b_3}    &  -1& 1 &-1 &-1 &-1 &1 &1 &1 &1  \\
\bm{\alpha} &  1& 1 &1 &1 &-1 &-1& -1& -1& -1 \\
\bm{\beta}  &  1 &1 &1 &1 &1 &-1& -1& -1& 1 \\
\bm{\gamma} &  1 &1 &1 &-1& 1& -1& -1& -1& -1 \\
\bm{\delta} &  1 &1 &1 &-1 &-1& -1 &1 &-1 &1
\end{array}$
} 
& $ Z = -24 + 2\bar q^{-1} +56q\bar q^{-1}  -672q +2048q^{\frac14}\bar q^{\frac14} +128q^{-\frac12}\bar q^{\frac12} -24576q^{\frac12}\bar q^{\frac12} +9216q^{-\frac14}\bar q^{\frac34} -411648q^{\frac34}\bar q^{\frac34} +133248\bar q +1437184q\bar q$ 
& $0.0191$ & 1,469 \\
\hline
{\scriptsize
$\begin{array}{c|ccccccccc}
 & \mathbf{1} & \bm{S} & \bm{b_1} & \bm{b_2} & \bm{b_3}
 & \bm{\alpha} & \bm{\beta} & \bm{\gamma} & \bm{\delta} \\ \hline
\mathbf{1}  & -1 &1 &-1 &-1 &-1 &-1 &-1 &-1& -1 \\
\bm{S}      &  1 &1 &1 &1 &1 &-1& -1& 1& 1 \\
\bm{b_1}    &  -1 &-1 &-1 &-1 &-1& -1 &-1 &1 &1 \\
\bm{b_2}    & -1 &-1 &-1& -1& -1& 1 &1 &1 &1  \\
\bm{b_3}    &  -1 &-1& -1& -1& -1& 1& -1& -1& 1  \\
\bm{\alpha} &  -1 &-1& -1& 1 &-1& 1 &-1 &-1& -1 \\
\bm{\beta}  &  -1 &-1 &-1& -1 &-1 &-1& 1 &-1& 1 \\
\bm{\gamma} &  -1 &1 &-1& 1 &-1& -1 &-1& 1 &-1 \\
\bm{\delta} &  -1 &1 &-1& -1& -1& -1& 1& -1& -1
\end{array}$
} 

& $ Z = 456 +2\bar q^{-1} +56q\bar q^{-1} -96q^{\frac12}\bar q^{-\frac12}  +5088q +2048q^{\frac14}\bar q^{\frac14} +128q^{-\frac12}\bar q^{\frac12} -43008q^{\frac12}\bar q^{\frac12} +6144q^{-\frac14}\bar q^{\frac34} -405504q^{\frac34}\bar q^{\frac34} +162048\bar q +1782784q\bar q$ 
& $-0.0653$ & 977 \\
\hline
{\scriptsize
$\begin{array}{c|ccccccccc}
 & \mathbf{1} & \bm{S} & \bm{b_1} & \bm{b_2} & \bm{b_3}
 & \bm{\alpha} & \bm{\beta} & \bm{\gamma} & \bm{\delta} \\ \hline
\mathbf{1}  & -1 &-1& 1 &-1 &-1 &-1 &-1& 1& -1 \\
\bm{S}      &  -1 &-1& 1& 1& -1 &-1& 1& 1& -1 \\
\bm{b_1}    &  1 &-1& 1 &-1 &-1 &1 &1 &-1& -1  \\
\bm{b_2}    & -1 &-1& -1 &-1& -1 &-1& 1& -1& 1  \\
\bm{b_3}    &  -1& 1& -1& -1& -1& -1& -1& -1& -1  \\
\bm{\alpha} &  -1& -1& 1& -1& 1& 1& -1& 1& -1 \\
\bm{\beta}  &  -1& 1& 1& -1& -1& -1& 1& 1& 1 \\
\bm{\gamma} &  1& 1 &1& -1& -1& 1& 1& -1& -1 \\
\bm{\delta} &  -1& -1& 1& -1& 1& -1& 1& -1& -1
\end{array}$
}

& $ Z = 456 + 2\bar q^{-1} +24q\bar q^{-1} -64q^{\frac12}\bar q^{-\frac12}  +5472q -51712q^{\frac12}\bar q^{\frac12} +9216q^{-\frac14}\bar q^{\frac34} -18432q^{\frac34}\bar q^{\frac34} +192768\bar q +2313216q\bar q$ 
& $-0.0580$ & 976 \\
\hline

\end{tabular}

\end{table}

It is interesting to note that only powers in multiples of $1/4$ appear in these partition functions unlike in the other example model classes which have multiples of $1/8$. This is due to there being no single $\bm{e}_i$ vectors in the basis set, which acts to half the number of mass levels.

\subsection{Example Model}

Taking an example GGSO phase matrix, we again look at which states contribute to the spectrum. Taking the second GGSO Matrix listed in Table \ref{freqD},
\begin{equation}
\small
\CC{\bm{v_i}}{\bm{v_j}}= 
\begin{blockarray}{ccccccccccc}
&\mathbf{1}& \bm{S} & \bm{b_1}&\bm{b_2}&\bm{b_3}& \bm{\alpha}& \bm{\beta} & \bm{\gamma} & \bm{\delta}\\
\begin{block}{c(rrrrrrrrrr)}
\mathbf{1}&-1 & 1 & 1&  1 & 1 & 1 & 1 & 1 & 1&\ \\
\bm{S}   &  1 & 1 & 1 &-1 &-1 &-1 &-1 & 1 & 1 & \ \\
\bm{b_1}&  1 &-1 & 1 &-1 &-1 & 1 &-1 & 1 &-1& \ \\
\bm{b_2}&   1 & 1 &-1 & 1 &-1 &-1 & 1 &-1 & 1& \ \\
\bm{b_3}&  1 & 1 &-1 &-1 & 1 &-1 &-1 & 1 & 1& \ \\ 
\bm{\alpha}& 1 &-1 & 1 &-1 & 1 &-1 &-1 &-1& -1 & \ \\
\bm{\beta}&  1 &-1 &-1 &-1 &-1 &-1 &-1 &-1& -1 & \ \\
\bm{\gamma}& 1 & 1 &-1 &-1 & 1 &-1 &-1 &-1 & 1& \ \\
\bm{\delta}& 1 & 1 & 1& -1& -1& -1& -1&  1 & 1 & \ \\
\end{block}
\end{blockarray} \text{  },
\end{equation}

all three Heavy Higgs states from the sectors, $\bm{S}+\bm{b}_i$, are present in the spectrum, along those from $\bm{V}_3$ and the three fermionic generations from $\bm{b}_i$.

\subsection{Triplet Free Models}
We can also analyse the case where all three vectorial $\bm{V}_i$ states are projected out, with all other phenomenological criteria satisfied. In this case there are no triplets in the model. We find that this restricted the allowed potentials further, given in Table \ref{freqE}.

\begin{table}[htbp]
\centering
\small
\renewcommand{\arraystretch}{1.2}
\caption{\label{freqE} Distribution of vacuum energy values of $10^4$ phenomenological $\cN=0$ Class 3) models, the assosiated partition function, and a representitive GGSO matrix, for completely fixed models free from all triplets.}
\begin{tabular}{|>{\centering\arraybackslash}p{6cm}|
>{\centering\arraybackslash}m{5cm}|
>{\centering\arraybackslash}m{1.5cm}|
>{\centering\arraybackslash}m{1.5cm}|}
\hline
Representitive GGSO Matrix & Partition Function & Vacuum Energy / $\mathcal{M}^{4}$ & Frequency\\
\hline

{\scriptsize
$\begin{array}{c|ccccccccc}
 & \mathbf{1} & \bm{S} & \bm{b_1} & \bm{b_2} & \bm{b_3}
 & \bm{\alpha} & \bm{\beta} & \bm{\gamma} & \bm{\delta} \\ \hline
\mathbf{1}  & -1& -1& -1& -1& -1& 1& -1& -1& 1 \\
\bm{S}      &  -1& -1& -1& 1& 1& 1& -1& 1& -1 \\
\bm{b_1}    &  -1& 1& -1& -1& -1& -1& -1& -1& -1 \\
\bm{b_2}    &  -1& -1& -1& -1& -1& 1& -1& 1& -1 \\
\bm{b_3}    &  -1& -1& -1& -1& -1& -1& 1& 1& -1 \\
\bm{\alpha} &  1& 1& -1& 1& 1& -1& 1& 1& 1 \\
\bm{\beta}  &  -1& -1& -1& 1& 1& 1& 1& 1& -1 \\
\bm{\gamma} &  -1& 1& 1& 1& 1& 1& 1& 1& 1 \\
\bm{\delta} &  1& -1& 1& 1& 1& 1& -1& 1& 1
\end{array}$
}

& $ Z =  296 + 2\bar q^{-1} +24q\bar q^{-1} -32q^{\frac12}\bar q^{-\frac12} +3552q -45568q^{\frac12}\bar q^{\frac12} +10240q^{-\frac14}\bar q^{\frac34} -20480q^{\frac34}\bar q^{\frac34} +183168\bar q +2198016q\bar q$ 
& $-0.0280$ & 4,262 \\
\hline

{\scriptsize
$\begin{array}{c|ccccccccc}
 & \mathbf{1} & \bm{S} & \bm{b_1} & \bm{b_2} & \bm{b_3}
 & \bm{\alpha} & \bm{\beta} & \bm{\gamma} & \bm{\delta} \\ \hline
\mathbf{1}  & -1& 1& -1& -1& -1& 1& -1& 1& -1 \\
\bm{S}      &  1& 1& -1& -1& 1& -1& -1& 1& 1  \\
\bm{b_1}    &  -1& 1& -1& -1& -1& -1& 1& 1& -1 \\
\bm{b_2}    &  -1& 1& -1& -1& -1& -1& 1& -1& 1 \\
\bm{b_3}    &  -1& -1& -1& -1& -1& -1& -1& 1& 1 \\
\bm{\alpha} &  1& -1& -1& -1& 1& -1& 1& -1& 1 \\
\bm{\beta}  &  -1& -1& 1& -1& -1& 1& 1& 1& -1 \\
\bm{\gamma} &  1& 1& -1& -1& 1& -1& 1& -1& -1  \\
\bm{\delta} &  -1& 1& 1& -1& -1& 1& -1& -1& -1
\end{array}$
}

& {\centering
$Z =  296 +2\bar q^{-1} +56q\bar q^{-1} -64q^{\frac12}\bar q^{-\frac12}  +3168q + 2048q^{\frac14}\bar q^{\frac14} +128q^{-\frac12}\bar q^{\frac12} -36864q^{\frac12}\bar q^{\frac12} +7168q^{-\frac14}\bar q^{\frac34} -407552q^{\frac34}\bar q^{\frac34} +152448\bar q +1667584q\bar q$ \arraybackslash
}
& $-0.0372$ & 4,288 \\
\hline

{\scriptsize
$\begin{array}{c|ccccccccc}
 & \mathbf{1} & \bm{S} & \bm{b_1} & \bm{b_2} & \bm{b_3}
 & \bm{\alpha} & \bm{\beta} & \bm{\gamma} & \bm{\delta} \\ \hline
\mathbf{1}  & -1& -1& -1& 1& 1& 1& 1& -1& -1 \\
\bm{S}      &  -1& -1& -1& 1& 1& 1& 1& 1& 1 \\
\bm{b_1}    &  -1& 1& -1& -1& -1& -1& -1& 1& 1 \\
\bm{b_2}    & 1& -1& -1& 1& -1& 1& 1& -1& -1  \\
\bm{b_3}    &  1& -1& -1& -1& 1& 1& -1& -1& 1  \\
\bm{\alpha} &  1& 1& -1& 1& -1& -1& 1& 1& -1 \\
\bm{\beta}  &  1& 1& -1& -1& -1& 1& -1& -1& 1 \\
\bm{\gamma} &  -1& 1& -1& -1& -1& 1& -1& 1& 1 \\
\bm{\delta} &  -1& 1& -1& 1& -1& -1& 1& 1& -1
\end{array}$
}

& $ Z = -24 + 2\bar q^{-1} +56q\bar q^{-1}  -672q +2048q^{\frac14}\bar q^{\frac14} +128q^{-\frac12}\bar q^{\frac12} -24576q^{\frac12}\bar q^{\frac12} +9216q^{-\frac14}\bar q^{\frac34} -411648q^{\frac34}\bar q^{\frac34} +133248\bar q +1437184q\bar q$ 
& $0.0191$ & 1,450 \\
\hline
\end{tabular}

\end{table}

\section{Heavy Higgs}\label{heavyhiggs}

It is noted that aside from Class 0), the other classes do not accommodate states that could be used as heavy Higgs representations to break the Pati--Salam subgroup along supersymmetric flat directions in the $N=1$ spacetime supersymmetric models, which warrants further discussion. This restriction does not arise in the non--supersymmetric models, in which heavy Higgs representations are obtained from the $\bm{S}+\bm{b_i}$ sectors. For the purpose of this discussion we resort to the ``old--school" model building in which the NAHE--set provides a starting point. Three additional basis vectors are added to the NAHE--set and their primary function is to reduce the degeneracy of the states from the $\bm{b_i}$ sectors, and producing one generation from each of the sectors $\bm{b_1}$, $\bm{b_2}$ and $\bm{b_3}$. Our discussion below focuses on the assignment of boundary conditions for the set of internal fermions $\{y,\omega\vert {\bar y},{\bar\omega}\}^{1,\dots,6}$, which fixes the number of light chiral generations. Additionally, an $\bm{x}$--like vector is assumed that fixes the $U(1)_{1,2,3}$ horizontal charges. We illustrate the issue by focusing on a few examples. We note that the first two added basis vectors, $\bm{b_4}$ and $\bm{b_5}$ preserve the $SO(10)$ in order to produce states that could serve as heavy Higgs representations. Additionally, the states arising from these sectors should not add to the net chirailty as three generations are already obtained by construction from the sectors $\bm{b_{1,2,3}}$. We emphasise that the discussion here is intended to illustrate the issue of heavy Higgs multiplets, rather than a complete analysis. 

\begin{eqnarray}
&\begin{tabular}{c|c|ccc|c|ccc|c}
 ~ & $\psi^\mu$ & $\chi^{12}$ & $\chi^{34}$ & $\chi^{56}$ &
        $\bar{\psi}^{1,...,5} $ &
        $\bar{\eta}^1 $&
        $\bar{\eta}^2 $&
        $\bar{\eta}^3 $&
        $\bar{\phi}^{1,...,8} $ \\
\hline
\hline
  $\bm{b_4}$     &  1 & 1&0&0 & 1~1~1~1~1 & 1 & 0 & 0 & 0~0~0~0~0~0~0~0 \\
  $\bm{b_5}$     &  1 & 0&1&0 & 1~1~1~1~1 & 0 & 1 & 0 & 0~0~0~0~0~0~0~0 \\
  $\bm{\alpha}$  &  0 & 0&0&0 & 1~1~1~0~0 & 0 & 0 & 0 & 1~1~1~1~0~0~0~0 \\
\end{tabular}
\nonumber\\
   ~  &  ~ \nonumber\\
   ~  &  ~ \nonumber\\
     &
\begin{tabular}{c|c|c|c}
 ~&   $y^3{y}^6$
      $y^4{\bar y}^4$
      $y^5{\bar y}^5$
      ${\bar y}^3{\bar y}^6$
  &   $y^1{\omega}^6$
      $y^2{\bar y}^2$
      $\omega^5{\bar\omega}^5$
      ${\bar y}^1{\bar\omega}^6$
  &   $\omega^1{\omega}^3$
      $\omega^2{\bar\omega}^2$
      $\omega^4{\bar\omega}^4$
      ${\bar\omega}^1{\bar\omega}^3$ \\
\hline
\hline
$\bm{b_4}$  & 1 ~~~ 0 ~~~ 0 ~~~ 1  & 0 ~~~ 0 ~~~ 1 ~~~ 0  & 0 ~~~ 0 ~~~ 1 ~~~ 0 \\
$\bm{b_5}$  & 0 ~~~ 0 ~~~ 1 ~~~ 0  & 1 ~~~ 0 ~~~ 0 ~~~ 1  & 0 ~~~ 1 ~~~ 0 ~~~ 0 \\
${\bm\alpha}$ & 1 ~~~ 0 ~~~ 0 ~~~ 0 & 1 ~~~ 0 ~~~ 0 ~~~ 0 & 1 ~~~ 0 ~~~ 0 ~~~ 0 \\
\end{tabular}
\label{modelex}  
\end{eqnarray}

In the model of eq. (\ref{modelex}) the twisted sectors $\bm{b}_{4, 5}$ give rise a priori to additional $16$ and $\overline{16}$ $SO(10)$ representations. As noted from eq. (\ref{modelex}), three left-- and right--moving internal real fermions are complexified in this model, giving rise to flavour $U(1)_{4,5,6}$. All the geometrical moduli are retained in this model which therefore belongs to Class 0), with $M=(4,4,4)$. The three untwisted triplet pairs are projected out due to the asymmetry of the $\bm{\alpha}$, whereas the corresponding untwisted bi--doublets are retained. However, we note from (\ref{modelex}) that the overlap of the internal periodic fermions $\{y,\omega\vert{\bar y}, {\bar\omega}\}^{1,\dots,6}$ between the two vectors is empty. This entails that the chirality of the states produced in the two sectors is the same \cite{NAHE2}. Hence, the two vectors can produce components of either the $16$ or the $\overline{16}$ $SO(10)$ representations, but not of the $16$ and the $\overline{16}$. Hence, they affect the net chirality and cannot produce non--chiral heavy Higgs multiplets. However this model is compatible with vectors $\bm{e}_2$, $\bm{e}_4$ and $\bm{e}_5$, the addition of which increases the number of spinorial sectors to 12. This additional freedom permits three generational SUSY models that contain heavy Higgs states.
This observation is compatible with the classification of class 0) models.  

To contrast this with classes of reduced moduli, consider the model in eq. (\ref{modelex2}), where only the boundary conditions of the internal fermions are displayed, 

\begin{eqnarray}
     &
\begin{tabular}{c|c|c|c}
 ~&   $y^3{y}^6$
      $y^4{\bar y}^4$
      $y^5{\bar y}^5$
      ${\bar y}^3{\bar y}^6$
  &   $y^1{\omega}^5$
      $y^2{\bar y}^2$
      $\omega^6{\bar\omega}^6$
      ${\bar y}^1{\bar\omega}^5$
  &   $\omega^2{\omega}^4$
      $\omega^1{\bar\omega}^1$
      $\omega^3{\bar\omega}^3$
      ${\bar\omega}^2{\bar\omega}^4$ \\
\hline
\hline
$\bm{b_4}$  & 0 ~~~ 1 ~~~ 1 ~~~ 0  & 0 ~~~ 0 ~~~ 1 ~~~ 0  & 0 ~~~ 0 ~~~ 1 ~~~ 0 \\
$\bm{b_5}$  & 0 ~~~ 0 ~~~ 1 ~~~ 0  & 0 ~~~ 1 ~~~ 1 ~~~ 0  & 0 ~~~ 1 ~~~ 0 ~~~ 0 \\
${\bm\alpha}$ & 0 ~~~ 1 ~~~ 0 ~~~ 1 & 0 ~~~ 1 ~~~ 0 ~~~ 1 & 1 ~~~ 0 ~~~ 0 ~~~ 0 \\
\end{tabular}
\label{modelex2}  
\end{eqnarray}
Here, the corresponding overlap is not empty and the sectors $\bm{b_{4,5}}$ produce non--chiral multiplets. This model belongs to Class 3) and all the geometrical moduli are projected out. The untwisted triplet pairs are projected out, whereas the untwisted bi--doublets are retained. However, it is seen that the product $\bm{b_1}\cdot\bm{\alpha}= -4$, whereas $\bm{b_4}\cdot\bm{\alpha}= -3$, and similarly for $\bm{b_2}$ and $\bm{b_5}$, and there is no way to reconcile this difference. In this case there are no compatible asymmetric shift vectors which are compatible with the model, and this problem can't be circumvented. Hence, in this case inclusion of heavy Higgs states cannot be reconciled with the modular invariance constraints. The contradiction may therefore stem from the projection of all the untwisted colour triplet pairs, which necessitate the pairing of real fermion from the set $\{y,\omega\vert{\bar y}, {\bar\omega}\}^{1,\dots,6}$ into three complex fermions and the asymmetric assignment in $\bm{\alpha}$ that projects all the untwisted triplets. To illustrate how this situation changes with only a single complexified fermion, which is the case in the models of ref. \cite{alr, lr}, we consider the model in eq. (\ref{modelex3})
\begin{eqnarray}
     &
\begin{tabular}{c|c|c|c}
 ~&   $y^3{\bar y}^3$
      $y^4{\bar y}^4$
      $y^5{\bar y}^5$
      ${y}^6{\bar y}^6$
  &   $y^1{\bar y}^1$
      $y^2{\bar y}^2$
      $\omega^5{\bar\omega}^5$
      ${\omega}^6{\bar\omega}^6$
  &   $\omega^2{\omega}^4$
      $\omega^1{\bar\omega}^1$
      $\omega^3{\bar\omega}^3$
      ${\bar\omega}^2{\bar\omega}^4$ \\
\hline
\hline
$\bm{b_4}$  & 0 ~~~ 1 ~~~ 1 ~~~ 0  & 0 ~~~ 0 ~~~ 0 ~~~ 1  & 0 ~~~ 0 ~~~ 1 ~~~ 0 \\
$\bm{b_5}$  & 0 ~~~ 0 ~~~ 1 ~~~ 1  & 0 ~~~ 1 ~~~ 0 ~~~ 0  & 0 ~~~ 1 ~~~ 0 ~~~ 0 \\
${\bm\alpha}$ & 0 ~~~ 1 ~~~ 0 ~~~ 0 & 1 ~~~ 1 ~~~ 0 ~~~ 0 & 1 ~~~ 1 ~~~ 0 ~~~ 0 \\
\end{tabular}
\label{modelex3}  
\end{eqnarray}
In the model generated by eq. (\ref{modelex3}) there is a single complexified fermion and therefore its spectrum contains two untwisted triplet pairs and a single bi--doublet Higgs representation. The overlap of real fermions between $\bm{b_4}$ and $\bm{b_5}$ is not empty and therefore the states arising from these sectors can have opposite chirality and give rise to heavy Higgs representations. 
We remark here that a similar situation is found in the Standard--Like models of ref. \cite{fny, Faraggi:1991jr, Faraggi:1991be, slm2}. In this models the heavy Higgs states are the Standard Model singlets in the $16$ and $\overline{16}$ $SO(10)$ representations, which break the $U(1)$ symmetries to the weak hypercharge $U(1)_Y$ combination. The scalar singlets in the $16$ representation are obtained in these models from the sectors $\bm{b_{1,2,3}}$. However, the corresponding $\overline{N}$ states from the $\overline{16}$ representation does not exist in these models. In the Standard--like Models, exotic states that arise from sectors that break the $SO(10)$ symmetry to the Standard Model are used to generate supersymmetric flat directions \cite{Faraggi:1991jr, slm2, cfn}. These states are Standard Model singlets and carry exotic charges with respect to the extra $SO(10)$ $U(1)$ symmetry, which is orthogonal to the Standard Model gauge symmetry. This solution is not afforded in the Pati--Salam string models since the heavy Higgs representations are obtained from standard $SO(10)$ representations. It is further noted that in the symmetric classification of the Standard--like Models, it was found that the vacua which admit the $SO(10)$ $\overline{N}$ state are rare and a special fertility algorithm is required to extract them \cite{slmclass}.

\section{Conclusion}\label{Conc}

In this paper we have extended the free fermionic $\ztwo$
classification programme to Pati--Salam heterotic string vacua with asymmetric
boundary conditions. The inclusion of asymmetric orbifold actions introduces a
layer of complexity absent from the symmetric case: different asymmetric
assignments of the Pati--Salam breaking vector $\Bga$ lead to inequivalent
residual moduli spaces, which in turn determine which additional basis vectors
are compatible with modular invariance. A systematic enumeration of all
modular-invariant choices of $\Bga$, yields 24 inequivalent classes organised by the number of
surviving real geometric moduli (12, 8, 4 or 0), the distribution of asymmetric twists and shifts across the three internal tori, and the number of retained untwisted Higgs doublets and triplets (Table \ref{tb:Bga_class_B_2}). Through constructing a general definition of the degeneracy of spinorial sectors, representative basis sets admitting three chiral generations have been constructed in Table \ref{tab:AddBasVec},
providing the starting point for phenomenological scans within each subspace.

The asymmetric Pati--Salam vector serves a triple role: it breaks the SO(10) GUT symmetry to the Pati--Salam subgroup, it freezes geometric moduli at the free fermionic point and it implements a doublet--triplet splitting mechanism in the untwisted (NS) sector. This last feature is of particular phenomenological
significance, since in the symmetric case, electroweak doublets are
generically forced to arise from twisted sectors. The doublet--triplet splitting
mechanism is governed by left--right asymmetric boundary conditions which can be associated with an asymmetric twist or shift action. It is
therefore a general consequence of asymmetric boundary conditions, and independent of moduli stabilisation.

The complete removal of untwisted colour triplets from the massless spectrum is found to be incompatible with the presence of the $\bm{x}$ vector and hence with
the GUT enhancement $\mathrm{SO}(10)\to E_6$, establishing a non-trivial link between the doublet--triplet splitting mechanism and the embedding structure of the GUT group. Furthermore, our analysis across the example model classes
suggests that projecting all three untwisted triplets is incompatible with the presence of a Pati--Salam-breaking Heavy Higgs state from the spinorial sectors, pointing to a tension between maximal triplet projection and the mechanism for further gauge symmetry breaking to the Standard Model. The interplay between these constraints in other constructions warrants further investigation.

Explicit phenomenological classifications have been carried out in four
representative model classes with 12, 8, 4 and 0 geometric moduli, scanning samples of $10^9$ GGSO phase configurations in each case.  In the Class 0) (12 geometric moduli) case, exophobic models satisfying all criteria were
found for both $\cN=1$ and $\cN=0$ vacua. 
By contrast, the basis sets employed
for the Class 1b) (8 moduli) and Class 2a) (4 moduli) examples both lacked phenomenological viable, SUSY vacua with a Pati--Salam breaking (Heavy) Higgs. This obstruction is not
intrinsic to the moduli classes but rather reflects a limitation of the
specific basis sets chosen. More generally, identifying the optimal basis sets within each of the 24 classes remains an important open problem requiring more in depth analysis, especially of how additional basis vectors impact on the phenomenological constraints for the presence of a Heavy Higgs and exotics. The Class 3) construction, in which composite vectors formed from sums of the simplest symmetric shifts in Table \ref{tab:AddBasVec} are used to circumvent exotic-sector obstructions, yields phenomenologically viable models and provides a concrete framework for further analysis.

A particularly striking finding is the pronounced degeneracy of partition functions among models satisfying the phenomenological constraints as asymmetric models with fewer geometric moduli are considered. In the Class 0) example, a sample of $10^9$ GGSO
configurations yielded 138 exophobic $\mathcal{N}=0$ models with 75 distinct
partition functions (computed at the free fermionic point). Moving to Class 3), $10^4$ phenomenologically viable models collapsed to just 5
distinct partition functions. This dramatic reduction suggests that the landscape of asymmetric (free fermionic) string vacua has a qualitatively different structure from the symmetric counterpart: rather than the GGSO phases (and therefore discrete torsions) defining a large space of distinct configurations, the
strong constraints imposed by asymmetric boundary conditions, channel the GGSO phase space into a small number of physically distinct theories. 
We stress, however, that matching partition functions to a given order in the $q$-expansion does not establish full equivalence of the underlying string vacua, which would require a detailed comparison of all state U(1) charges and couplings.

This collapse of the phenomenologically viable landscape to a small
number of distinct theories also has implications for the choice of
computational search strategy. In symmetric constructions, where the
space of viable vacua is large and diverse, machine learning
approaches trained on GGSO phase matrices are well suited to
recognising patterns and classifying the statistical features of the
landscape. In the asymmetric case, however, the phenomenological
constraints are sufficiently restrictive that the problem shifts from
one of classification within a rich solution space to one of
constraint satisfaction within a tightly constrained one. For such
problems, targeted algorithms such as SAT/SMT
solvers \cite{fpsw} and genetic algorithms \cite{genalgo}, which
are designed to navigate efficiently toward solutions of constraint systems, are likely better adapted than
broad-sampling approaches.

Several directions for future work present themselves. First, the
classification of asymmetric actions performed here for the Pati--Salam
subgroup can be extended to other phenomenologically relevant SO(10) subgroups, in particular to Standard-like and to Left--Right Symmetric models. Second, translating the 24 cases of Table \ref{tb:Asymm_Bga_actions} to the bosonic orbifold language, using the dictionaries developed
in \cite{z2z25,florakis2023freefermionicconstructionsorbifolds}, would identify the Narain
T-fold compactification data and the generalised geometry underlying each class,
following the framework of \cite{SGNPKV}. Such a translation may also help clarify the origin of the partition function degeneracy observed here, and what equivalences it points to of the underlying T-fold. Third, this translation to the bosonic description would also allow for the residual moduli dependence and type of supersymmetry breaking to be analysed along the lines of \cite{fr1,fr2,ADM,AFMP1}. Furthermore, it must always be remembered that the dilaton vacuum expectation value remains unfixed in all models, generating a runaway potential and destabilising the background of the theory. Any discussion of vacuum stability for non-supersymmetric string is fraught with uncertainties, however we can mention the non-perturbative racetrack mechanism \cite{Denef:2004dm} as a possible way we may stabilise the dilaton. An alternative approach may be the Banks-Zaks stabilisation explored in \cite{Abel:2024vov} and it was also highlighted in \cite{AFMP1} that the additional tadpoles from the presence of an anomalous $U(1)$ will generically further modify the stability analysis. 
One tangential direction to this would be to examine the (one-loop)
vacuum energies of the $\cN=0$ models uncovered here in light of
the various non-supersymmetric swampland conjectures, such as the
AdS instability conjecture, which posits that all
non-supersymmetric AdS vacua are at best
metastable \cite{ooguri2017nonsupersymmetricadsswampland}. 

\section*{Acknowledgements}
The work of LD is supported in part by STFC grant 2890873, and in part by 
LIV.INNO grant ST/W006766/1. 
The work of AEF is supported by the STFC Consolidated Grant ST/X000699/1.

\clearpage
\appendix

\section{Modular invariance rules}\label{app:ABKrules}

Basis vectors must preserve modular invariance, and so are constructed according to the rules:
\begin{align}
\begin{split}
\label{eq:ABKbvecs}
N_{i} \bm{v_i} \cdot \bm{v_i} &= 0 \text{  mod } 8 \text{,}\\
N_{ij} \bm{v_i} \cdot \bm{v_j} &= 0 \text{  mod } 4 \text{,}\\
\sum_{f}  \bm{v_i}(f)  \bm{v_j}(f)  \bm{v_k}(f)  \bm{v_l}(f) &= 0 \text{  mod } 2 
\end{split}
\end{align}
where \( N_i \) is the smallest positive integer such that  
\(N_i \bm{v_i} = 0\) and \( N_{ij} \) is the lowest common multiple of \( N_i \) and \( N_j \). In the case of real boundary conditions \( N_i = N_{ij} = 2\) for all \(i,j\).

We have also defined the Lorentzian dot product is defined as
\beq
\bm{v_i} \cdot \bm{v_j} 
= \left\{ 
   \left( \sum_{\text{cx. Left}} + \tfrac{1}{2}\sum_{\text{real Left}} \right) 
   - \left( \sum_{\text{cx. Right}} + \tfrac{1}{2}\sum_{\text{real Right}} \right) 
  \right\} 
  \bm{v_i}(f)\, \bm{v_j}(f).
\eeq
Modular invariance imposes the following rules on the GGSO phases
\begin{align}
C\!\begin{bmatrix} \bm{\alpha} \\ \bm{\beta} \end{bmatrix}
&= \delta_{\bm{\alpha}}\, e^{2\pi i\, n_{N_{\bm{\beta}}}}
= \delta_{\bm{\beta}}\, e^{2\pi i\, m_{N_{\bm{\alpha}}}} \,
e^{\,\tfrac{i\pi}{2}\, \bm{\alpha} \cdot \bm{\beta}}, \\
C\!\begin{bmatrix} \bm{\alpha} \\ \bm{\alpha} \end{bmatrix}
&= -\, e^{\,\tfrac{i\pi}{4}\, \bm{\alpha} \cdot \bm{\alpha}}\,
C\!\begin{bmatrix} \bm{\alpha} \\ \bm{1} \end{bmatrix}, \\
C\!\begin{bmatrix} \bm{\alpha} \\ \bm{\beta} \end{bmatrix}
&= e^{\,\tfrac{i\pi}{2}\, \bm{\alpha} \cdot \bm{\beta}}\;
C\!\begin{bmatrix} \bm{\beta} \\ \bm{\alpha} \end{bmatrix}^{\!*}, \\
C\!\begin{bmatrix} \bm{\alpha} \\ \bm{\beta} + \bm{\gamma} \end{bmatrix}
&= \delta_{\bm{\alpha}}\;
C\!\begin{bmatrix} \bm{\alpha} \\ \bm{\beta} \end{bmatrix}\;
C\!\begin{bmatrix} \bm{\alpha} \\ \bm{\gamma} \end{bmatrix}.
\end{align}

\clearpage

\clearpage 
\printbibliography[heading=bibintoc]

\end{document}